\documentclass[preprint2]{exoplanet}
\usepackage{amsmath}
\usepackage{times}
\usepackage{graphicx}
\usepackage{longtable}
\usepackage[abs]{overpic}
\usepackage{hyperref}
\hypersetup{colorlinks=true,citecolor=blue,linkcolor=blue}

\newcommand{\M}{M_*}
\newcommand{\cve}{x}
\newcommand{\cveq}{x^2}
\newcommand{\vv}[1]{\boldsymbol{#1}}

\def\apjl{Astrophys.\ J. }
\def\apjs{Astrophys.\ J.\ (Supp.) }

\def\aap{Astron.\ Astrophys. }

\begin{document}

\title{\textbf{\LARGE Spin evolution of Earth-sized exoplanets, \\ 
   including atmospheric tides and core-mantle friction}}

\author {\textbf{\large Diana Cunha}}
\affil{\small\em Centro de Astrof\'{i}sica da Universidade do Porto, \\ Rua das Estrelas, 4150-762 Porto, Portugal.}
\affil{\small\em Departamento de F\'{i}sica e Astronomia, Faculdade de Ci\^encias, \\ Universidade do Porto, Portugal.}

\author {\textbf{\large Alexandre C.M. Correia}}
\affil{\small\em Departamento de F\'{i}sica, I3N, Universidade de Aveiro, \\ Campus de Santiago, 3810-193 Aveiro, Portugal.}
\affil{\small\em ASD, IMCCE-CNRS UMR8028, Observatoire de Paris, \\ 77 Av. Denfert-Rochereau, 75014 Paris, France.}

\author {\textbf{\large Jacques Laskar}}
\affil{\small\em ASD, IMCCE-CNRS UMR8028, Observatoire de Paris, \\ 77 Av. Denfert-Rochereau, 75014 Paris, France.}

\begin{abstract}
\begin{list}{ }
{\rightmargin 0.4in}
\baselineskip = 11pt
\parindent=1pc
{\small 
Planets with masses between $0.1 - 10 \,M_\oplus$ are believed to host 
dense atmospheres. These atmospheres can play an important role on the planet's spin evolution,  since thermal atmospheric tides, driven by the host star, may counterbalance gravitational tides.
In this work we study the long-term spin evolution of Earth-sized exoplanets.
We generalize previous works by including the effect of eccentric orbits and obliquity.
We show that under the effect of tides and core-mantle friction, the obliquity of the planets evolve either to $0^\circ$ or $180^\circ$.
The rotation of these planets is also expected to evolve into a very restricted number of equilibrium configurations.
In general, none of this equilibria is synchronous with the orbital mean motion.
The role of thermal atmospheric tides becomes more important for Earth-sized planets in the habitable zones of their systems, so they cannot be neglected when we search for their potential habitability.
 \\~\\~\\~}
 
\end{list}
\end{abstract}

\section{Introduction}

In \citeyear{Mayor_Queloz_1995}, \citeauthor{Mayor_Queloz_1995} detected the first exoplanet orbiting a Sun-like star. This was just the first of many, and ten years later
 more than 150 exoplanets were already known. It  was at that time that a planet with a mass lower than 
$10\,M_\oplus$ was found orbiting the low-mass star GJ\,876 \citep{Rivera_etal_2005}. This was the first ``Earth-sized'' planet to be found, also called ``Super-Earth''.

This new class of planets with mass lower than $10\,M_\oplus$,
 is believed to have an icy/rocky core surrounded by  a dense gaseous envelope \citep{Alibert_etal_2006}.
 As an example, for a planet with a mass of $8\,M_\oplus$, \citet{Rafikov_2006} 
showed that $\sim 87.5\%$ of its mass may be made of  rock,  while $\sim12.5\%$ may be a H$_2-$He  gaseous envelope. 
We may compare these results with the values of the planet Earth,
 whose atmosphere represents only  $0.0001\%$ of the total mass, and for  Venus, 
whose atmosphere/mass ratio is  $ 0.01\%$. 

There are many differences between the so-called ``Super-Earths''  and the telluric planets of the Solar System, and also much to  be discovered about these new worlds.
 Still, the use of our knowledge on the telluric planets to study the properties of similar  exoplanets is of major interest. 
In particular, because  some of these planets may be orbiting in the habitable zone (HZ)  \citep[see e.g.][]{Udry_etal_2007}, 
the study of their spin evolution  to infer their possible climate becomes important in the search of life elsewhere in the Universe.
Although most of these planets will be quite different from the Earth, in the remaining of the paper, the term {\it Earth-sized planet}   denotes
a planet with mass smaller than $10\,M_\oplus$ orbiting in the vicinity of the  HZ, or closer to their parent star.

The HZ is defined as a range of distance to the star where the insolation received by an
Earth-sized planet is adequate to allow its surface to maintain liquid water \citep{Kasting_1993b, Kopparapu_etal_2013, Kopparapu_etal_2013e}. 
But to say that a planet is in the HZ is not the same as to  say a planet is habitable. 
There are other factors, as the obliquity, eccentricity, atmosphere composition, or tidal effects, that should also be considered when assessing habitability. 

Several efforts have been made to explore these and other factors that may contribute to
the habitability of an exoplanet.
As an example, \citet{Barnes_2013} show that tidal heating can induce a runaway greenhouse on explanets orbiting low-mass stars,
which may cause all the hydrogen to escape, and so may all the water.
 On the other hand, modeling by \citet{Barnes_etal_2008} show that Earth-sized exoplanets orbiting in the HZ of an M-dwarf can support life. 
These planets may have strong enough magnetic fields that might protect their atmospheres and surfaces. 
But to infer about the habitability of a planet, there is an effect that should not be neglected: tidal dissipation. 
Tidal interactions will contribute to the planetary spin evolution, having an impact on the final obliquity, spin rate, and eccentricity \citep[e.g.][]{Heller_2011}. 

Since Earth-sized  exoplanets are believed to have dense atmospheres, 
when studying their spin evolution, we have to consider two kinds of tidal effects: the traditional bodily tides of gravitational origin (gravitational tides), but also thermal atmospheric tides. 
The former tends to despin the planet, while the latter may counteract the gravitational tidal effect on the planet's rotation \citep{Gold_Soter_1969}.
  In the Solar System, Venus and the Earth are  the only telluric planets having a significant atmosphere.
From these two, only Venus is believed to have  reached a final equilibrium rotation rate \citep{Dobrovolskis_1980,Correia_Laskar_2001,Correia_Laskar_2003I}.
 Because all  known  Earth-sized  exoplanets are  closer to the parent star than Venus is to the Sun, it is expected that they have also reached rotational equilibrium \citep{Laskar_Correia_2004,Correia_Levrard_2008}. 
Therefore, their environments are probably more similar to Venus than to the Earth.
We can thus use the  model developped for the undestanding of the spin evolution of Venus \citep{Correia_etal_2003, Correia_Laskar_2003I}  for investigating the possible spin evolution of the Earth-sized exoplanets.

In this work we expect to be able to infer about the present rotation of Earth-sized planets. 
Unlike Venus, whose orbit is almost circular, most of these exoplanets have non-zero eccentricities. Thus, the above mentioned  model for Venus' rotation needs to be  generalized in order to include the effect of the orbital eccentricity.
We also assess  if these planets can only evolve to  final obliquity of $0^\circ$ and $180^\circ$ \citep{Correia_etal_2003}, or if they can present  intermediate stable obliquities.
In section~2 we present the equations of motion  that describe the long-term spin evolution of a terrestrial planet. 
We also describe the contribution of the main dissipative effects: the gravitational tides, thermal atmospheric tides, and core-mantle friction.
In section~3 we  present a dynamical analysis for the spin evolution and for the final equilibrium rotation states.
In section~4 we present numerical simulations for the spin evolution starting with different initial rotation periods and obliquities.
We apply our model to the known Earth-sized planets as well as fictitious Earth-sized planets in the HZ of Sun-like stars. 
We end this study by presenting our conclusions in section~5.


\section{Equations of motion}

\subsection{Conservative motion}
 

\begin{figure}[t!]
\begin{center}
\includegraphics[width=.96\columnwidth]{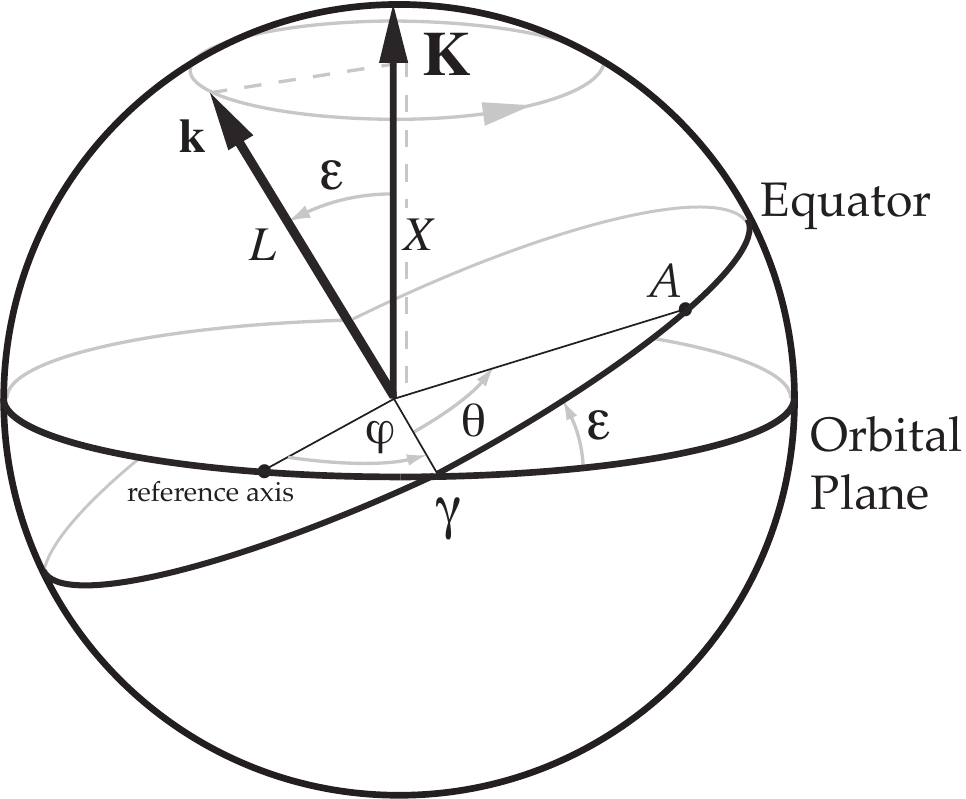}
  \caption{\label{variables} Andoyer's canonical variables. $ L $ is the projection of the total rotational angular momentum vector $ \vv{L} $ on the principal axis of inertia $ \vv{\mathrm{k}} $, and $ X $ the projection of the angular momentum vector on the normal to the orbit $ \vv{\mathrm{K}} $. The angle between the line of nodes $\gamma$ and a fixed point of the equator $ A $ is the hour angle, $ \theta $, while the angle between a reference point in the orbit and the line of nodes $\gamma$ is the precession angle, $\varphi $.}  
\end{center}
\end{figure}

The  Hamiltonian of the motion can be written using the classical canonical Andoyer's action  
and angle  variables \citep{Andoyer_1923, Kinoshita_1977} but these variables becomes degenerate 
due to a simplifying assumption that is done here:
The planet is considered to be a rigid body with moments of inertia $A = B<C$, 
and we merge the figure axis 
with the direction of the angular momentum (gyroscopic approximation)\footnote{The figure axis of a solid is its principal axis of inertia with maximum momentum of inertia. In the case of the Earth, 
the angle between the axis of figure and the angular momentum is of the order of $7\times 10^{-7}$ radians. }.
We are then left with only two action variables  $(L,X)$ and their conjugate angles $(\theta, \varphi)$ \citep{Surgy_Laskar_1997}\footnote{In the section 2 in \citet{Surgy_Laskar_1997}, the actions are the same $(L,X)$, but the notations  for the conjugate angles are $(l,-\psi)$ instead of $(\theta, \varphi)$.}. The quantity  $L=C\omega$, with rotation rate $\omega$, is the  the angular momentum along the $C$ axis, and $X= L\cos \varepsilon$, where $\varepsilon$ is the obliquity,  is  its projection  on the normal to the quasi-inertial ecliptic plane. The variable 
$\theta$ is the hour angle between the equinox 
and a fixed point of the equator, and $\varphi$ is the precession angle (Figure~\ref{variables}).
Averaging the Hamiltonian over the rotation angle and the mean anomaly,
we get\footnote{Since $A=B$, we neglect the effect from spin-orbit resonances. 
\citet{Correia_Laskar_2003I} have shown that thermal tides drive the spin away from these equilibria. 
A full description of the Hamiltonian with $A \ne B$ can be found in \citet{Correia_Laskar_2010}.} 
\citep{Kinoshita_1977, Surgy_Laskar_1997,Correia_Laskar_2010B}
\begin{eqnarray}
 \overline {\cal H} & = & \frac{L^2}{2 C} - \alpha \frac{X^2}{2 L} \ , \label{130418a}
\end{eqnarray} 
where
\begin{equation}
 \alpha= \frac{3G\M}{2 a^3(1-e^2)^{3/2}}\frac{E_d}{\omega}\simeq \frac{3}{2}\frac{n^2}{\omega} \left(1 + \frac{3}{2} e^2 \right) E_d
\end{equation}
is the ``precession constant'', while quantities
$G$, $\M$, $a$, $n$, and $e$ are the gravitational constant, stellar mass, semi-major axis, mean motion, and eccentricity, respectively. 
Quantity $E_d$ is the dynamical ellipticity\footnote{When $A\neq B$, after averaging over the fast rotation angles we can replace $A$ by $(A+B)/2$ \citep[e.g.][]{Boue_Laskar_2006}.},
\begin{equation}
 E_d=\frac{C-A}{C}=\frac{k_f R^5}{3GC}\omega^2 + \delta E_d \ ,
\end{equation}
where $R$ is the planet radius, and $k_f$ is the fluid Love number. The first part of this expression corresponds to the 
flattening in hydrostatic equilibrium \citep{Lambeck_1980}, 
and the second corresponds to the departure from this equilibrium. 

Since Andoyer's variables are canonical, the spin equations of motion are easily obtained from the mean Hamiltonian (Eq.\ref{130418a}) as
\begin{equation}\label{eq:dLdt}
\frac{dL}{dt}=- \frac{\partial  \overline {\cal H}}{\partial \theta} \ , \quad \frac{dX}{dt}=-\frac{\partial  \overline {\cal H}}{\partial \varphi} \ , \quad \frac{d\varphi}{dt}= \frac{\partial  \overline {\cal H}}{\partial X} \ ,
\end{equation}
which gives
\begin{equation}\label{130418b}
\frac{dL}{dt}=  \frac{dX}{dt}= 0 \ , \quad \mathrm{and} \quad \frac{d\varphi}{dt}= -\alpha \cos \varepsilon \ ,
\end{equation}
that is, the rotation rate and the obliquity are constant, and the planet precesses at a constant rate.

 \subsection{Tidal effects}
 Tidal effects arise from planetary differential and inelastic deformations caused by a perturbing body. 
There are two types of tidal effects: the gravitational tides and the thermal atmospheric tides. The estimations for both  
effects are based on a general formulation of the tidal potential, initiated by \citet{Darwin_1880}. 

 \subsubsection{Gravitational tides}\label{sec:eqmotiomgrav}

Gravitational tides are raised on the planet by a perturbing body because of 
the gravitational gradient across the planet.
The force experienced by the side facing the perturbing body is stronger than
that experienced by the far side.
These tides are mainly important upon the solid (or liquid) part of the
planet, and are independent of the existence of an atmosphere.

Since the planets are not perfectly rigid, there
is a distortion that gives rise to a tidal bulge.
This redistribution of mass modifies the gravitational potential
generated by the planet in any point of the space.
The additional amount of potential, the tidal potential $ U_g $, is responsible
for the modifications in the planet's spin (and orbit), and
it is given by\footnote{We neglect terms in $(R/r)^4$.} \citep[e.g.][]{Lambeck_1980}:
\begin{equation}
U_g = - k_2 \frac{G \M^2}{R} \left(
\frac{R}{r_*} \right)^3 \left( \frac{R}{r} \right)^3 P_2 (\cos
S) \, , \label{V6} 
\end{equation} 
where  $r $ and $ r_* $ are the distances from the planet's center of mass to
a generic point and to the star, respectively, $ S $ is the angle between these two directions,  $ P_2 $ are the second order Legendre polynomials, and
$ k_2 $ is the second potential Love number. 

Expressing the tidal potential given by expression (\ref{V6}) in terms of Andoyer angles $( \theta, \varphi )$, we can obtain the contribution to the spin evolution from expressions (\ref{eq:dLdt}) using $ U_g $ at the place of $ \overline {\cal H} $.
As we are interested here in the study of the secular evolution of the spin, we also average $ U_g $ over the periods of mean anomaly and longitude of the periapse of the orbit. 
This work is done with the help of the algebraic manipulator TRIP \citep{Laskar_1989t,Laskar_1994t}, which expands the potential in Fourier series, as in
\citet{Kaula_1964}  and \citet{Correia_Laskar_2010B}.
For a planet orbiting its host star, where the star is both the perturbing and interacting body ($r=r_*$), we then find for the averaged equations of motion:

\begin{equation}
\frac{d L}{d t} = K_g \sum _\sigma b_g (\sigma)\Lambda_\sigma^g(x,e) \ , 
\label{eq:variacaospinL}
\end{equation}
\begin{equation}
\frac{d X}{d t} = K_g \sum _\sigma b_g (\sigma)\Upsilon_\sigma^g(x,e) \ , 
\label{eq:variacaospinX}
\end{equation}
where 
\begin{eqnarray}
K_g = -\frac{G \M^{2} R^5}{a^6} \ , \label{130613a}
\end{eqnarray}
and  $x=X/L=\cos \varepsilon$.
The coefficients $\Lambda_ \sigma ^g $ and $\Upsilon_ \sigma ^ g$ are polynomials
in the eccentricity \citep{Kaula_1964}. 
For a planet with moderate eccentricity, we may neglect terms in $e^4$ and greater, and obtain the following expansion for the previous equations:
\begin{eqnarray}
\frac{1}{K_g} \frac{d L}{d t}&=& b_g\left(2\omega+3n\right)\frac{147}{128}e^2\left(1 -x\right)^4\nonumber\\
&+&  b_g\left(2\omega+2n\right)\frac{3}{32}\left(1-5e^2\right)\left(1-\cve\right)^4\nonumber\\
&+& b_g\left(2\omega+n\right)\frac{3}{128}e^2\left(37+70\cve+37\cveq\right)\left(1-\cve\right)^2\nonumber\\
&+& b_g\left(2\omega\right)\frac{3}{8}\left(1+3e^2\right)\left(1-\cveq\right)^2\nonumber\\
&+& b_g\left(2\omega-n\right)\frac{3}{128}e^2\left(37-70\cve+37\cveq\right)\left(1+\cve\right)^2\nonumber\\
&+& b_g\left(2\omega-2n\right)\frac{3}{32}\left(1-5e^2\right)\left(1 +\cve\right)^4\nonumber\\
&+& b_g\left(2\omega-3n\right)\frac{147}{128}e^2\left(1 +\cve\right)^4\nonumber\\
&+& b_g\left(\omega+3n\right)\frac{147}{64}e^2\left(1-\cveq\right)\left(1-\cve\right)^2\nonumber\\
&+& b_g\left(\omega+2n\right)\frac{3}{16}\left(1-5e^2\right)\left(1-\cveq\right)\left(1-\cve\right)^2\nonumber\\
&+& b_g\left(\omega+n\right)\frac{3}{64}e^2\left(1-2\cve+37\cveq\right)\left(1-\cveq\right)\nonumber\\
&+& b_g\left(\omega\right)\frac{3}{4}\left(1+3e^2\right)\left(1-\cveq\right)\cveq\nonumber\\
&+& b_g\left(\omega-n\right)\frac{3}{64}e^2\left(1+2\cve+37\cveq\right)\left(1-\cveq\right)\nonumber\\
&+& b_g\left(\omega-2n\right)\frac{3}{16}\left(1-5e^2\right)\left(1-\cveq\right)\left(1+\cve\right)^2\nonumber\\
&+& b_g\left(\omega-3n\right)\frac{147}{64}e^2\left(1-\cveq\right)\left(1+\cve\right)^2 \ ,
\label{eq:Tg1}
\end{eqnarray}
and

\begin{eqnarray}
\frac{1}{K_g} \frac{dX}{dt}&=& - b_g\left(2\omega+3n\right)\frac{147}{128}e^2\left(1-\cve\right)^4\nonumber\\
&-& b_g\left(2\omega+2n\right)\frac{3}{32}\left(1-5e^2\right)\left(1-\cve\right)^4\nonumber\\
&-& b_g\left(2\omega+n\right)\frac{3}{128}e^2\left(1-\cve\right)^4\nonumber\\
&+& b_g\left(2\omega-n\right)\frac{3}{128}e^2\left(1+\cve\right)^4\nonumber\\
&+& b_g\left(2\omega-2n\right)\frac{3}{32}\left(1-5e^2\right)\left(1+\cve\right)^4\nonumber\\
&+& b_g\left(2\omega-3n\right)\frac{147}{128}e^2\left(1+\cve\right)^4\nonumber\\
&-& b_g\left(\omega+3n\right)\frac{147}{32}e^2\left(1-\cveq\right)\left(1-\cve\right)^2\nonumber\\
&-& b_g\left(\omega+2n\right)\frac{3}{8}\left(1-5e^2\right)\left(1-\cveq\right)\left(1-\cve\right)^2\nonumber\\
&-& b_g\left(\omega+n\right)\frac{3}{32}e^2\left(1-\cveq\right)\left(1-\cve\right)^2\nonumber\\
&+& b_g\left(\omega-n\right)\frac{3}{32}e^2\left(1-\cveq\right)\left(1+\cve\right)^2\nonumber\\
&+& b_g\left(\omega-2n\right)\frac{3}{8}\left(1-5e^2\right)\left(1-\cveq\right)\left(1+\cve\right)^2\nonumber\\
&+& b_g\left(\omega-3n\right)\frac{147}{32}e^2\left(1-\cveq\right)\left(1+\cve\right)^2\nonumber\\
&-& b_g\left(3n\right)\frac{441}{64}e^2\left(1-\cveq\right)^2\nonumber\\
&-& b_g\left(2n\right)\frac{9}{16}\left(1-5e^2\right)\left(1-\cveq\right)^2\nonumber\\
&-& b_g\left(n\right)\frac{9}{64}e^2\left(1-\cveq\right)^2 \ .
\label{eq:Tg1b}
\end{eqnarray}
The coefficients $ b_g (\sigma) $ are related to the dissipation of the mechanical energy
of tides in the planet's interior,
responsible for a time delay $ \Delta t_g (\sigma) $ between the position of
``maximal tide'' and the sub-stellar point.  
They are related to the geometric lag $ \delta_g (\sigma) $ as:
\begin{equation}
b_g (\sigma) =  k_2  \sin 2 \delta_g (\sigma) = k_2  \sin
\left( \sigma \Delta t_g(\sigma)  \right) \ . \label{130422z}
\end{equation}
Dissipation equations (\ref{eq:Tg1}) and (\ref{eq:Tg1b}) must be invariant under the transformation $(\omega, x) $ by $ (- \omega, -x) $ which implies that $ b (\sigma) = - b (- \sigma) $. 
That is, $ b (\sigma) $ is an odd function of $\sigma$ \citep{Correia_etal_2003}.
Although mathematically equivalent, the couples $ (\omega, x) $ and $ (-\omega, -x) $ correspond to two different physical situations \citep[see][]{Correia_Laskar_2001}.

 \subsubsection{Thermal atmospheric tides}
 \label{sec:eqmotiomatm}

The differential absorption of the Solar heat by the planet's atmosphere gives 
rise to local variations of temperature and consequently to pressure gradients. 
The mass of the atmosphere is then permanently redistributed, adjusting for an
equilibrium position.
More precisely, the particles of the atmosphere move from the high temperature
zone (at the sub-solar point) to the low temperature areas \citep[e.g.][]{Arras_Socrates_2010}.
Observations on Earth show that the pressure redistribution is
essentially a superposition of two pressure waves: a diurnal tide of small amplitude and a strong semi-diurnal tide \citep{Bartels_1932,Haurwitz_1964, Chapman_Lindzen_1970}.

As for gravitational tides, the redistribution of mass in the atmosphere gives
rise to an atmospheric bulge that modifies the gravitational potential generated
by the atmosphere in any point of the space.
The tidal potential $ U_a $ responsible for the spin changes is given by\footnote{We did not include the diurnal surface pressure variations, because they correspond to a displacement of the center of mass of the atmosphere bulge, which has no dynamical implications. We also neglect terms in $(R/r)^4$.}
\citep[e.g.][]{Correia_Laskar_2003JGR}:
\begin{equation} 
U_a = - \frac{3}{5} \frac{\tilde{p}_2}{\bar{\rho}} \left( 
\frac{R}{r} \right)^3 P_2(\cos S) \, , \label{A5} 
\end{equation}
where $ \tilde{p}_2 $ is the second order surface pressure variations, 
and $\bar\rho$ is the mean density of the planet.

To find the contributions to the spin we use expressions (\ref{eq:dLdt}) together with the  averaging method over fast varying angles (mean anomaly and longitude of the periapse), which gives:
\begin{equation}
\frac{d L}{d t}= K_a \sum _\sigma b_a (\sigma)\Lambda_\sigma^a(x,e) \ , 
\label{eq:variacaospin1}
\end{equation}
\begin{equation}
\frac{d X}{d t}=K_a \sum _\sigma b_a (\sigma)\Upsilon_\sigma^a(x,e) \ , 
\label{eq:variacaospin2}
\end{equation}
where 
\begin{equation}
K_a= -\frac{3 \M R^3}{5\bar\rho a^3} \ . \label{130613b}
\end{equation} 
The $b_a(\sigma)$ factor is now:
\begin{eqnarray}
b_a (\sigma) &=& \tilde p_2(\sigma)  \sin
\left( \sigma \Delta t_a(\sigma)  \right)= \tilde p_2(\sigma)   \sin 2 \delta_a (\sigma)\nonumber\\
 &=&|\tilde p_2|\sin2(\delta _a(\sigma)+\pi/2)=-|\tilde p_2|\sin 2\delta_a(\sigma) \ , 
\label{130422y}
\end{eqnarray}
where $\Delta t _a$ is the atmosphere's delayed response to the stellar heat excitation, and $\delta _a$ is the corresponding geometric lag.
The amplitude of the bulge, $\tilde p_2$, is the second order surface pressure variations \citep{Chapman_Lindzen_1970}:
\begin{equation}\label{eq:p2}
 \tilde p_2 (\sigma) = i \frac{\gamma}{\sigma}\tilde p_0\left(\vv\nabla\cdot\vv\upsilon_\sigma
-\frac{\gamma-1}{\gamma}\frac{J_\sigma}{gH_0}\right)= i \frac{\mathcal{P}_\sigma}{\sigma} \ ,
\end{equation}
 where $\gamma =7/5$ for a perfect diatomic gas, $\tilde p_0$ is the mean surface pressure,
 $\vv\upsilon_\sigma$ is the tidal winds velocity, $J_\sigma$ is the amount of heat absorbed or emitted by unit of mass of air  
per unit time, and $H_0$ is the scale height at the surface.  
The imaginary number in equation  (\ref{eq:p2}) causes the pressure variations to lead the star ($\mathrm i=e^{\mathrm i \pi/2}$).

The coefficients $\Lambda_\sigma^a $ and $\Upsilon_\sigma^ a$ are also polynomials in the eccentricity, but different from their analogs for gravitational tides (Eqs.~\ref{eq:Tg1} and \ref{eq:Tg1b}). 
However, for zero eccentricity they become equal, i.e., $\Lambda_ \sigma^a (e=0) = \Lambda_\sigma^g (e=0)$, and $\Upsilon_ \sigma^a (e=0) = \Upsilon_\sigma^g (e=0)$.
Once more, for a planet with moderate eccentricity,  we can neglect terms in $e^4$ and greater, and obtain the following expansion for equations (\ref{eq:variacaospin1}) and (\ref{eq:variacaospin2}):
\begin{eqnarray}
\label{eq:Ta1}
\frac{1}{K_a} \frac{d L}{d t}&=& b_a\left(2\omega+3n\right)\frac{27}{32}e^2\left(1-\cve\right)^4\nonumber\\
&+& b_a\left(2\omega+2n\right)\frac{3}{32}\left(1-3e^2\right)\left(1-\cve\right)^4\nonumber\\
&-& b_a\left(2\omega+n\right)\frac{3}{32}e^2\left(1-\cve\right)^4\nonumber\\
&+& b_a\left(2\omega\right)\frac{3}{8}\left(1+5e^2\right)\left(1-\cveq\right)^2\nonumber\\
&-& b_a\left(2\omega-n\right)\frac{3}{32}e^2\left(1+\cve\right)^4\nonumber\\
&+& b_a\left(2\omega-2n\right)\frac{3}{32}\left(1-3e^2\right)\left(1+\cve\right)^4\nonumber\\
&+& b_a\left(2\omega-3n\right)\frac{27}{32}e^2\left(1+\cve\right)^4\nonumber\\
&+& b_a\left(\omega+3n\right)\frac{27}{16}e^2\left(1-\cveq\right)\left(1-\cve\right)^2\nonumber\\
&+& b_a\left(\omega+2n\right)\frac{3}{16}\left(1-3e^2\right)\left(1-\cveq\right)\left(1-\cve\right)^2\nonumber\\
&-& b_a\left(\omega+n\right)\frac{3}{16}e^2\left(1-\cveq\right)\left(1-\cve\right)^2\nonumber\\
&+& b_a\left(\omega\right)\frac{3}{4}\left(1+5e^2\right)\cveq\left(1-\cveq\right)\nonumber\\
&-& b_a\left(\omega-n\right)\frac{3}{16}e^2\left(1-\cveq\right)\left(1+\cve\right)^2\nonumber\\
&+& b_a\left(\omega-2n\right)\frac{3}{16}\left(1-3e^2\right)\left(1-\cveq\right)\left(1+\cve\right)^2\nonumber\\
&+& b_a\left(\omega-3n\right)\frac{27}{16}e^2\left(1-\cveq\right)\left(1+\cve\right)^2 \ ,
\label{eq:Ta1}
\end{eqnarray}
and

\begin{eqnarray}
\frac{1}{K_a} \frac{dX}{dt}&=& - b_a\left(2\omega+3n\right)\frac{27}{32}e^2\left(1-\cve\right)^4\nonumber\\
&-&  b_a\left(2\omega+2n\right)\frac{3}{32}\left(1-3e^2\right)\left(1-\cve\right)^4\nonumber\\
&+&  b_a\left(2\omega+n\right)\frac{3}{32}e^2\left(1-\cve\right)^4\nonumber\\
&-&  b_a\left(2\omega-n\right)\frac{3}{32}e^2\left(1+\cve\right)^4\nonumber\\
&+&  b_a\left(2\omega-2n\right)\frac{3}{32}\left(1-3e^2\right)\left(1+\cve\right)^4\nonumber\\
&+&  b_a\left(2\omega-3n\right)\frac{27}{32}e^2\left(1+\cve\right)^4\nonumber\\
&-&  b_a\left(\omega+3n\right)\frac{27}{8}e^2\left(1-\cveq\right)\left(1-\cve\right)^2\nonumber\\
&-&  b_a\left(\omega+2n\right)\frac{3}{8}\left(1-3e^2\right)\left(1-\cveq\right)\left(1-\cve\right)^2\nonumber\\
&+&  b_a\left(\omega+n\right)\frac{3}{8}e^2\left(1-\cveq\right)\left(1-\cve\right)^2\nonumber\\
&-&  b_a\left(\omega-n\right)\frac{3}{8}e^2\left(1-\cveq\right)\left(1+\cve\right)^2\nonumber\\
&+&  b_a\left(\omega-2n\right)\frac{3}{8}\left(1-3e^2\right)\left(1-\cveq\right)\left(1+\cve\right)^2\nonumber\\
&+&  b_a\left(\omega-3n\right)\frac{27}{8}e^2\left(1-\cveq\right)\left(1+\cve\right)^2\nonumber\\
&-&  b_a\left(3n\right)\frac{81}{16}e^2\left(1-\cveq\right)^2\nonumber\\
&-&  b_a\left(2n\right)\frac{9}{16}\left(1-3e^2\right)\left(1-\cveq\right)^2\nonumber\\
&+&  b_a\left(n\right)\frac{9}{16}e^2\left(1-\cveq\right)^2 \ .
\label{eq:Ta1b}
\end{eqnarray}
Neglecting the tidal winds $\vv\upsilon_\sigma$, from expression (\ref{eq:p2}) we have
\begin{equation}
\mathcal{P}_\sigma \propto J_\sigma \propto r^{-2} \label{140125a} \ .
\end{equation}
As a consequence, in the computation of the averaged expressions for $\Lambda_\sigma^a$ and $\Upsilon_\sigma^a$ we used a factor $r^{-5}$, corresponding to the $r^{-2}$ from $\mathcal{P}_\sigma$ together with the contribution in $r^{-3}$ from the tidal potential (Eq.\,\ref{A5}).

\subsubsection{Tidal models}

\label{021024j}

\begin{figure}[t!]
\begin{center}
\includegraphics[width=1.00\columnwidth]{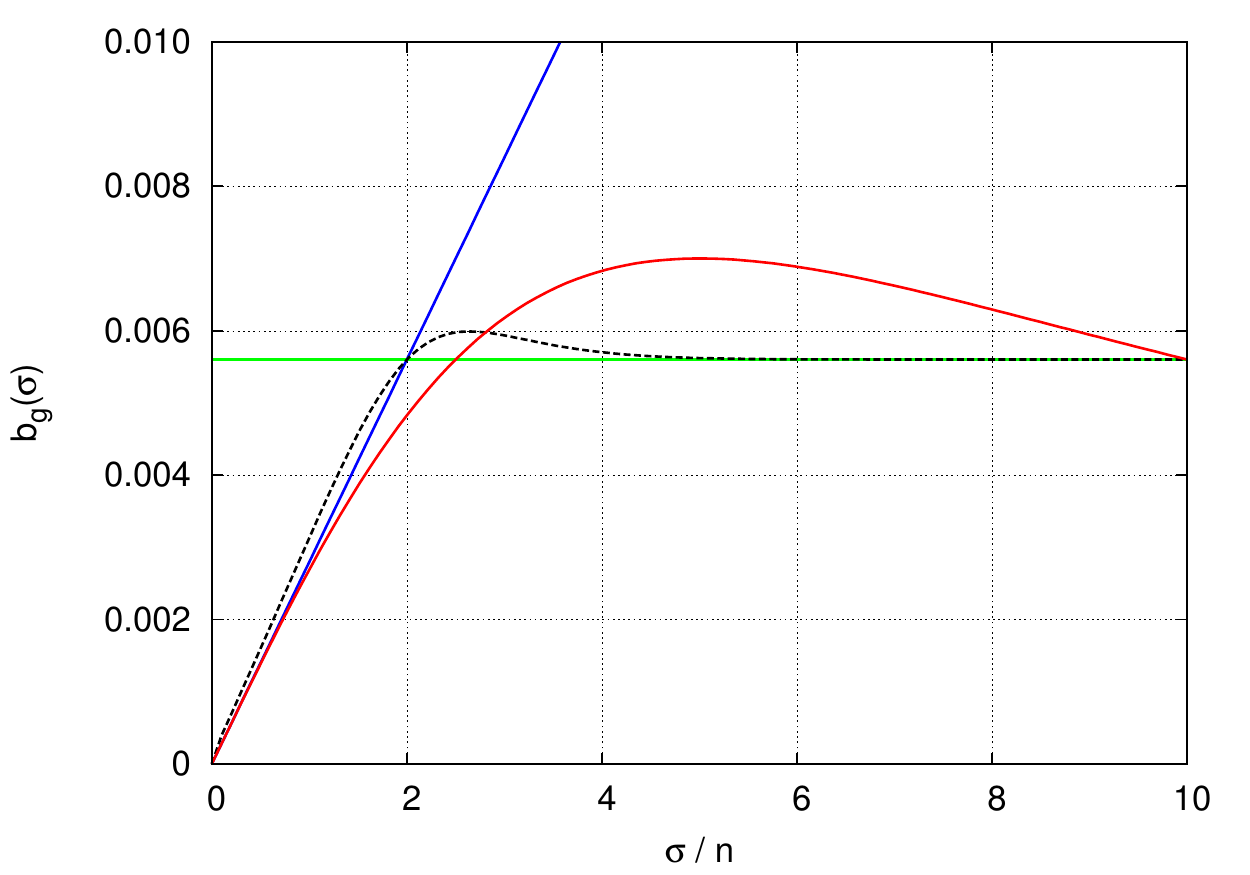}
\caption{\label{figmodels} Commonly used models for frequency dependence of tides: 
 visco-elasctic (Eq.\,\ref{090419f}, \textbf{red}), viscous or linear (Eq.\,\ref{091126a}, \textbf{blue}), constant$-Q$ (Eq.\,\ref{090407a}, \textbf{green}), and an interpolated model between the linear and the constant one (Eq.\,\ref{130612a}, \textbf{dashed line}).}
\end{center}
\end{figure}

The dissipation of the mechanical energy of tides in the planet's interior is
responsible for the phase lags $ \delta (\sigma) $.
A commonly used dimensionless measure of tidal damping is the quality factor $
Q $, defined as the inverse of the ``specific''
dissipation and related to the phase lags by 
 \citep[][Eq.\,141]{Efroimsky_2012}
\begin{equation} 
Q^{-1} (\sigma) = \frac{\Delta E}{2 \pi E} = | \sin 2 \delta_g (\sigma) |  \ ,
\label{090419e}
\end{equation}
where $ E $ is the total tidal energy stored in the planet, 
and $ \Delta E $ is the energy dissipated per cycle. We can rewrite expression (\ref{130422z})
as: 
\begin{equation} 
b_g (\sigma) = \mathrm{sign}(\sigma) \frac{k_2}{Q (\sigma)} \ .
\end{equation} 
The present $ Q $ value for the planets in the Solar system
can be estimated from orbital measurements,
but as rheology of the planets is  badly known,
the exact dependence of $ b_g (\sigma)$ on the tidal frequency $ \sigma $
is unknown.
Many different authors have studied the problem and several models for $ b_g
(\sigma)$  have been developed so far, from the simplest ones to the more
complex \citep[for a review see][]{Efroimsky_Williams_2009}.
The huge problem in validating one model better than the others is the difficulty
to compare the theoretical results with the observations, as the effect
of tides are very small and can only be detected efficiently after long periods
of time.
Therefore, here we only describe the most commonly models that are used (Fig.\,\ref{figmodels}).

\paragraph{The visco-elastic model}

\label{021120y}

\citet{Darwin_1908} assumed that the planet behaves like a Maxwell
solid\footnote{A Maxwell body
behaves like an elastic body over short time scales, but flows like a fluid over long periods of time. It is characterized by a homogenous rigidity (or shear modulus) $ \mu_e $ and by a viscosity $ \upsilon_e $.}
of constant density $ \rho $, and found:
\begin{equation} 
b_g (\sigma) = k_f \frac{\tau_b - \tau_a}{1 + (\tau_b \, \sigma)^2} \sigma \ ,
\label{090419f}
\end{equation}
where $ \tau_a = \upsilon_e / \mu_e $ and $ \tau_b $ are the time constants for
damping of the body tides,
\begin{equation} 
\tau_b = \tau_a ( 1 + 19 \mu_e R / 2 G m \rho ) \ . \label{090419g}
\end{equation}
The Maxwell solid model is usually accepted as a good approximation of the planet's response to tidal perturbations, 
although more complete visco-elastic models exist \citep[for a review see][]{Henning_etal_2009}.
However, it depends on many uncertain parameters and it is not of practical use when we aim to do simple approximations and considerations on the tidal evolution.
In addition, while it can be appropriate to simultaneously describe the elastic and viscous response of the solid body, it is questionable whether it is still valid for the atmospheric  deformation is response to thermal gradients.

\paragraph{The viscous or linear model}

\label{021120a}

In the viscous model, 
it is assumed that the response time delay to the perturbation is 
independent of the tidal frequency, i.e., the position of the ``maximal
tide'' is shifted from the sub-stellar point by a constant time-lag $ \Delta t_g $
 \citep[e.g][]{Mignard_1979}. As usual we have  $ \sigma \Delta t_g \ll 1 $,
the viscous model becomes linear:
\begin{equation} 
b_g (\sigma) = k_2 \sin (\sigma \Delta t_g) \approx k_2 \, \sigma \Delta t_g \ .
\label{091126a}
\end{equation}  
The viscous model is a particular case of the visco-elastic model and is
specially adapted to describe the behavior of planets in slow rotating regimes
($ \omega \sim n $).

\paragraph{The constant-$Q$ model}

\label{021024i}

For the Earth, $ Q $ changes by slightly more than an order of magnitude between the
Chandler wobble period (about 440 days) and seismic periods of a few
seconds \citep[e.g.][]{Anderson_Minster_1979,Karato_Spetzler_1990}.
 Thus, it is also common to treat the specific dissipation as independent of frequency:
\begin{equation} 
b_g (\sigma) \approx \mathrm{sign}(\sigma) k_2 / Q \ . \label{090407a}
\end{equation}
The constant-$Q$ model can be used for long periods of time where the tidal dissipation does not change much, as is the case for moderately rotating planets. However, for slow rotating planets, the constant-$Q$ model is not appropriate as it gives rise to discontinuities for $ \sigma = 0 $.

\subsection{Core-mantle friction}

We assume that the Earth-sized planets have internal structures similar to that of our planet, that is, these planets are composed of a solid mantle and a liquid core.
We suppose that the liquid core is inviscid, incompressible and homogeneous.
We also suppose that  the internal structure of the planets is unchanged in time, since the core formation time (or condensation velocity) is poorly  known. 

If there is slippage between the liquid core and the mantle, an additional source of
dissipation of rotational energy results from friction occurring at the
core-mantle boundary. 
Indeed, because of their different shapes and densities, the core and the mantle
do not have the same dynamical ellipticity and the two parts tend to precess
at different rates \citep{Poincare_1910}.
This tendency is more or less counteracted by different interactions produced at
their interface: the torque $ \vv{N} $ of non-radial inertial pressure forces of
the mantle over the core provoked by the non-spherical shape of the interface;
the torque of the viscous (or turbulent) friction between the core and the
mantle; and the torque of the electromagnetic friction, caused by the interaction
between electrical currents of the core and the bottom of the magnetized mantle.

\citet{Poincare_1910} classic study shows that the core responds to the precession with a rotation velocity $\vv{\omega_c}$, whose vector is inclined by a small angle $ \chi$ with respect to the mantle velocity vector $\vv {\omega}= \omega \, \vv{\mathrm{k}}$ \citep{Rochester_1976,Correia_2006}:

\begin{equation}
\vv{\delta}=\vv{\omega}-\vv{\omega_c} \ ,\end{equation}
and
\begin{equation}
\sin\chi=\left|\frac{\vv{\delta}\times\vv{\mathrm{k}}}{\omega_c}\right| \ .
\end{equation}
\citet{Sasao_etal_1980} demonstrated that  the pressure inertial torque may be expressed in a general way by:
\begin{equation}
\vv N=\vv{\omega_c} \times \vv{L_c}-\vv{P_c},
\end{equation}
where $\vv{P_c}$ is the precession torque over the core and $\vv{L_c}$ its angular moment:
\begin{equation}
\vv{L_c}=  \vv{\mathcal{\tilde  I}}_c\cdot\vv{\omega_c},
\end{equation}
with
\begin{equation}
\vv{\mathcal{\tilde  I}}_c=
\begin{pmatrix}A_c & 0 & 0 \\
0 & A_{c} & 0 \\
0 & 0 & C_{c} \\
\end{pmatrix}.
\end{equation}
$C_c$ and $A_c$ are the principal inertial moments of the core. The core's dynamical ellipticity is then given by $E_c=(C_c-A_c)/C_c$. 
The lower part of the mantle has irregularities that prevent the border between the two layers from being a perfect ellipsoid. According to \citet{Hide_1969} the height
of these bumps may reach a few kilometers. Thus, for the dynamical ellipticity of the mantle, we establish that $E_c=E_{c_h}+\delta E_c$, that is, $E_c$ has a hydrostatic component  and a non-hydrostatic component. For Earth, \citet{Herring_etal_1986} provide
the value $\delta E_c=1.2\times10^{-4}$.

The two types of friction torques (viscous and electromagnetic) depend
on the differential rotation between the core and the mantle and can be
expressed by a single effective friction torque, $\vv \Phi$.
As a general expression for this torque we adopt
\citep{Rochester_1976,Mathews_Guo_2005}
\begin{equation}
\vv\Phi=-(\kappa+\kappa' \,\vv{\mathrm{k}} \times ) \, \vv{\delta} \ ,
\end{equation}
where $ \kappa $ and $ \kappa' $ are effective coupling parameters.
Both parameters account for viscous and electromagnetic stresses at the core-mantle interface, that can be written as: $ \kappa = \kappa_\mathrm{vis} + \kappa_\mathrm{em} $, and $ \kappa' = \kappa_\mathrm{vis}' + \kappa_\mathrm{em}' $.
Estimations for these coefficients can be found in the works of
\citet{Mathews_Guo_2005} and \citet{Deleplace_Cardin_2006}.
In the simplified case of no magnetic field, the coupling parameters are only
given by the viscous friction contributions, which can be simplified as
\citep{Noir_etal_2003,Mathews_Guo_2005}: 
\begin{equation}
\kappa_\mathrm{vis} = 2.62 \sqrt{\nu | \omega | } / R_c  \quad \mathrm{and} \quad  
\kappa_\mathrm{vis}' = 0.259 \sqrt{\nu | \omega | } / R_c \ , \label{V14a} 
\end{equation}
where $ R_c $ is the core radius and $ \nu $ the kinematic viscosity, which
is poorly known. Even in the case of the Earth, the uncertainty in $ \nu $
covers about 13 orders of magnitude \citep{Lumb_Aldridge_1991}, the best
estimate so far being $ \nu \simeq 10^{-6} \, \mathrm{m}^2 \mathrm{s}^{-1} $
\citep{Gans_1972,Poirier_1988,deWijs_etal_1998}.

The contributions for  spin variations of a planet  may be obtained writing the equations of angular momentum conservation for the core and mantle:

\begin{eqnarray}
\frac{d\vv{L_m}}{dt} &=&\vv{P_m}-\vv{N} -\vv{\Phi} \ , \\
 \frac{d\vv{L_c}}{dt} &=&\vv{P_c}+\vv{N}+\vv{\Phi} \ ,
\end{eqnarray}
where the $\vv{P_m}$ is the precession torque over the mantle and $\vv{L_m}$ is its angular momentum.

A general formulation for the equations of motion, valid for the both fast and slow rotation regimes of the planet, can be written as \citep{Correia_2006}: 
\begin{eqnarray}
\frac{dL}{dt}&\simeq&-\frac{\kappa A_c C_c \omega\alpha^2\cos^2\varepsilon\sin^2\varepsilon}{(C_cE_c\omega)^2+\kappa^2}  \label{V13a} \ ,  \\
 \frac{dX}{dt} &\simeq& 0 \label{V13} \ .
 \end{eqnarray}

\section{Dynamical Analysis}
\label{dynanal}

\subsection{Obliquity Evolution }
\label{sec:oblvariation}

Until now, we have been expressing the
variations of the spin in  Andoyer's variables.
Despite their practical use, these variables do not give a clear view of the
obliquity variation.
Since $ x = \cos \varepsilon = X/L $, one obtains:
\begin{equation} 
\frac{d x}{d t} = \frac{1}{L} \left( \frac{d X}{d t} -
x \frac{d L}{d t} \right) \ . \label{V30} 
\end{equation}

For tidal effects ($ \tau = g $ or $ \tau = a $), we express $ d x / dt $ using the eccentricity series for $ d L / d t $ and $ d X / d t $ (Eqs.\ref{eq:Tg1}, \ref{eq:Tg1b} and \ref{eq:Ta1}, \ref{eq:Ta1b}):
\begin{eqnarray} 
\frac{d x}{d t} & = & - \frac{K_\tau}{\omega} \sum_\sigma b_\tau
(\sigma) \left(  \Upsilon_\sigma - x \Lambda_\sigma \right) \nonumber \\ 
& = & K_\tau \frac{1-x^2}{\omega} \sum_\sigma b_\tau
(\sigma) \Theta_\sigma (x,e) \ .  \label{V31N} 
\end{eqnarray} 
We thus have
\begin{equation}
\frac{d x}{dt}\propto\frac{1-x^2}{\omega}\frac{d \omega}{dt} \ , \label{eq:dedw}
\end{equation}
meaning that the obliquity variations are smaller than the rotation rate variations for initially fast rotating planets, and that $\dot \varepsilon \propto - \sin \varepsilon$.

For the core-mantle friction effect, the variation of $ \varepsilon $ is easily computed from expression (\ref{V30}), since $ d X / d t \simeq 0 $ (Eq.\ref{V13}).
\begin{equation} 
 \frac{d x}{d t} \simeq -\frac{x}{L} \frac{d L}{d t} = -\frac{x}{\omega} \frac{d \omega}{d t} 
  \label{V32a} 
\end{equation}
It also follows that
 \begin{equation} 
X = cte. \quad \Rightarrow \quad 
x  \, \omega = cte. \label{V32b} 
\end{equation}

\subsection{Gravitational tides alone}
\label{sec:Velocityvariation}

From expression (\ref{eq:p2}), we notice that for the initial stages of the evolution ($\sigma \gg n$) atmospheric tides are weak. 
The same is true for core-mantle friction (Eq.\,\ref{V13a}).
As a consequence, gravitational tides dominate the spin evolution for fast rotating rates, thermal tides and core-mantle friction only playing a role for slow rotations \citep{Correia_etal_2003}.
Therefore, most studies on the spin evolution only consider gravitational tides.
For a better comparison with our study, we recall here the main consequences of this effect alone.

If we assume a fast initial rotation, planets spend
most of their evolution in the $\omega \gg n$ regime. 
In this case, ${d\omega}/{dt}$  evolves independently  of the  dissipative model,
 because all the terms in expression (\ref{eq:Tg1})  have the same sign. Thus, in this regime gravitational tides always decrease the rotation rate, since $K_g < 0 $ (Eq.\,\ref{130613a}).
 
When the planet arrives in the slow rotation regime ($\omega \sim n$) some of the terms in equation (\ref{eq:Tg1}) 
become negative and we can no longer generalize our conclusions to all dissipative models. 
However, in the vicinity of tidal frequencies near zero ($\sigma \approx 0$), except for the constant$-Q$ model,  all dissipative models can be made linear (Fig.\,\ref{figmodels}). 
Then, adopting the linear dissipation model described  in section \ref{021024j}, we can write
\begin{equation}
b_g(\sigma)\simeq k_2 \sigma\Delta t_g=\frac{k_2}{Q_n}\frac{\sigma}{2 n} \ , 
\end{equation}
where $Q_n^{-1} = 2 n \Delta t_g $ is the specific  dissipation factor for small  frequencies.
For small eccentricity, the highest tidal frequency is $\sigma = 2 \omega + 3 n$ (Eq.\,\ref{eq:Ta1}), so this approximation is valid\footnote{Assuming $\omega = n$ and $\sin \delta \approx \delta$ for $\delta<0.5$.} for $Q_n \gtrsim 5$.

In the limit of slow rotation rates,  we can then simplify expression (\ref{eq:Tg1}):
\begin{equation}
\frac{d \omega}{dt}= - K_0 \left[ \omega \left( \frac{1+\cveq}{2} \right)-n\left(1+6 e^2\right)\cve\right] \ , \label{eq:dwdtGef}
\end{equation}
where
\begin{equation}
K_0=- \frac{3K_gk_2\Delta t_g}{C}\left(1+\frac{15}{2}e^2\right) \ .
\end{equation}
Thus, for each obliquity there is an equilibrium value for the rotation rate $\omega_e$, 
obtained when $d\omega/dt=0$: 
\begin{equation}\frac{\omega_{e}}{n}= \left(1+6 e^2\right) \frac{2\cve}{1+\cveq} \ .
\label{eq:weq}\end{equation}
For $\omega>\omega_e$ the rotation rate decreases, while for  $\omega<\omega_e$ it increases.


\subsection{Equilibrium damping time}

In order to check if the spin of a given planet is still evolving or if it can be found at its equilibrium rotation, we can compute an evolutionary characteristic time and then compare it to the age of the host star.
Since gravitational tides are the dominant effect for the most part of the evolution, we can use this effect alone to estimate the characteristic time $\tau_{eq}$ needed to reach the equilibrium rotation.
Thus, from expression (\ref{eq:dwdtGef}) we get:
\begin{equation}
\tau_{eq}\sim 
 \frac{1}{K_0}
= \frac{GC}{3k_2\Delta t_gR^5}\frac{1}{n^4}=\frac{ma^6}{9G\M^{2}k_2\Delta t_gR^3},
\label{eq:tau_eq}
\end{equation}
with $C\simeq m R^2/3 $ and $n^4={G^2\M^2}/{a^6}$. 
In Table~\ref{tbl:1} we computed the characteristic times for existing Earth-sized  exoplanets. 
We notice that, except for GJ\,667C\,g, all known planets have a $\tau_{eq}$ lower than the age of the system,
 and so it is believed that they have already reached an equilibrium rotation state.


\subsection{Thermal tides effect}

\label{sec:DynEv}

We now include in our analysis the effect from thermal atmospheric tides to the spin evolution (section~\ref{sec:eqmotiomatm}).
As shown earlier, 
the averaged variation of the spin can be expressed by equation (\ref{eq:Tg1}) for gravitational tides
and by equation (\ref{eq:Ta1}) for  atmospheric tides.
Considering that the planet's obliquity is small ($\varepsilon \sim 0$),
 we can neglect terms of order $\varepsilon^2$ or higher in equations (\ref{eq:Tg1}) and (\ref{eq:Ta1}). 
In this case, the variation of the spin caused by gravitational tides is then \citep{Correia_Levrard_2008}:
\begin{eqnarray}
T_g=\left.\frac{dL}{dt}\right]_g&=&\frac{3}{2}K_g \left[(1-5e^2)b_g(2\omega-2n)\phantom{\frac{1}{4}}\right.\nonumber \\
&&\left.+\frac{1}{4}e^2b_g(2\omega-n)+\frac{49}{4}e^2 b_g(2\omega-3n)\right] \ .
\label{eq:Tg}
\end{eqnarray}
In the same way, for thermal atmospheric tides we obtain
\begin{eqnarray}
T_a=\left.\frac{dL}{dt}\right]_a&=&\frac{3}{2}K_a \left[(1-3e^2)b_a(2\omega-2n)\phantom{\frac{1}{4}}\right.\nonumber \\
&&\left.\phantom{\frac{1}{4}}-e^2b_a(2\omega-n)+9e^2 b_a(2\omega-3n)\right] \ .
\label{eq:Ta}
\end{eqnarray}
As the time-lag from gravitational and atmospheric tides is poorly known, in this section we use again the viscous model.
Since $\sigma \Delta t$ is usually small, then:
\begin{equation}
 b_g(\sigma)\simeq k_2 \sigma\Delta t_g\quad \mathrm{and}\quad b_a(\sigma)\simeq-|\tilde p_2|\sigma\Delta t_a.
 \label{130506a}
\end {equation}

For atmospheric tides it is also necessary to consider the response of the surface pressure variations 
to tidal frequency (Eq.\,\ref{eq:p2}). We use the ``heating at the ground'' model, 
described in \citet{Dobrovolskis_Ingersoll_1980}. It is supposed that all the stellar flux absorbed by the ground, 
$F_s$, is immediately deposited in a thin layer of atmosphere at the surface. 
The heating distribution is then written as a delta-function just above the ground ($J_\sigma=gF_s/\tilde p_{0}$). 
Besides the good agreement with the observations, this approximation is justified because tides in the upper 
atmosphere are decoupled from the ground by the disparity between their rotation rates. 
 Neglecting $\vv\upsilon_\sigma$ over the thin heated layer, equation (\ref{eq:p2}) becomes:
\begin {equation}
\mathcal P_\sigma=F_s/(8H_0)\propto L_*/a^2
\label{eq:psigma}
,\end{equation}
where $L_*$ is the star luminosity and $a$ the semi-major axis.

The average evolution of the rotation rate can then be  obtained  adding the effects of both tidal torques 
acting on the planet:

\begin {equation}
\frac{dL}{dt}=C\dot \omega=(T_g+T_a).
\label{eq:omegadot1}
\end{equation}
Substituting in previous expression  $T_g$ and $T_a$ given by equations (\ref{eq:Tg}) and (\ref{eq:Ta}) together with the viscous dissipative model (Eq.\,\ref{130506a}), we have\footnote{Eqs.~\ref{eq:omega.K0} and \ref{eq:omegas1} are similar to Eqs.~5 and~6 in \citet{Correia_Levrard_2008}. However, in the present work there is a factor $21/2$ instead of $3$ in the term $e^2 \mathrm{sign}(\omega-n)$, because we have included the effect from $\Omega(e)$ in the expression of $\dot \omega$ and removed it from the definition of $\omega_s$. In \citet{Correia_Levrard_2008} there was also a factor 1/2 missing in the expression of $\omega_s$.}:
\begin{eqnarray}
\frac{\dot \omega}{-K_0}&=&\omega-(1+6e^2)n-\omega_s\left[\left(1-\frac{21}{2}e^2\right)\mathrm{sign}(\omega-n)\right. \nonumber \\
&&\left. \phantom{\frac{21}{2}} -e^2\mathrm{sign}(2\omega-n)+9e^2\mathrm{sign}(2\omega-3n)\right] \ ,
\label{eq:omega.K0}
\end{eqnarray}
with
\begin{equation}
\omega_s=\frac{K_a F_s\Delta t_a}{16H_0K_gk_{2}\Delta t _g}\propto\frac{L_* R}{\M m}a \ .
\label{eq:omegas1}
\end{equation}

\subsection{Equilibrium final states for the rotation rate}

\label{130510b}

\begin{figure}[ht!]
\begin{center}
 \begin{tabular}{c}
\begin{overpic}[width=.98\columnwidth]{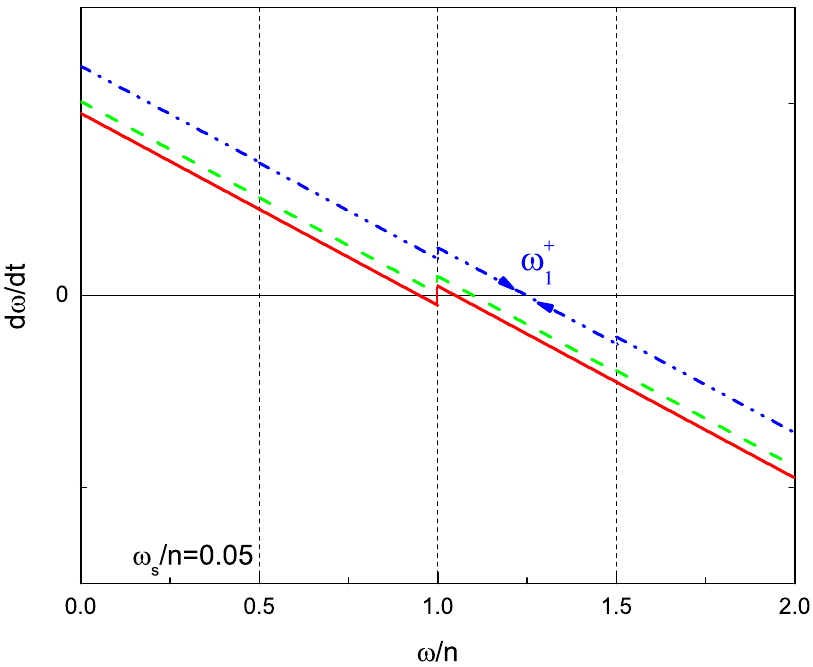}\put(205,173){\bf \Large (a)}\end{overpic} \\
\begin{overpic}[width=.98\columnwidth]{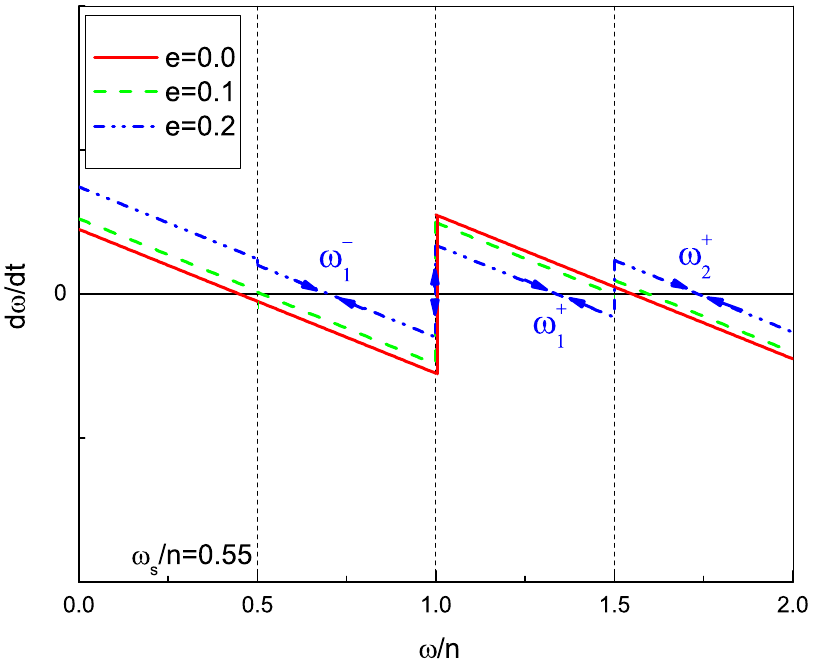}\put(205,173){\bf \Large (b)}\end{overpic} \\
\begin{overpic}[width=.98\columnwidth]{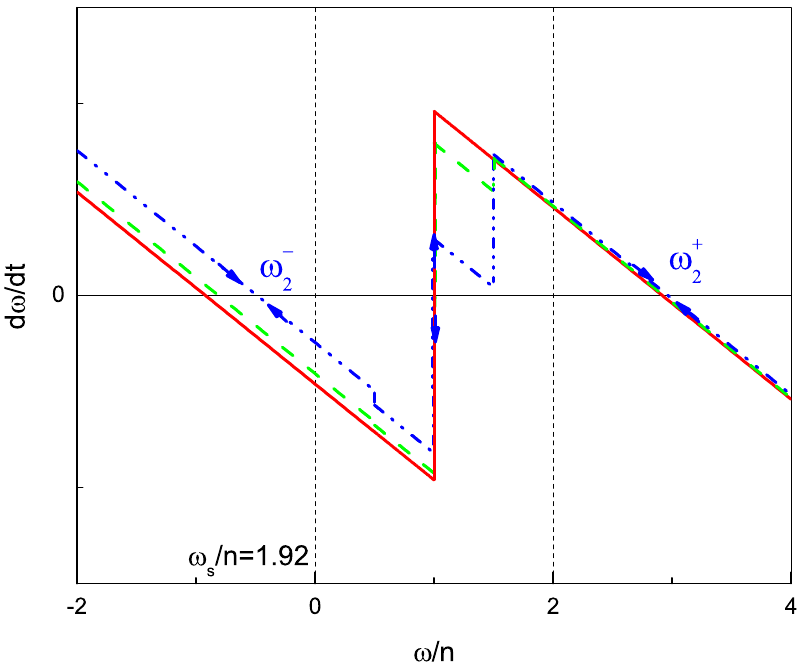}\put(205,173){\bf \Large (c)}\end{overpic} 
 \end{tabular}
\caption{\label{fig:wn} Variation of $\dot \omega$ with $\omega/n$ (Eq.\,\ref{eq:omega.K0}) for \textbf{(a)} $\omega_s/n =0.05$,
\textbf{(b)} $\omega_s/n =0.55$, and \textbf{(c)} $\omega_s/n =1.92$, using different eccentricities (e=0.0, 0.1, 0.2). 
The equilibrium rotation rates are given by $\dot \omega=0$ and the arrows indicate whether it is a stable or unstable equilibrium position.
}
\end{center}
\end{figure}

An equilibrium final state occurs when $\dot \omega=0$. From equation (\ref{eq:omegadot1}) 
this happens when $T_g=-T_a$.  When $e=0$ \citep[see][]{Correia_Laskar_2001},
\begin{equation}
f(\omega-n)=-\frac{T_g}{T_a}=-\frac{K_gb_g(2\omega-2n)}{K_ab_a(2\omega-2n)}=1 \ ,
\label{130510a}
\end{equation}
where $f(x)$ is an even function of $x$. For the dissipative model in use, we can assume that $f(x)$ is monotonic close to equilibrium, and so we have\footnote{In the case $e=0$, we see that $\omega_s$ is also the synodic frequency.}:
\begin{equation}
|\omega-n|=f^{-1}(1)=\omega_s \ .
\label{def:ws}
\end{equation}
This means that there are two final possibilities for the equilibrium rotation of the planet:
\begin{equation}
\omega^{\pm}=n\pm\omega_s.
\label{eq:w+-para e=0}
\end{equation}
If $\omega_s <n$, all final states correspond to prograde final rotation rates (Fig.\,\ref{fig:wn}a,b).
 Retrograde rotation appears  only if the planet  evolves to the $\omega^-$ state together with $\omega_s >n$.
 As we can see in Table~\ref{tbl:1}, this is the case of planet Venus (Fig.\,\ref{fig:wn}c).

For moderate values of the eccentricity and using equations (\ref{eq:Tg}) and (\ref{eq:Ta}),  equation (\ref{130510a}) 
can be rewritten as:
\begin{equation}
f(\omega-n)=1-e^2\left(2+\frac{g(\omega)}{b_a(2\omega-2n)}\right)
\end{equation}
where
\begin{eqnarray}
g(\omega)&=&\frac{K_g}{4K_a}\Big[ b_g(2\omega-n)+49b_g(2\omega-3n)\Big] \nonumber\\
&&-b_a(2\omega-n)+9b_a(2\omega-3n) \ .
\end{eqnarray}
When we compute the inverse function of previous equation and use 
 $\omega_s$ from expression (\ref{def:ws})  we get:
\begin{equation}
|\omega-n|=f^{-1}(1)-e^2\left(2+\frac{g(\omega)}{b_a(2\omega-2n)}\right)\left.\frac{\partial f^-1}{\partial x}\right|_{x=1},
\end{equation}
which means that  we now have  four final possibilities for the equilibrium rotation of the planet:
\begin{equation}
\omega_{1,2}^{\pm}=n\pm\omega_s + e^2 \delta_{1,2}^{\pm}
\ ,
\end{equation}
with \begin{equation}
\delta_{1,2}^\pm
=\left(2+\frac{g(\omega)}{|b_a(2\omega-2n)|}\right)\left.\frac{\partial f^-1}{\partial x} \right|_{x=1} \ .
\label{delta12.1}
\end{equation}
Adopting the dissipation model from equation (\ref{130506a}) and using expression (\ref{eq:omega.K0}), we find that 
 $\omega_2^+$ occurs when $\omega>3n/2$ and:
\begin{eqnarray}
{\omega_2^+}&=&(1+6 e^2) n + \omega_s \left[ 1 + e^2 \left( 9 - \frac{21}{2} - 1\right) \right]\nonumber\\
&=&n+\omega_s+e^2 \left(6n-\frac{5}{2}\omega_s\right) \ .
\label{w2+}
\end{eqnarray}
Similarly, we find the $\omega_1^+$ state when   $n<\omega<3n/2$
%
\begin {equation}
{\omega_1^+}={n+{\omega_s}+e^2\left(6n-\frac{41}{2}\omega_s\right)},\label{eq:w1+}
\end {equation}
when $ n/2<\omega<n$ we have the $\omega_1^-$ state
\begin {equation}
{\omega_1^-}={n-{\omega_s}+e^2\left(6n+\frac{1}{2}\omega_s\right)},\label{eq:w1-}
\end {equation}
and finally for
$\omega<n/2$ we have
\begin {equation}
{\omega_2^{-}}={n-{\omega_s}+e^2\left(6n+\frac{5}{2}\omega_s\right)}.\label{eq:w2-}
\end {equation}
Then, expression (\ref{delta12.1}) can be rewritten as:
\begin{equation}\delta_{1,2}^-=6n+\left(\frac{3}{2}\pm1\right)\omega_s \end{equation}
and
\begin{equation}\delta_{1,2}^+=6n-\left(\frac{23}{2}\pm9\right)\omega_s \ . \end{equation}
Because the set of final states $\omega_{1,2}^{\pm}$ must also verify
\begin{equation}
\omega_2^-<n/2<\omega_1^-<n<\omega_1^+<3n/2<\omega_2^+,
\label{cond:omega}
\end{equation}
in general, and depending on the values of $\omega_s$ and $e$, these four equilibrium rotation states cannot coexist all simultaneously.
In particular, when we adopt the viscous tidal model (Eq.\,\ref{130506a}), the final states $\omega^-_1$ and $\omega^+_1$ never coexist with $ \omega^-_2 $. 
At most three different equilibrium states are therefore possible, obtained when
$\omega_s/n$ is close to $1/2$, or more precisely, when
\begin{equation}
\frac{1/2- 6 e^2}{1-5 e^2/2} < \frac{\omega_s}{n} < \frac{1/2- 6 e^2}{1- 41 e^2/2} \ .
\label{130617z}
\end{equation}
Conversely, we have that $\omega^+_1$ is the single final state that exists whenever $\omega_s / n < 6 e^2 ( 1 + e^2/2 )$.

In Figure~~\ref{fig:wn} we plot some examples for the rotation rate evolution for some $\omega_s$ and eccentricity values.
We see that when $\omega_s/n$ is close to zero only the final  equilibrium state $\omega_1^+$ is possible (Fig.\,\ref{fig:wn}a).
For $\omega_s/n=1.92$, the Venus value, two final equilibrium states are possible:
\ the $\omega_2^-$ and the  $\omega_2^+$ (Fig.\,\ref{fig:wn}c).
However, when $\omega_s/n=0.55$  we still have two possible states for $e=0.0$
($\omega_1^-$ and $\omega_1^+$), but there are three possible final states for $e>0.1$
 ($\omega_1^-,\, \omega_1^+$ and $\omega_2^+$) (Fig.\,\ref{fig:wn}b).

\begin{figure}[t!]
\begin{center}
 \begin{tabular}{c}
\begin{overpic}[width=.95\columnwidth]{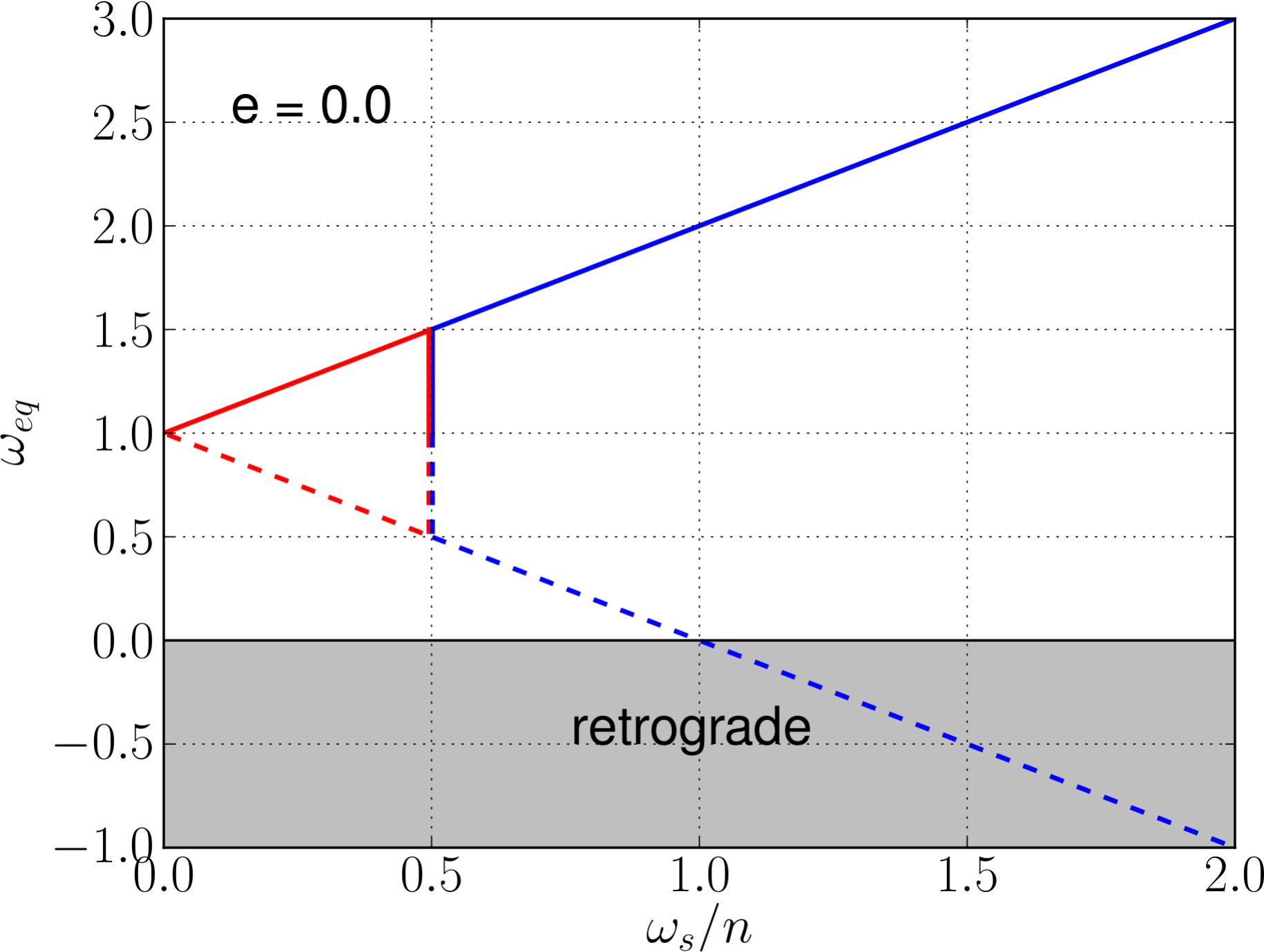}\put(211,160){\bf \Large (a)}\end{overpic} \\
\begin{overpic}[width=.95\columnwidth]{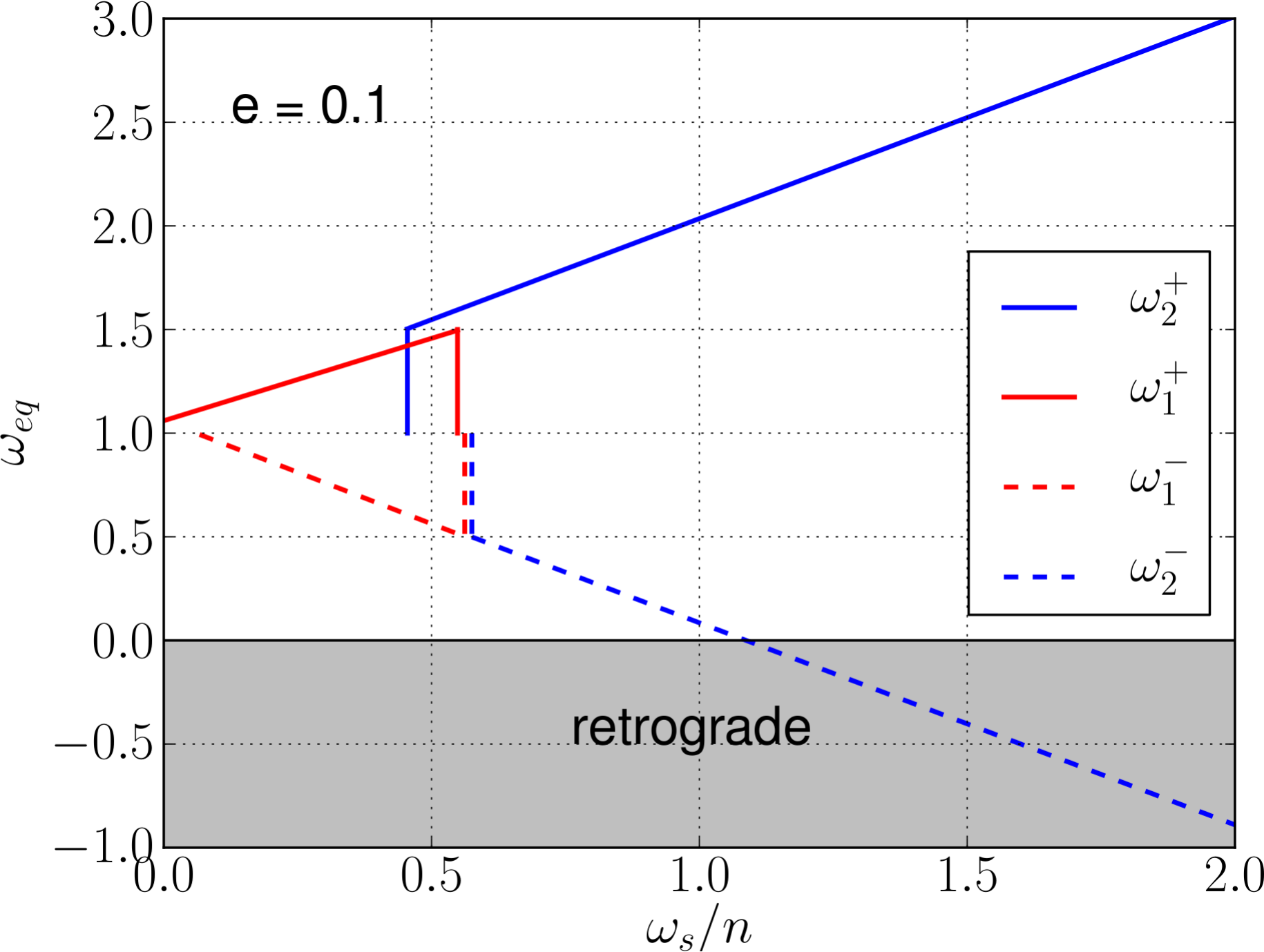}\put(211,160){\bf \Large (b)}\end{overpic} \\
\begin{overpic}[width=.95\columnwidth]{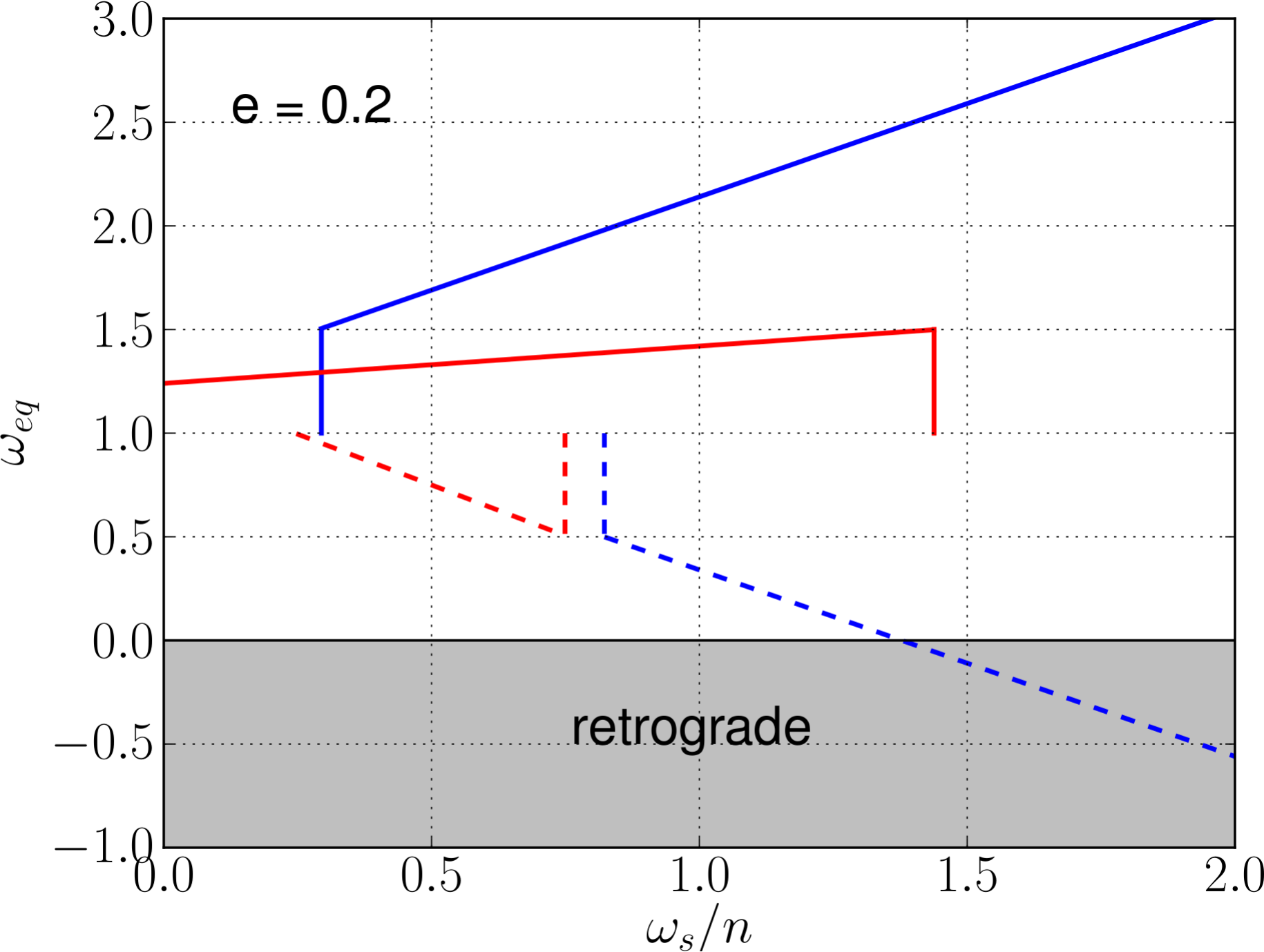}\put(211,160){\bf \Large (c)}\end{overpic} 
 \end{tabular}
\caption{\label{wsn}Equilibrium positions of the rotation rate as a function of the ratio $\omega_s/n$ for three different values of the eccentricity: \textbf{(a)} $e=0.0$, \textbf{(b)} $e=0.1$, and \textbf{(c)} $e=0.2$. The solid red line corresponds to the $\omega_1^+$ state, the dotted red to the $\omega _1^-$ state, the solid blue line to the $\omega_2^+$ state, and the dotted blue line to the $\omega_2^-$ state. }
\end{center}
\end{figure}

 In Figure~\ref{wsn} we plot the possible equilibrium rotation states as a function of $\omega_s/n$ for eccentricities of $e=0.0$,  0.1, and 0.2. 
Depending on  $\omega_s/n$ and the eccentricity values, the planet may have one, two, or three possibilities to evolve.
If  $\omega_s/n$ is close to 0.5 the planet has more possible final states (Eq.\,\ref{130617z}): 
if the eccentricity is zero, the planet has  two possible final states, but if the eccentricity of the planet is 0.1 or 0.2 there are three possibilities. 
We can also note that  for $e=0$, retrogade states appear 
when $\omega_s/n \ge 1$, but for $e=0.1 $ and $e=0.2$ the rotation rate must be slightly higher.
More exactly, it is needed that $\omega_s/n \ge (1 + 17e^2/2) $,
i.e.,  $\omega_s/n \ge 1.09$ for $e=0.1$, and $\omega_s/n  \ge1.34$ for $e=0.2$.

\subsection{Application to already known exoplanets}

Although we have witnessed in the last few years  an increase in the discovery of  Earth-sized exoplanets, the physical data for these planets are still scarce. 
Hence, we can only do some assumptions that allow us to infer some  constraints on their final spin evolution. 
We assume here that these planets are rocky with a dense atmosphere like Venus \citep{Alibert_etal_2006,Rafikov_2006}. 
We can call these Earth-sized exoplanets as {\it V-type planets}. 
Using the empirical mass-luminosity relation $L_*\propto \M^4 $ 
\citep{Cester_etal_1983} and the mass-radius relation for terrestrial planets\footnote{For transiting close-in planets with low density, the coefficient in the mass-radius relation appears to be higher than 0.274 \citep[e.g.][]{Lissauer_2011}, but as the density approaches the Earth's value, this relation fits more correctly \citep[e.g.][]{Pepe_etal_2013}.} 
$R\propto m^{0.274}$ \citep{Sotin_etal_2007}, we get from equation (\ref{eq:omegas1}):
\begin{equation}\frac{\omega_s}{n}\propto\frac{L_*R}{\M m}\frac{a}{n}=k(a \M)^{2.5}m^{-0.726} \ .
\label{eq:wsn}
\end{equation}
$k$ is a coefficient of proportionality  containing the parameters that we cannot constrain 
for these planets ($H_0$, $F_s$, $k_2$, $\Delta t_g$ and $\Delta t _a$).
Assuming that all unknown properties of the Earth-sized planets are the same as for Venus 
($2\pi/\omega_s=116.1$~day and so $\omega_s/n=1.9255$), we compute\footnote{We found here a slightly different value for $k$ than in \citet{Correia_Levrard_2008}, 
where the value was $k= 3.32\,(\mathrm{AU\,M}_\odot)^{-2.5}\,m_\oplus^{0.726}$. 
This difference was probably a misprint, 
because we found no problem with the expressions involving $k$ presented in that paper. 
}:
\begin{equation}
\label{k}
k=\frac{\omega_s}{n}(a\M)^{-2.5}m^{0.726}=3.73\,(\mathrm{AU}\,M_\odot)^{-2.5}\,m_\oplus^{0.726}
\end{equation}
Using the previous equation, we can estimate the $\omega_s/n$ ratio for all known Earth-sized planets (Table~\ref{tbl:1}).

\begin{figure}[ht!]
\begin{center}
 \begin{tabular}{c}
\begin{overpic}[width=.95\columnwidth]{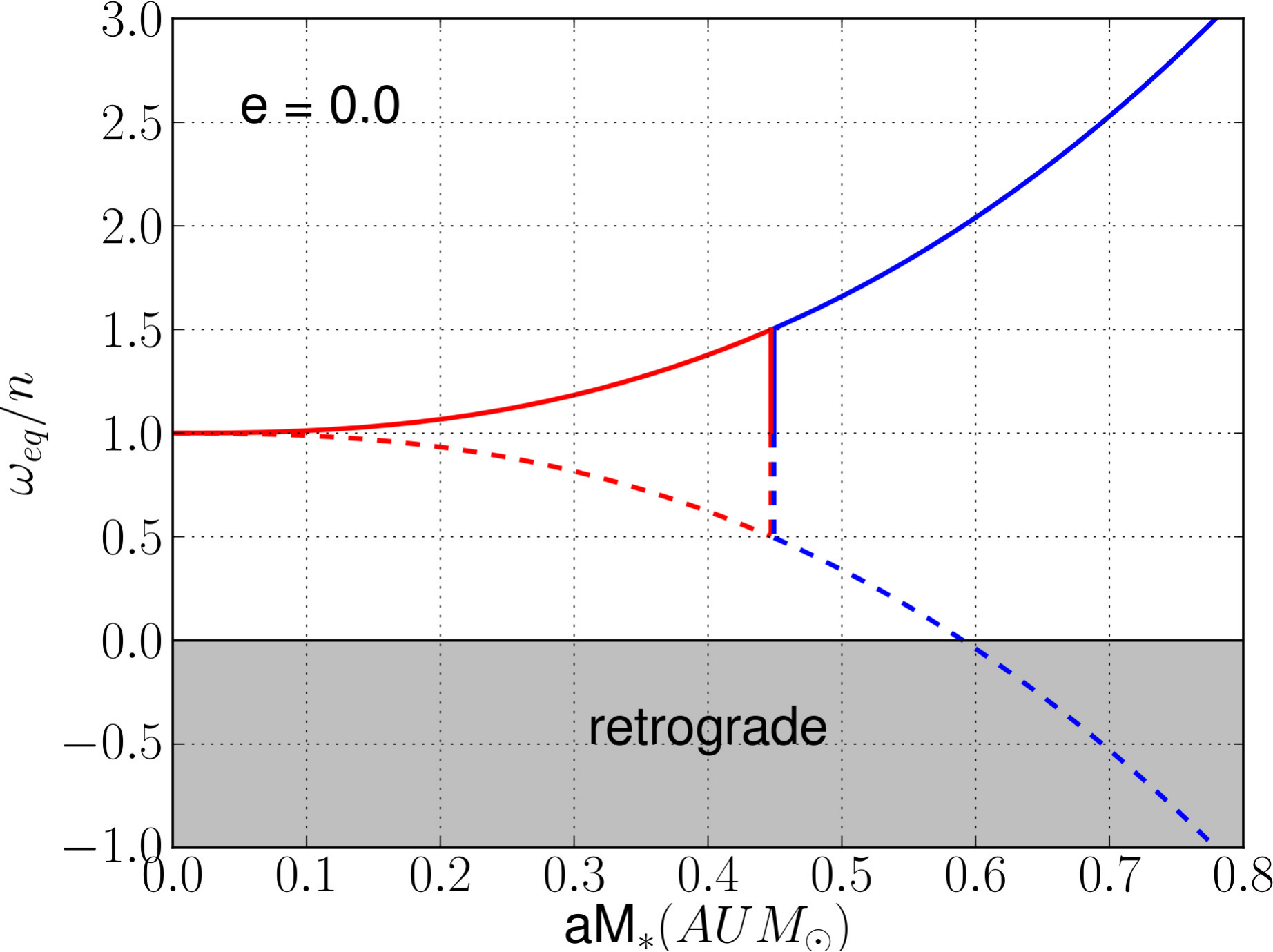}\put(211,160){\bf \Large (a)}\end{overpic} \\
\begin{overpic}[width=.95\columnwidth]{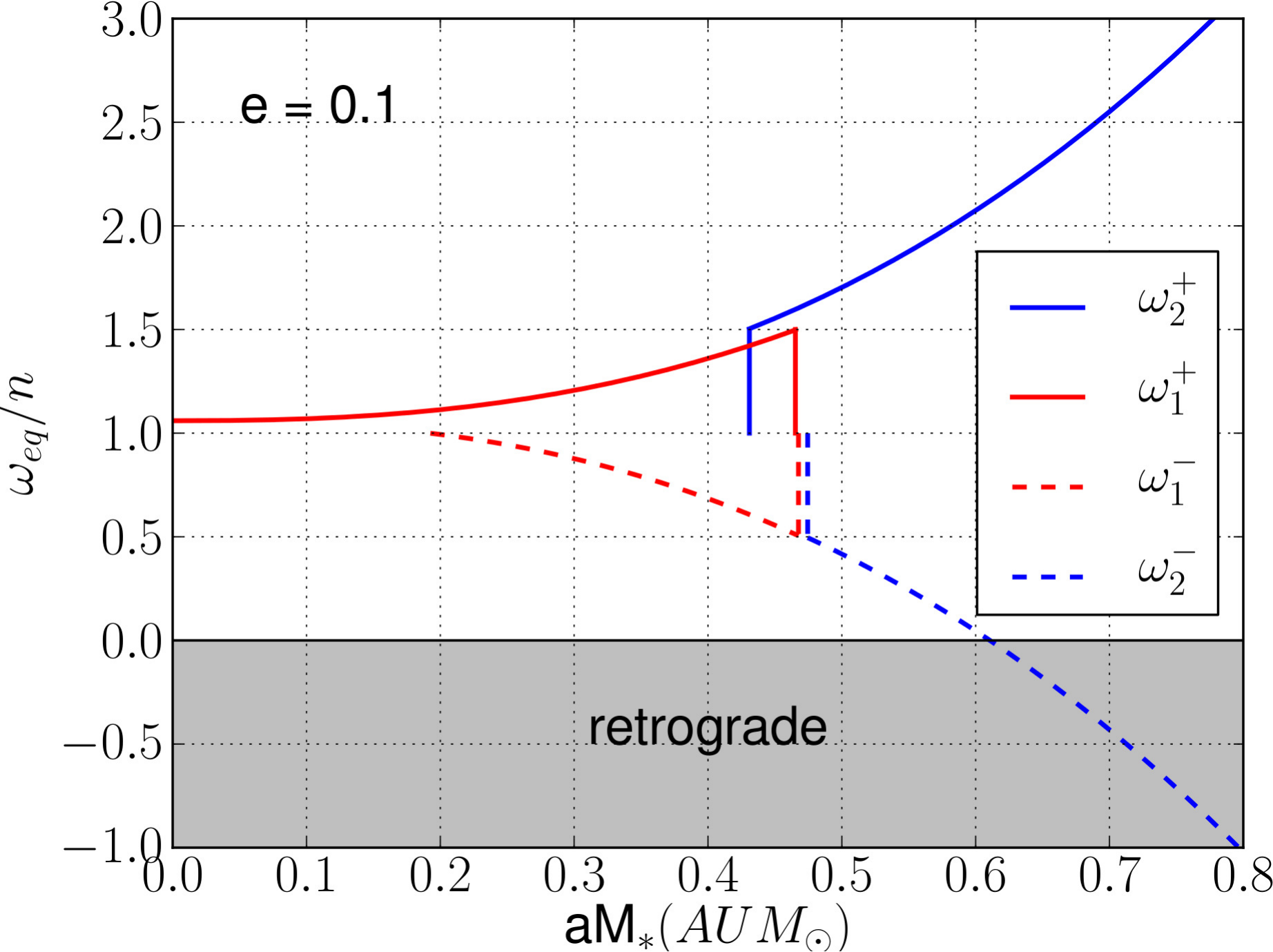}\put(211,160){\bf \Large (b)}\end{overpic} \\
\begin{overpic}[width=.95\columnwidth]{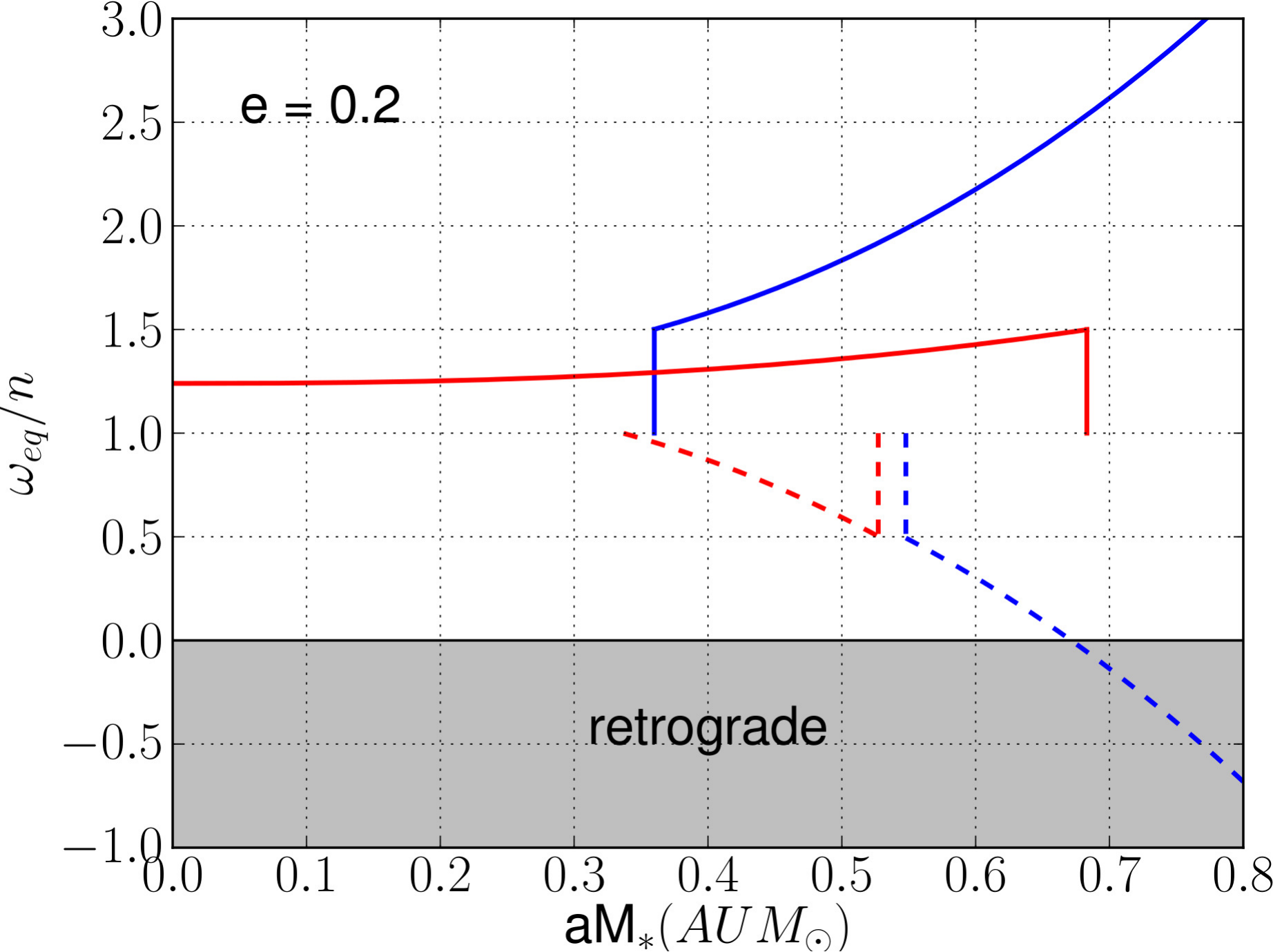}\put(211,160){\bf \Large (c)}\end{overpic} 
 \end{tabular}
\caption{\label{ecc}Equilibrium positions of the rotation rate as a function of the product $aM_{*}$ for three different values of the eccentricity: \textbf{(a)} $e=0.0$, \textbf{(b)} $e=0.1$ and \textbf{(c)} $e=0.2$. The solid red line corresponds to the $\omega_1^+$ state, the dotted red line to the $\omega _1^-$ state, the solid blue line to the $\omega_2^+$ state, and the dotted blue line to the $\omega_2^-$ state.}
\end{center}
\end{figure}

\begin{figure}[t!]
\begin{center}
 \begin{tabular}{c}
\hskip0.3cm\begin{overpic}[width=.95\columnwidth]{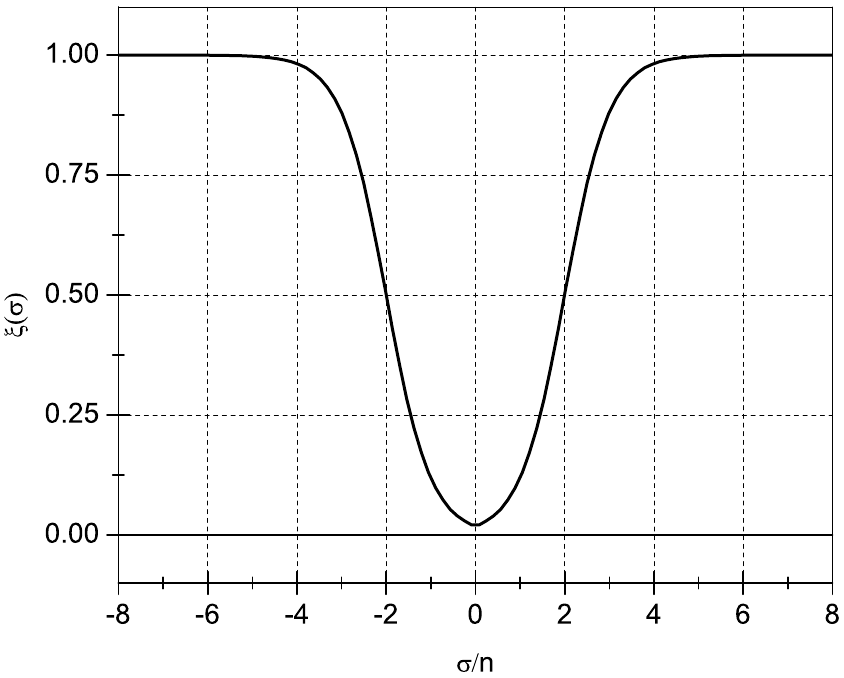}\put(40,174){\bf \Large (a)}\end{overpic} \\
\begin{overpic}[width=.98\columnwidth]{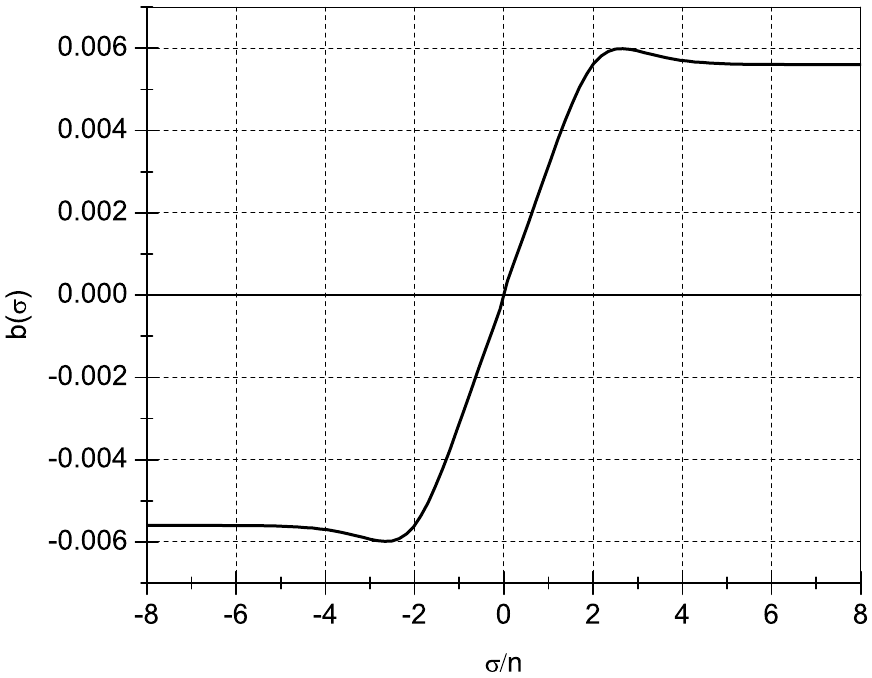}\put(48,174){\bf \Large (b)}\end{overpic} 
 \end{tabular}
\caption{\label{zero_function} 
\textbf{(a)} $\xi (\sigma)$ vs $\sigma$ with $\sigma_c / n = 2$;  \textbf{(b)} behavior of $b_g(\sigma)$ using the interpolated model smoothed  by $ \xi\,(\sigma)$.}
\end{center}
\end{figure}

As shown in  equation (\ref {eq:wsn}), $\omega_s/n$ is a function of the product $a\M$.
To see how the equilibrium rotation rate states depend on these parameters, 
in Figure~\ref{ecc} we plot the number of values of the equilibrium rotation states as a function of $a\M$, for eccentricities $e=0.0$,  0.1, and 0.2.
We can compare  Figure~\ref{ecc} with Table~\ref{tbl:1}, where the planet's parameters and the  possible final equilibrium 
states are presented. 
The actual rotation state of the planet Venus corresponds to the $\omega_2^-$ equilibrium state. 
For the remaining exoplanets here studied, only one equilibrium rotation state is possible, 
the $\omega_1^+$ state.
There is only one exception, planet HD\,40307\,$f$, which has two possible equilibrium states ($\omega_1^+$ and $\omega _1^-)$, but their values are too close to be 
distinguished in a future direct observation of the planet. 

The $a\M$ value for all planets listed in Table~\ref{tbl:1} is small
as a consequence of the present limitations of the detection techniques:
at present we are only  able  to detect Earth-sized planets orbiting 
very close to low-mass stars, for which gravitational tides dominate.

\subsection{Generalization to Earth-sized planets}
\label{App.Venus}
With the continuous increase in the precision of detection methods, we expect to soon find  planets orbiting Sun-like stars as distant from them as the Earth or Venus are from the Sun \citep[e.g.][]{EXPRESSOCODEX2007}.
Foreseeing this situation, we applied the previous calculations to long-period Earth-sized planets, that may eventually be found in the HZ of these systems.

Assuming that the planet's  mass and  radius are the same  of Venus and a host star with the mass of the Sun, we made  calculations
 for three different values of the eccentricity
($e=0$.0, 0.1, and 0.2), varying for each one the value of the semi-major axis $a$.
As expected, more possible equilibrium rotation states appear (Table~\ref{tbl:2}). 
For the same eccentricity
 we note that the difference in the rotation period of the several equilibrium  states is enlarged by the distance to the star.
 We can also see that for higher
values of $a\M$ retrograde states may be possible. Setting $ \omega
_2^- < 0$ (Eq.\,\ref{eq:w2-}) and substituting $\omega_s/n $ by equation (\ref{eq:wsn}), we may find the value of rotation 
rate needed for a retrograde state:
\begin{equation}
aM_{*}\ge \left[\frac{1}{3.73}\left(1+\frac{17}{2}e^2\right)\right]^{0.4}.
\label{130422a}
\end{equation}
Thus, for zero eccentricity  a retrograde state is possible when $a\M \gtrsim 0.59$, 
for $e=0.1$ a possible retrograde states appears when  $aM_{*} \gtrsim 0.61$,
while for an eccentricity of $e=0.2$, we need $a\M \gtrsim0.66$  for retrograde rotation.
This is also confirmed in Figure\,\ref{ecc}, where we can see that planets with eccentricities of $e=0.1$ and 0.2 have a retrograde equilibrium state
for a $a\M$  product higher than 0.6.
 Table~\ref{tbl:2} and Figure~\ref{ecc} are also in agreement on the number
 of states when $a\M$ is around 0.45~AU~M$_\odot$. Although only two states can be found when the eccentricity is zero,
 for eccentricities of 0.1 and 0.2 we  find three equilibrium states.

\section{Numerical Simulations}

\subsection{An Interpolated Dissipative Model}

\label{sec:newmodel}

During the spin evolution of the planets, the tidal frequency varies and 
so does the dissipation factors $b(\sigma)$ (Eqs.\,\ref{130422z}, and \ref{130422y}).
Because the rheology of the planets is poorly known, the exact dependence of the dissipation on the tidal frequency is unknown.
In section~\ref{021024j} we described the most commonly used dissipation models (Fig.\,\ref{figmodels}).
However, the visco-elastic model seems inappropriate for atmospheric tides, while the viscous one is not realistic for $\sigma \gg n$, and the  constant$-Q$ for $\sigma \ll n$.
More complex models exist \citep[e.g.][]{Efroimsky_Williams_2009}, but for simplicity we adopt here an interpolated model that behaves like the linear model for small values of $\sigma$ and like the constant$-Q$ model when $\sigma$ increases \citep[e.g.][]{Cuk_Stewart_2012}.
This model is also in agreement with the analytical simplifications from section~\ref{dynanal}.
It is also very close to the models obtained by \citet{Remus_etal_2012b} in their construction of a tidal model for small and moderate spin rates based on hydrodynamical equations \citep[see Fig.\,4 in][]{Remus_etal_2012b}.

For gravitational tides the time-lag is then given by
\begin{equation}
b_g(\sigma) = k_2 \sigma \Delta t_g(\sigma)= \frac{k_2}{2 Q_n}  \left[ \frac{\sigma}{n} +\left(2 \, \mathrm{sign}(\sigma)-\frac{\sigma}{n} \right)\xi(\sigma)\right] \ , \label{130612a}
\end{equation}
where  $\xi(\sigma) $ is a function varying between 0 and 1, used to
make a smooth passage between the two regimes:
\begin{equation}\xi\, (\sigma)=
\frac{1}{2}+\frac{1}{2} \tanh\left(\frac{|\sigma|}{n}-2\right).
\label{eq:zeroone}
\end{equation}
In Figure~\ref{zero_function} we plot the above expressions as a function of the tidal frequency. 
The transition of regime occurs for $\sigma_c = 2 n $, and $Q_n$ corresponds to the dissipation $Q-$factor during the constant phase.

For thermal atmospheric tides\footnote{Actually, expression (\ref{eq:p2}) is only valid as long as $|\tilde p_2| \ll \tilde p_0 $ (the average pressure at the ground). To avoid discontinuities when $\sigma=0$, in our numerical simulations we adopt an interpolate function for $\tilde p_2 (\sigma)$ as in \citep[][Eq.\,46]{Correia_etal_2003}.}, the time-lag $\Delta t_a(\sigma)$ is simply obtained from expression (\ref{eq:omegas1}):
\begin{equation}
\Delta t_a (\sigma) = \Delta t _g (\sigma) \frac{16 H_0 K_g k_2}{K_a F_s} \omega_s
\ . \label{130422x}
\end{equation}

\subsection{Choice of Parameters}

Since much is unknown about  exoplanets, we need to make some assumptions in our simulations. 
As we already stated, we consider here {\it V-type planets}.
We assume that these planets with masses lower than $10\,M_\oplus$ have a structure similar to the terrestrial planets of the Solar System, and a dense  atmosphere capable of influencing their spin evolution.
Thus, for a given  exoplanet, we adopt for the unknown parameters the corresponding value from Venus\footnote{The Earth and Venus' parameters are very similar, in particular when we want to extrapolate for exoplantes with masses ranging from 0.1 to 10~$M_\oplus$. We prefered to use Venus' parameters in order to better compare with previous studies on the Solar System.}.
Therefore, in our simulations we take the potential Love number $k_{2}  = 0.28$, $Q_n   = 50$, the  mean density 
$\bar  \rho= 5.24\times10^3\,\mathrm {Kg\,m}^{-3}$, and the stellar energy that reaches  the planet surface $F_s = 100\,\mathrm{Wm}^{-2}$.
For the mantle dynamic ellipticity we adopt  $\delta E_d= 1.3\times10^{-5} $ and for the kinematic viscosity $\nu=10^{-6} \, \mathrm{m}^2 \mathrm{s}^{-1} $.
For the core radius $R_c=3.1\times10^6\,\mathrm m$,  the planet structure coefficient $ C/mR^2= 0.336$,
 and the ratio between the  core moment of inertia and moment of inertia $C_c/C = 0.084$ \citep{Yoder_1995,Yoder_1997}.  



 In our simulations we start with $P_{in}=1$ or 2~day, a rotation period that is faster than  the predicted present rotation. 
Since some of these planets orbit very close to the star, and hence end up with fast equilibrium rotations 
(e.g. planet 55\,Cnc\,$e$, whose final rotation is estimated to be 0.733~day), we also tested the possibility of 
planets evolved from a slower rotation to a faster one, with
 $P_{in}=25\,\mathrm{day}$.
The results of these simulations are presented in the next two sections.

\subsection{Application to already known exoplanets}

\label{realplanets}

\begin{figure*}[h!tb]
\centering
\begin{tabular}{c c c}
\begin{overpic}[angle=-90, width=0.9\columnwidth]{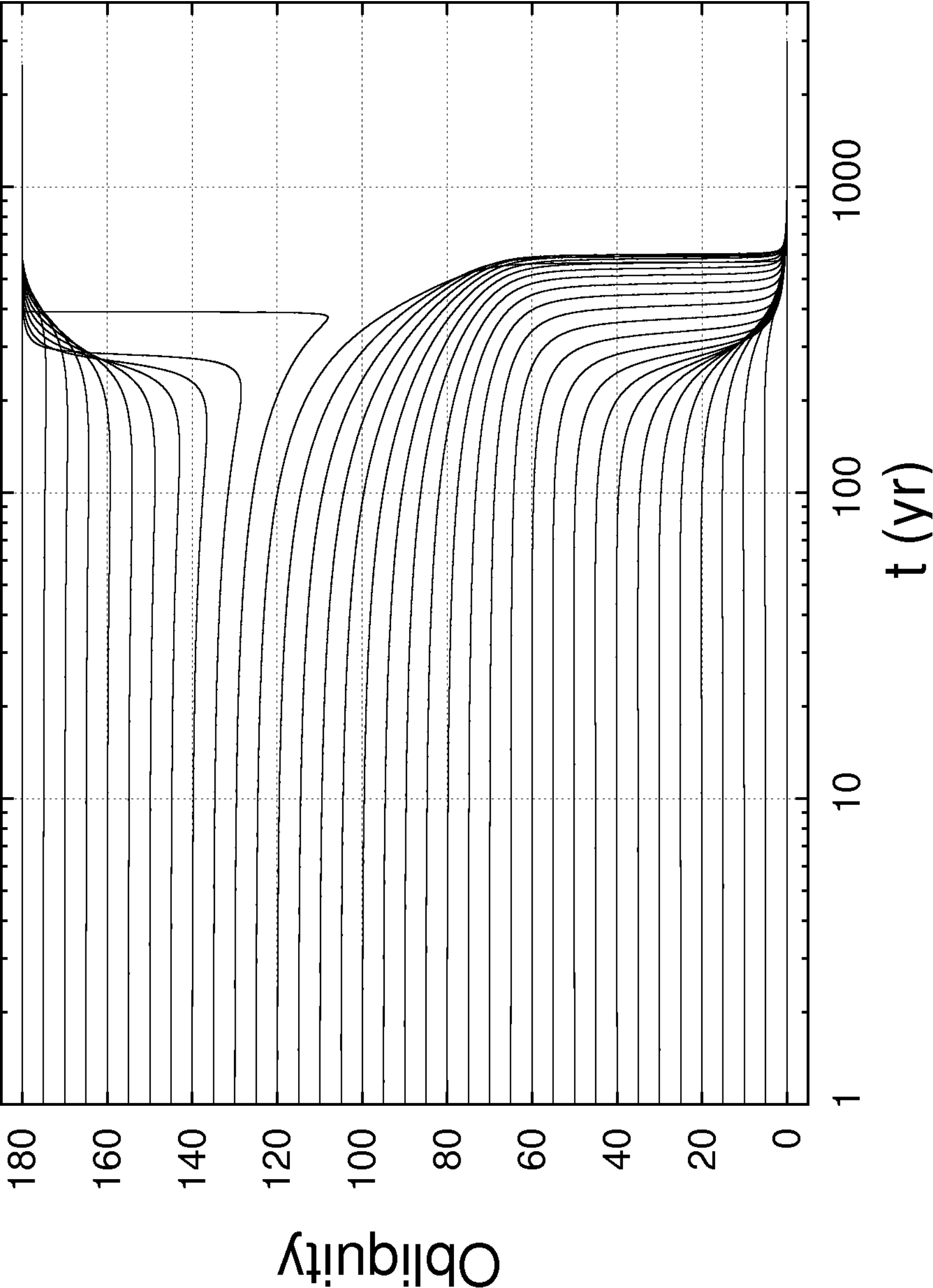}\put(40,140){\bf \Large (a)}\end{overpic} & &
\begin{overpic}[angle=-90, width=0.9\columnwidth]{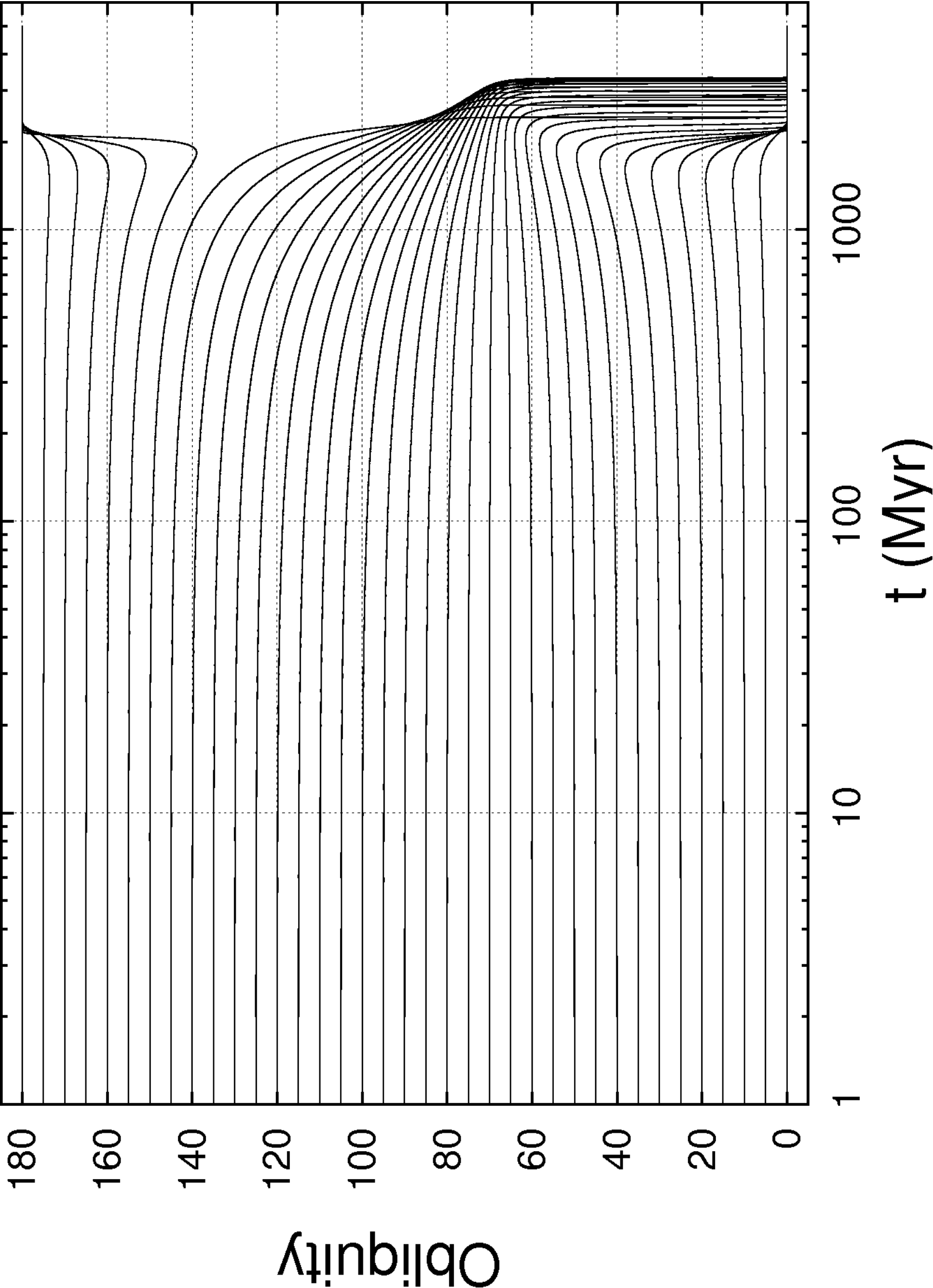}\put(40,140){\bf \Large (b)}\end{overpic} \\
\begin{overpic}[angle=-90, width=0.9\columnwidth]{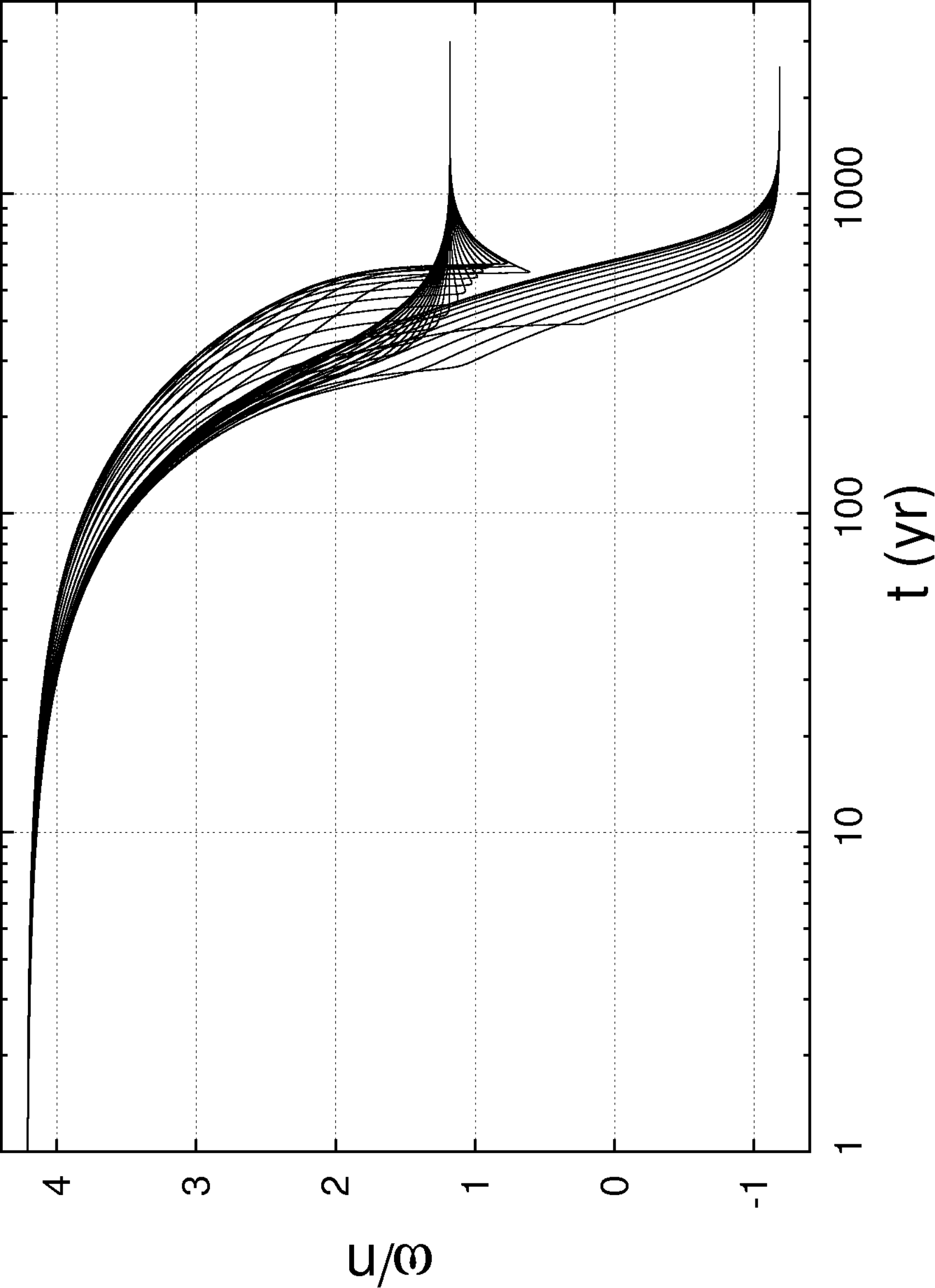}\put(35,140){\bf \Large (c)}\end{overpic} & &
\begin{overpic}[angle=-90, width=0.9\columnwidth]{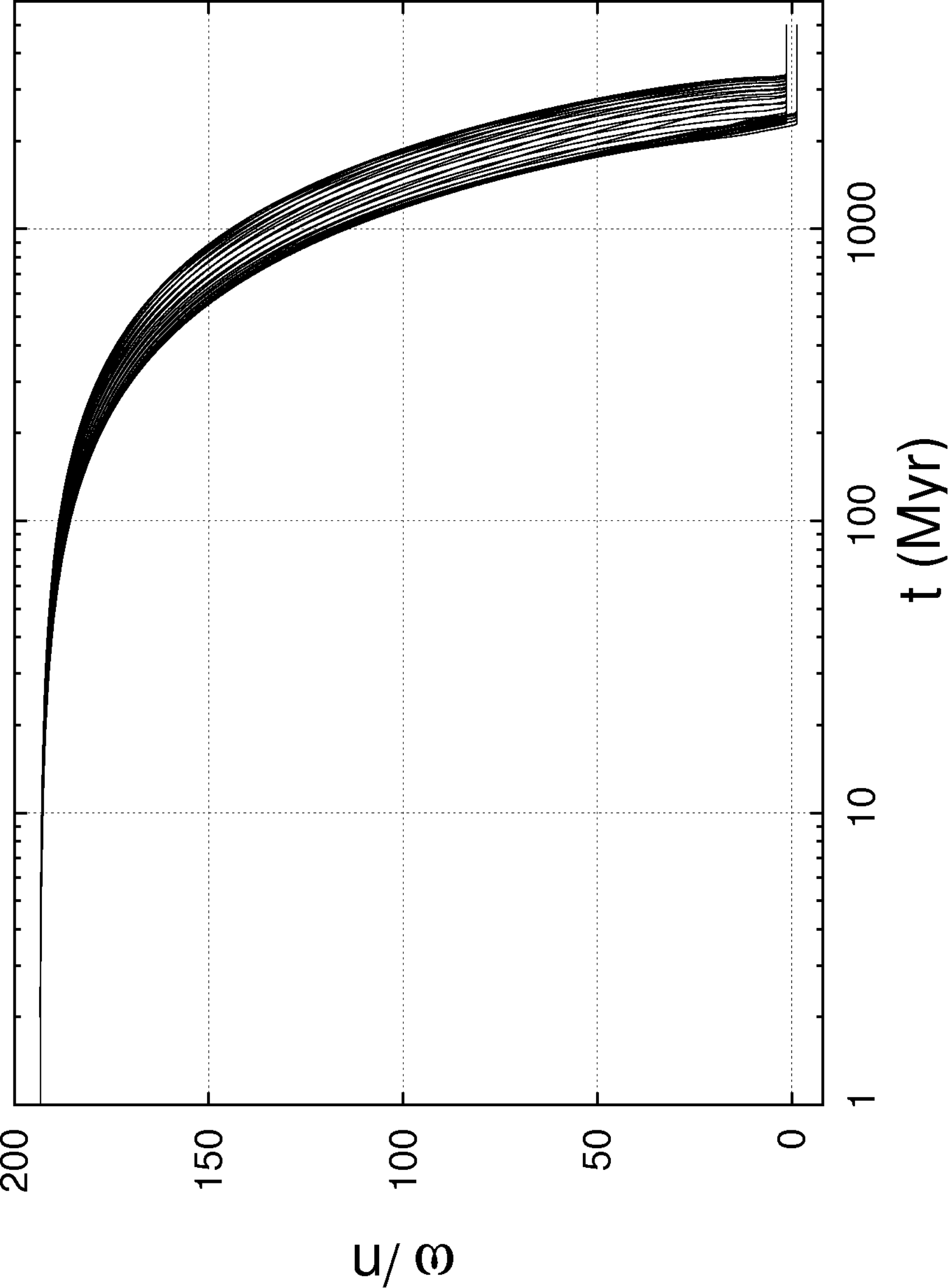}\put(35,140){\bf \Large (d)}\end{overpic} \\
\end{tabular}
\caption{Spin evolution with time for the planets HD\,40307\,$b$ \textbf{(left)} and HD\,40307\,$g$ \textbf{(right)}, with $P_{in}=1\,\mathrm{day}$ and initial obliquities ranging from $\varepsilon=0^\circ \;\mathrm{to}\;180^\circ$. 
We plot the obliquity \textbf{(top)} and $\omega/n$ \textbf{(bottom)} evolution.
Each line represents a different initial obliquity value. 
The lower lines in the $\omega/n$ plot corresponds to the initial obliquities closer to $180^\circ$.}
\label{HD-obl.t}
\end{figure*}

In Figure~\ref{HD-obl.t} we show the spin evolution for the planets HD\,40307\,$b$ and $g$ with time. We adopted $P_{in}=1\,\mathrm{day}$ and initial obliquities ranging from  $0^\circ$ to $180^\circ$, with a step of 5$^\circ$.
Although the initial rotation rate is the same for planets $b$ and $g$, the initial ratio $\omega/n$ is much higher for planet $g$, because the mean motion $n$ is smaller.
As a  consequence, the characteristic evolution time-scale is about 100~yr for planet $b$, while it is 1~Gyr for planet $g$. Those values are in agreement with the ones computed using equation (\ref{eq:tau_eq}) (Table\,\ref{tbl:1}).
After that time all trajectories reach a final equilibrium for the spin, but
depending on the initial value of the obliquity, the exact time can be different.
Contrary to what we could expect, initial obliquities close to 180$^\circ$,
usually take less time to reach the equilibrium than lower initial values.
This can be understood, since for high obliquities there are more harmonics with significant value that contribute to the spin evolution (Eqs.\,\ref{eq:Tg1}, \ref{eq:Tg1b}).

At first glance it may seem that there are two final equilibrium rotation states: one for final obliquities equal to 0$^\circ$, and another for 180$^\circ$. 
However, all initial obliquities evolve into the $2 \pi / \omega_1^+$ 
final rotation state, the only possibility for these two planets (Table\,\ref{tbl:1}).
For the initial obliquities close $0^\circ$, the rotation rate is always positive and the obliquity is decreased to zero degrees.
For initial obliquities close $180^\circ$ the obliquity evolves into this value, while the rotation rate slows down until zero and then increases in the other way, until it stabilizes at a negative value $-2 \pi / | \omega_1^+ | $. 
This is the same as having a planet with zero obliquity rotating prograde, that is, the couple ($-\omega$, $ \pi-\varepsilon$) is mathematically equivalent to ($\omega$, $ \varepsilon$).
Therefore, there are two different paths, but they lead to the same final equilibrium.
This has also been described for Venus \citep{Correia_Laskar_2001,Correia_Laskar_2003I}.

It is also interesting to note that for initial obliquities around 90$^\circ$, the rotation rate decreases to a value lower than that  of equilibrium, and then it increases again until equilibrium is reached.
This behavior is in agreement with expression (\ref{eq:weq}).

Comparing the rotation rate and the obliquity evolutions shown in Figure\,\ref{HD-obl.t} we also note that the obliquity is initially slightly constant and takes more time to reach the equilibrium than the rotation rate.
The explanation for this behavior is given in section \ref{sec:oblvariation}.
As we can see from expression (\ref{eq:dedw}), the obliquity variation is inversely proportional to the rotation rate. 
Thus, for a fast initial rotation, the obliquity does not change much.
This is particularly visible for HD\,40307\,$g$, since the initial rotation $\omega \gg n$.
However, as the rotation rate decreases we observe a strong variation in the obliquity.
Finally, the obliquity slowly decreases into zero, since $dx/dt \propto (1-x^2) \Leftrightarrow d \varepsilon / dt \propto - \sin \varepsilon $ (Eq.\,\ref{V31N}).

In Figure~\ref{EvolucaoExoplanets1-1} we show the obliquity evolution with the rotation rate $\omega/n$ for many systems listed in Table\,\ref{tbl:1} (including HD\,40307\,$b$ and $g$), always starting with an initial rotational period $P=1$\,day.
We observe again the dichotomy in the obliquity evolution: it can only end up in $0^\circ$ or $180^\circ$. This result was demonstrated in \citet{Correia_etal_2003} for Venus' parameters, but it seems to remain valid for all Earth-sized planets, even for those in eccentric orbits.
This is in agreement with the assumption $\varepsilon \simeq 0^\circ$ that we made when looking for final states (section {\ref{130510b}).

As a result of the obliquity's dichotomy, it appears that all planets have two possible final states, one prograde and another retrograde.
For most of them only one final rotation state is possible, we are just observing the equivalent couples ($\omega$, $\varepsilon$) and ($-\omega$,  $\pi-\varepsilon$).
However, in one case, for HD\,40307\,$f$ (Fig.\,\ref{EvolucaoExoplanets1-1}e), we can indeed observe two final rotation possibilities, the states $\omega_1^+$ and $\omega_1^-$ (Table\,\ref{tbl:1}).
Because these two states are so close to each other, it is nonetheless very difficult to distinguish them in the figure.

For all planets shown in Figure~\ref{EvolucaoExoplanets1-1} we also observe that the obliquity evolves into its final position at $0^\circ$ or $180^\circ$ always before $\omega/n = 0$.
The reason is that for very slow rotation rates, core-mantle friction becomes dominating over tidal effects (Eq.\,\ref{V32a}), quickly driving the obliquity to its final position.
Therefore, the transition from positive into negative rotation rates only occurs at obliquities very close to $0^\circ$ or $180^\circ$.

In the case of 55\,Cnc\,$e$ (Fig.\,\ref{EvolucaoExoplanets1-1}g) we initially have $\omega/n \approx 0.76$, that is, the initial rotation period of the planet is slower than the orbital period.
As a consequence, contrarily to the remaining examples in Figure~\ref{EvolucaoExoplanets1-1}, the rotation rate needs to increase its value to reach the equilibrium value near the synchronous rotation.
We indeed observe this behavior for initial obliquities close to $0^\circ$. 
However, for initial obliquities higher than $30^\circ$, the rotation rate still decreases to a lower ratio  $\omega /n < 0.76$, before evolving into to the final equilibrium rotation.
The explanation for this behavior can be found in equation (\ref{eq:weq}):
for non-zero obliquities, the equilibrium rotation rate is below the synchronous rotation, so as long as the obliquity remains high we have $0 < \omega/n < 1$.

\begin{figure*}
\begin{center}
\begin{tabular}{c c c}
\begin{overpic}[angle=-90, width=0.62\columnwidth]{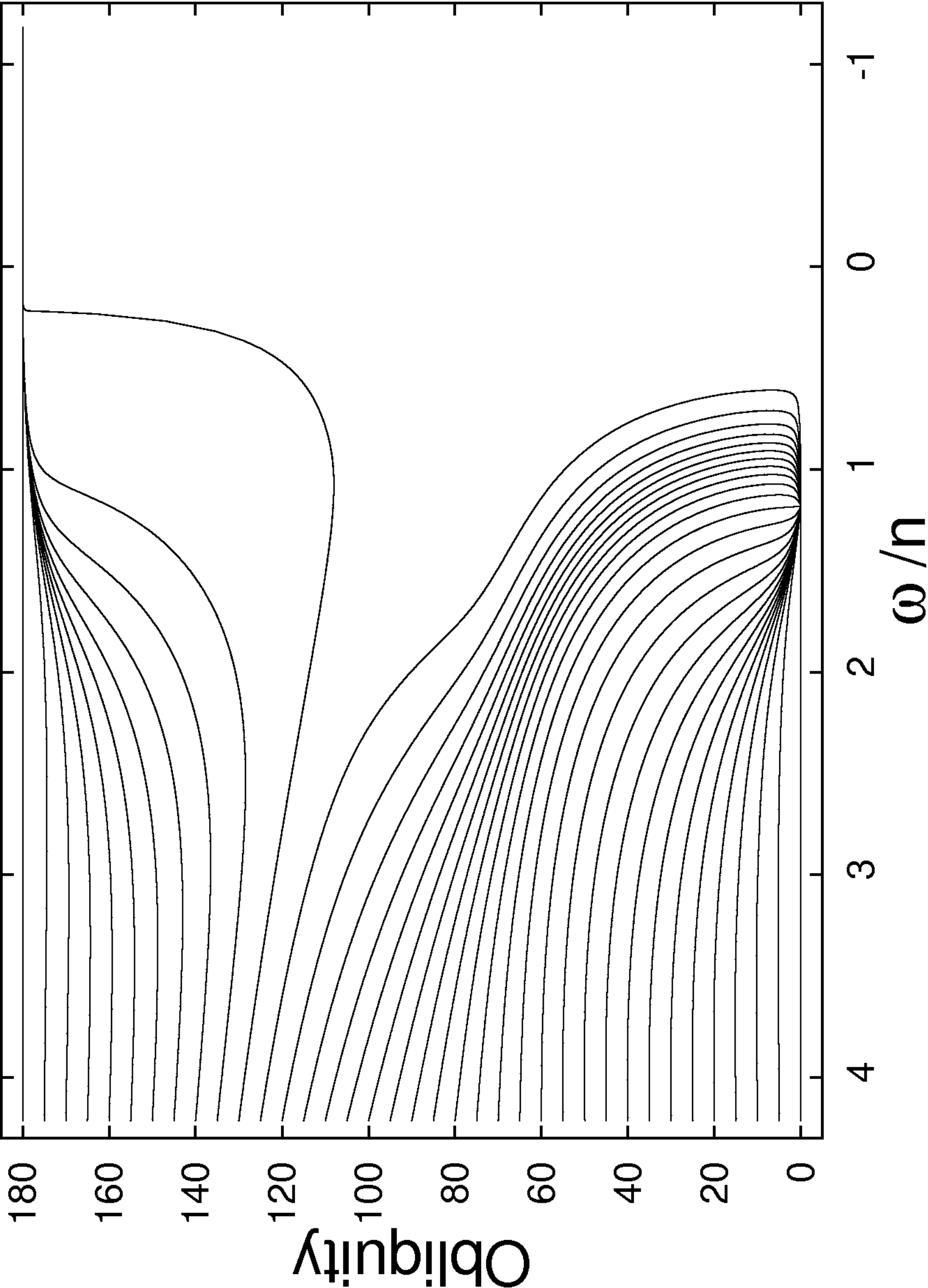}\put(23,100){\bf \large (a)}\end{overpic} &
\begin{overpic}[angle=-90, width=0.62\columnwidth]{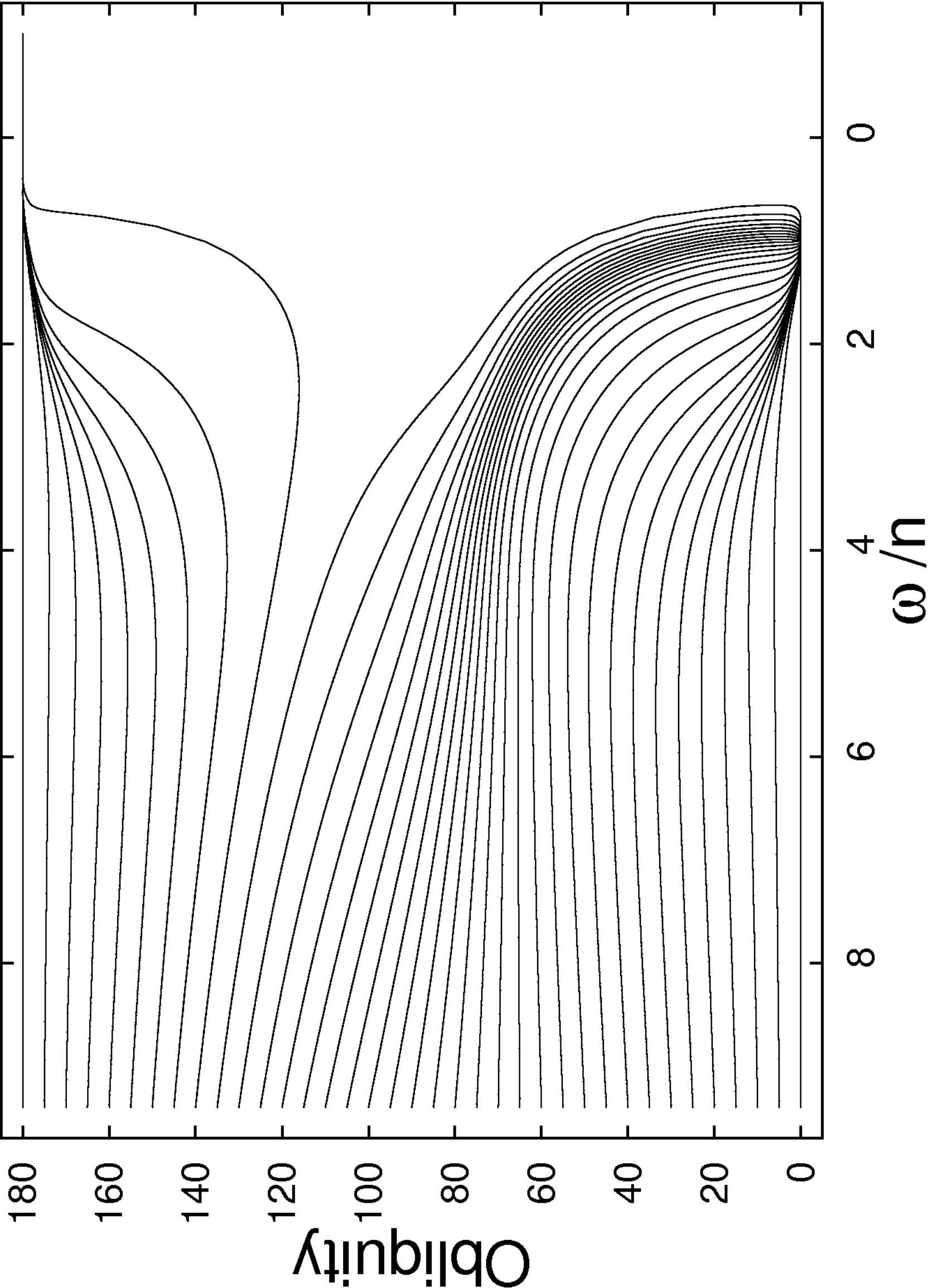}\put(23,100){\bf \large (b)}\end{overpic} &
\begin{overpic}[angle=-90, width=0.62\columnwidth]{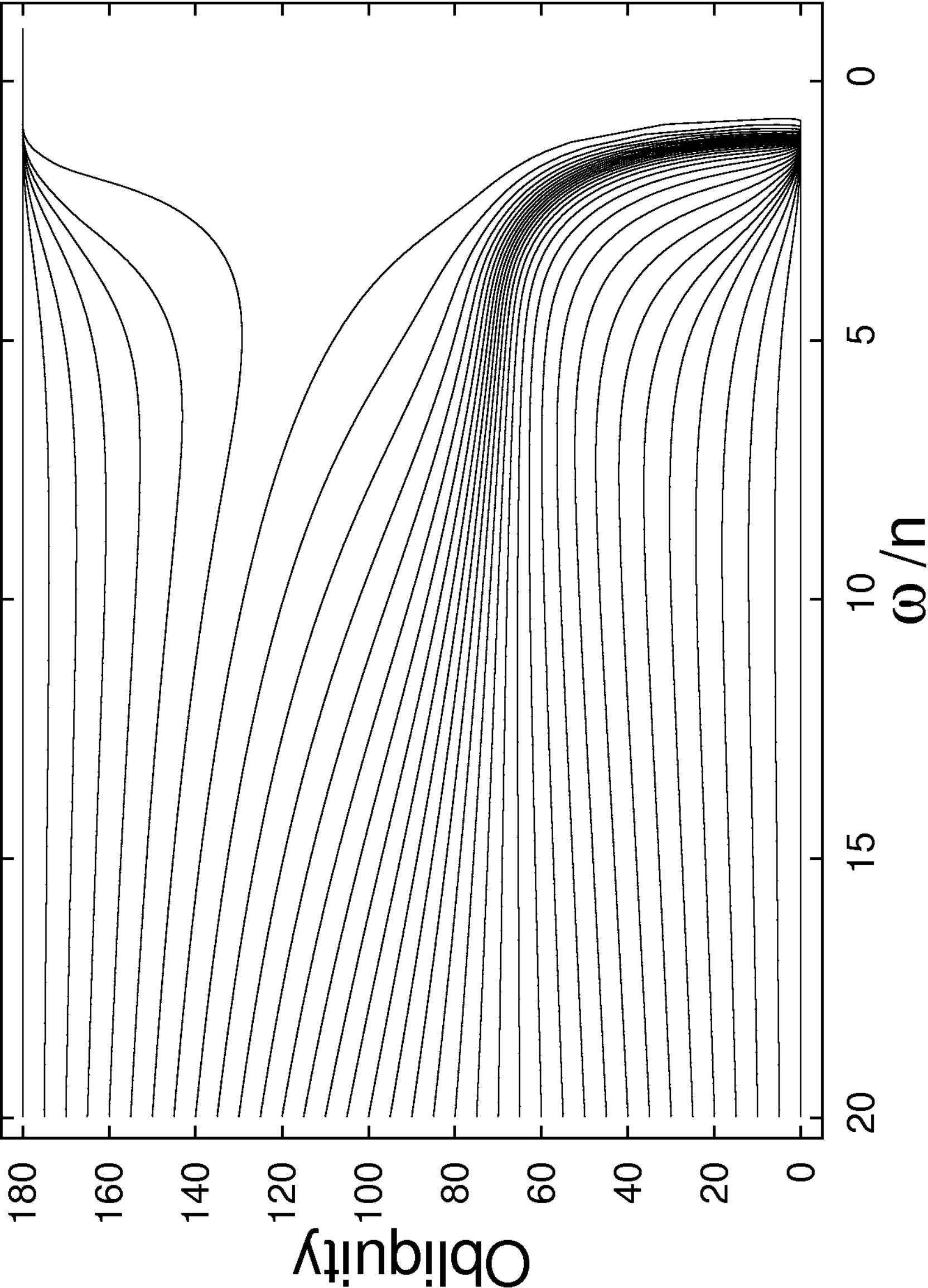}\put(23,100){\bf \large (c)}\end{overpic} \\
\begin{overpic}[angle=-90, width=0.62\columnwidth]{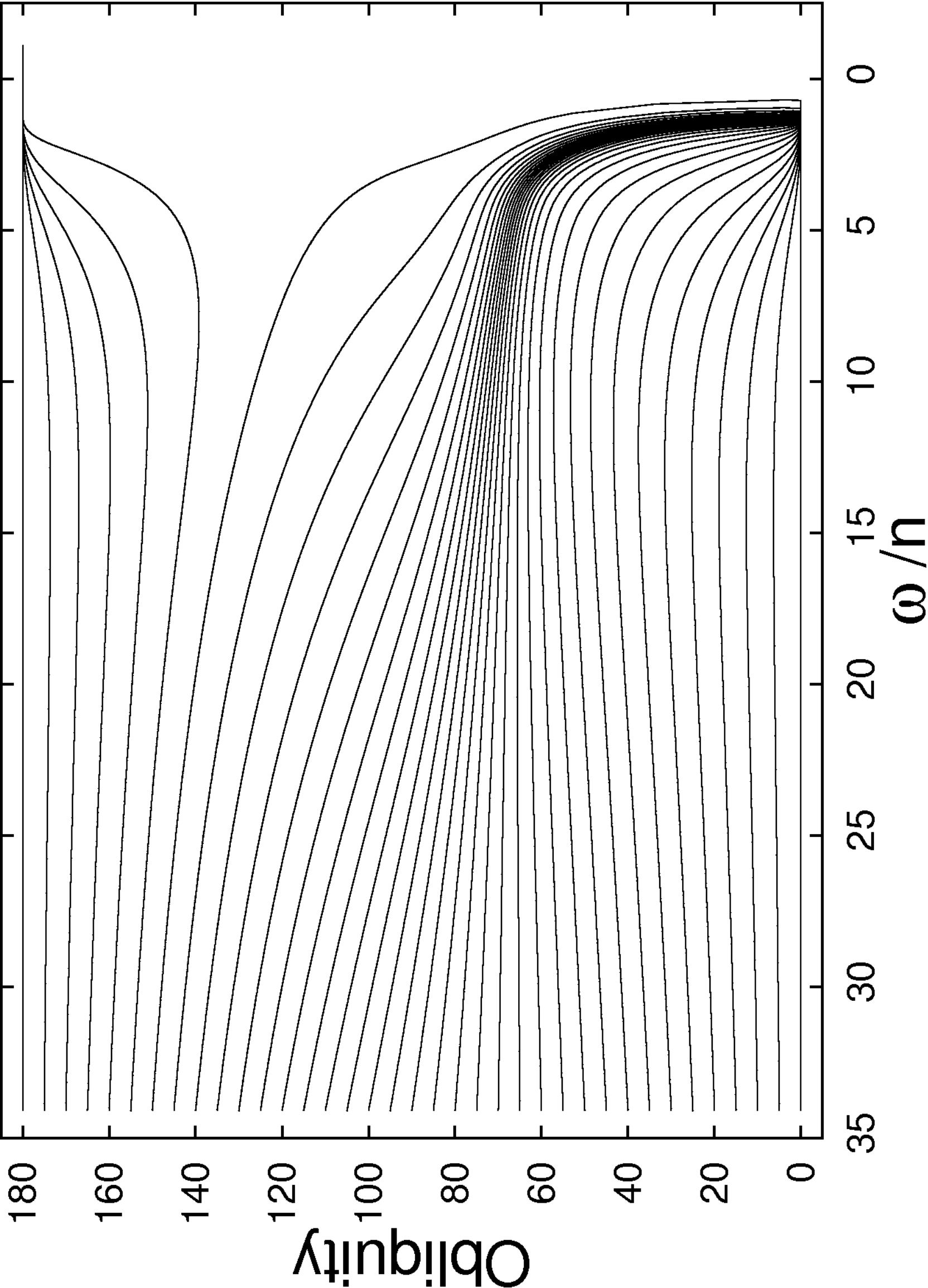}\put(23,100){\bf \large (d)}\end{overpic} &
\begin{overpic}[angle=-90, width=0.62\columnwidth]{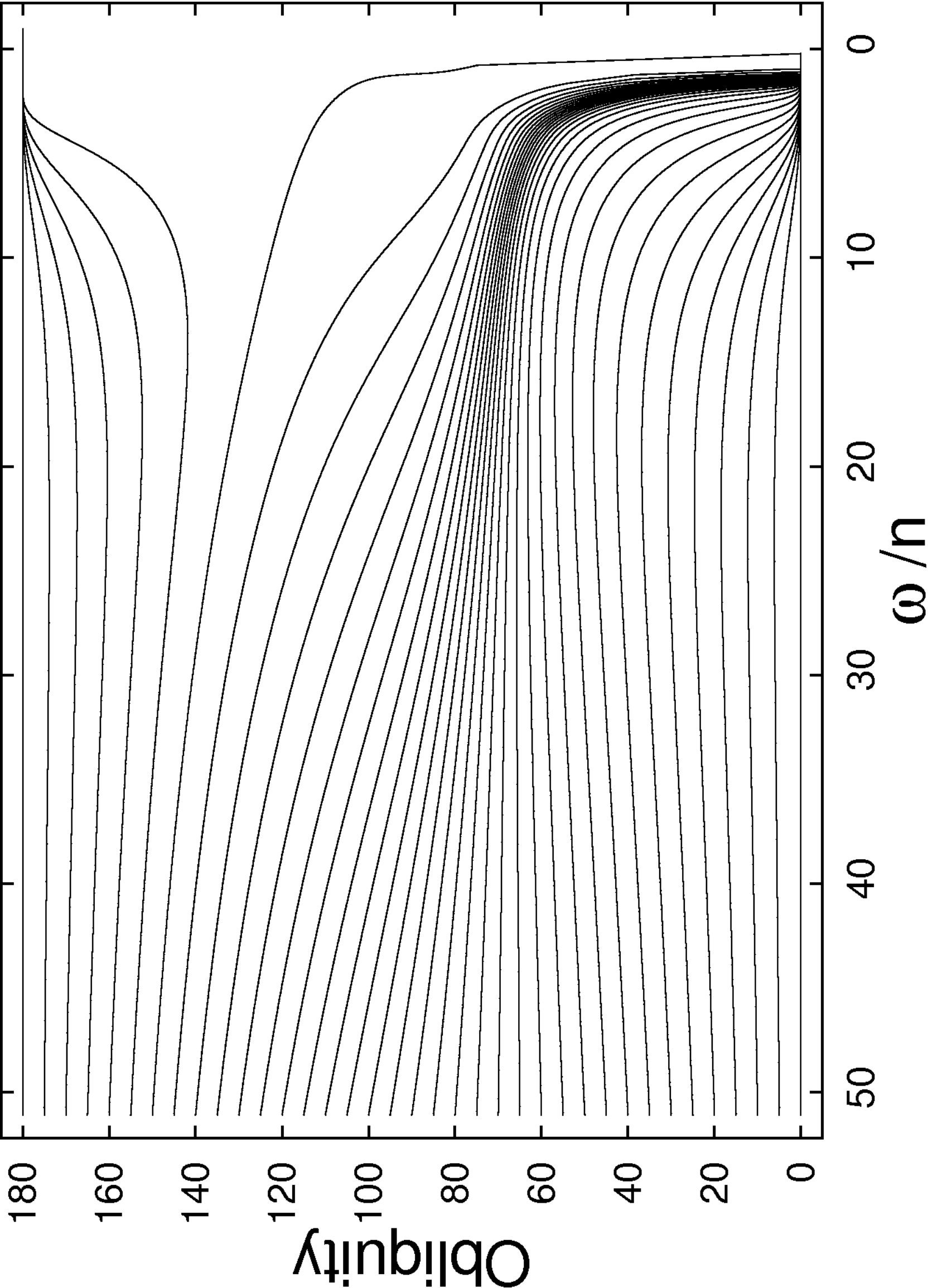}\put(23,100){\bf \large (e)}\end{overpic} &
\begin{overpic}[angle=-90, width=0.62\columnwidth]{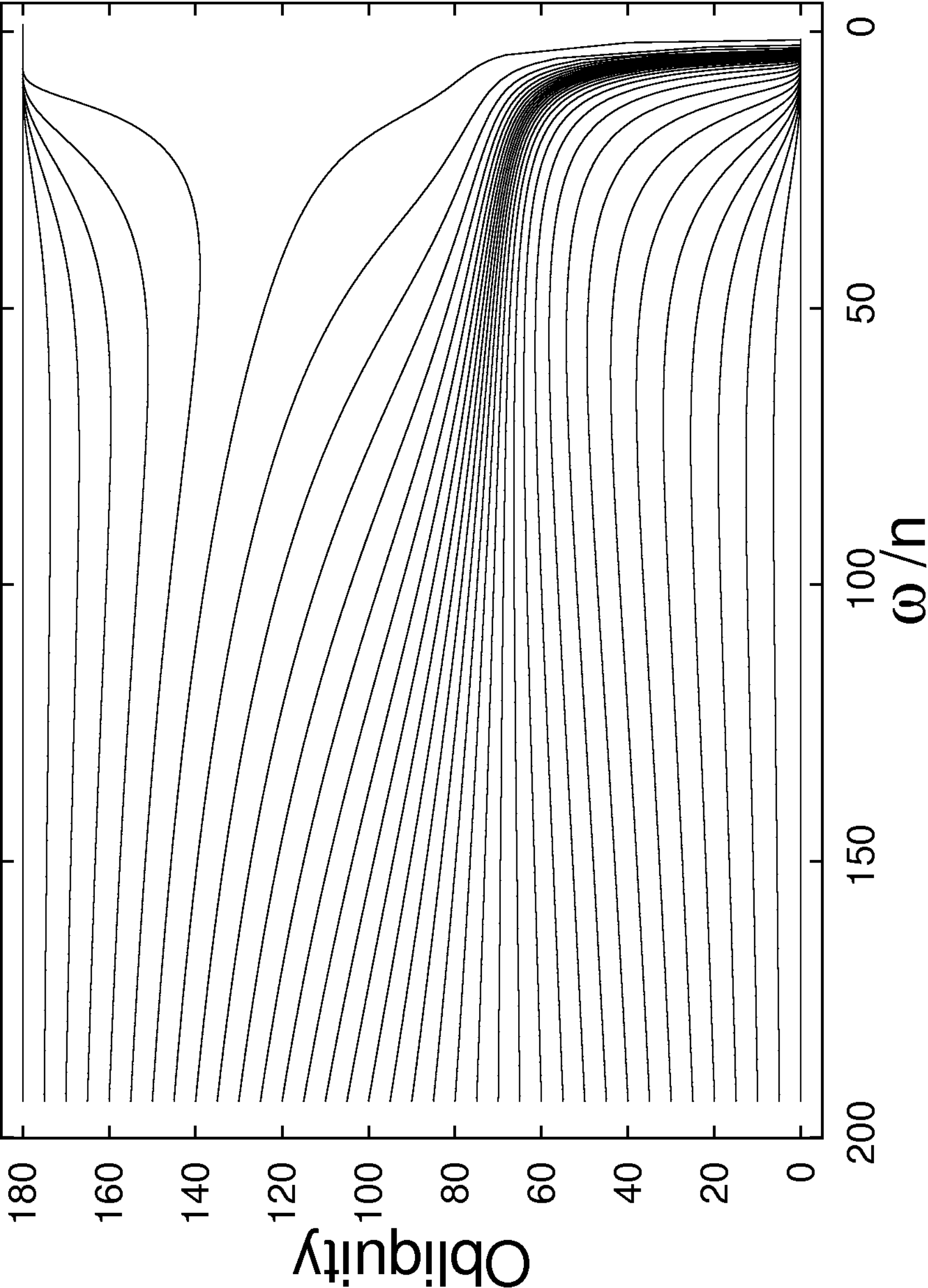}\put(23,100){\bf \large (f)}\end{overpic} \\
\begin{overpic}[angle=-90, width=0.62\columnwidth]{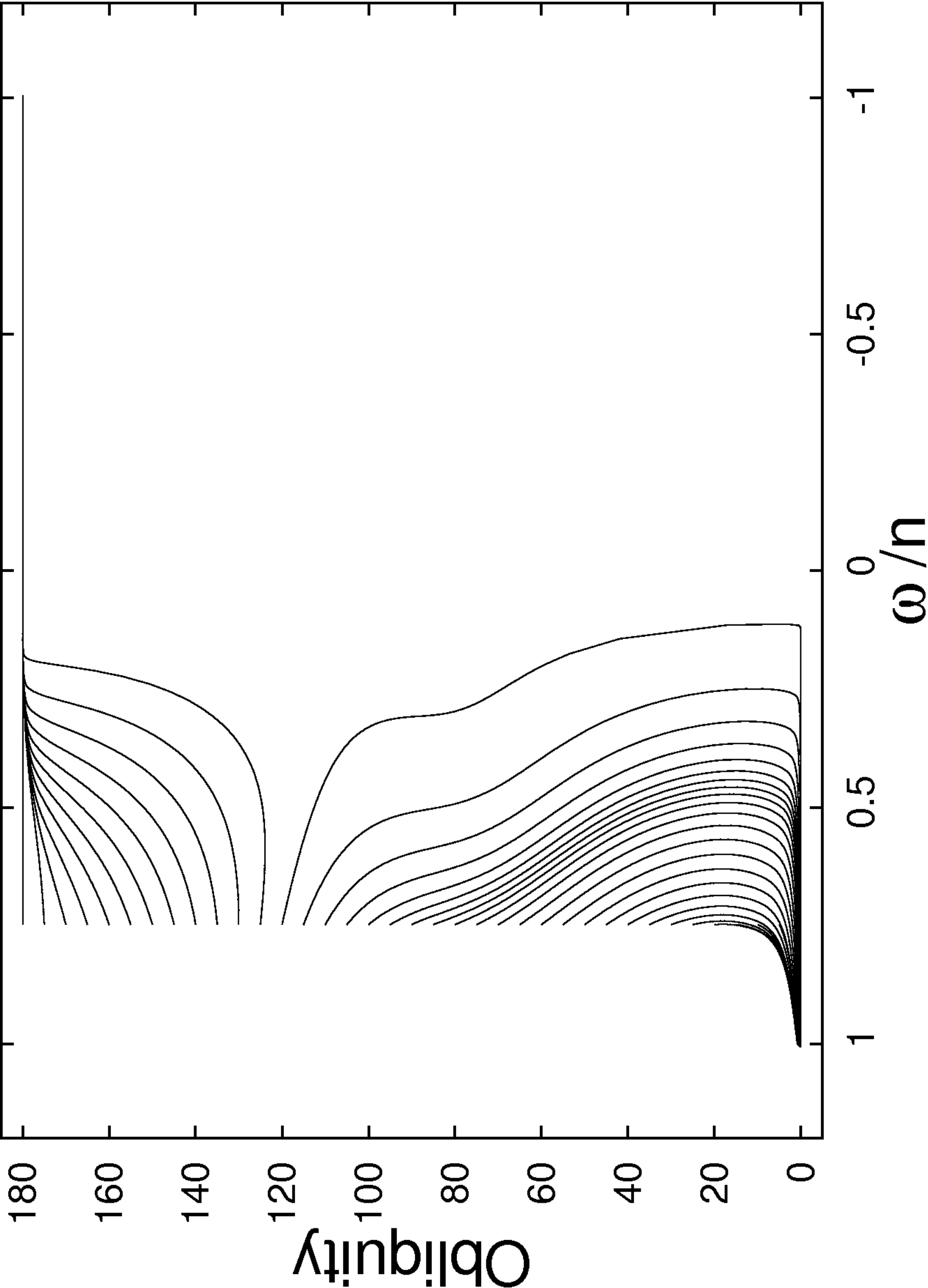}\put(23,100){\bf \large (g)}\end{overpic} &
\begin{overpic}[angle=-90, width=0.62\columnwidth]{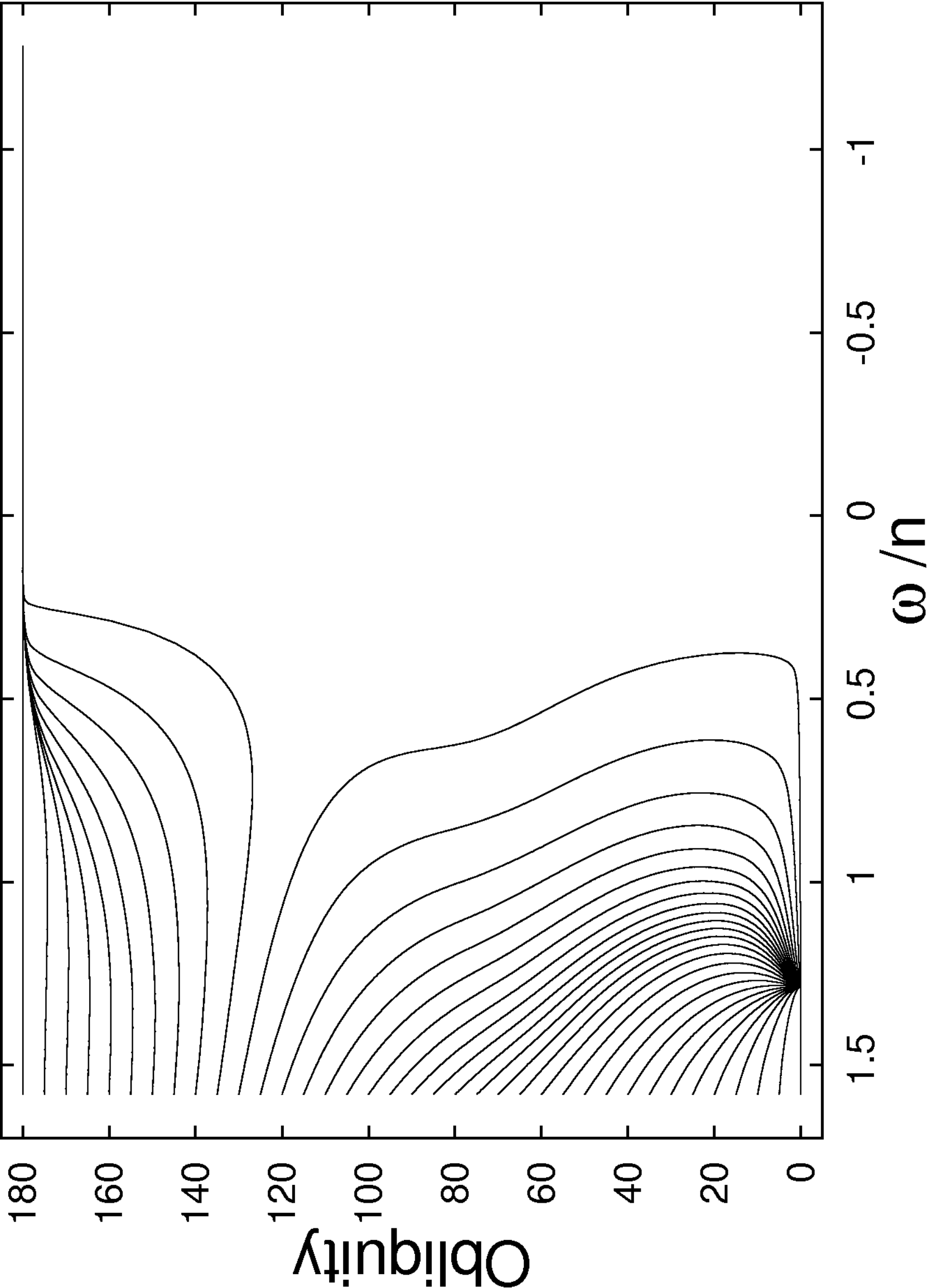}\put(23,100){\bf \large (h)}\end{overpic} &
\begin{overpic}[angle=-90, width=0.62\columnwidth]{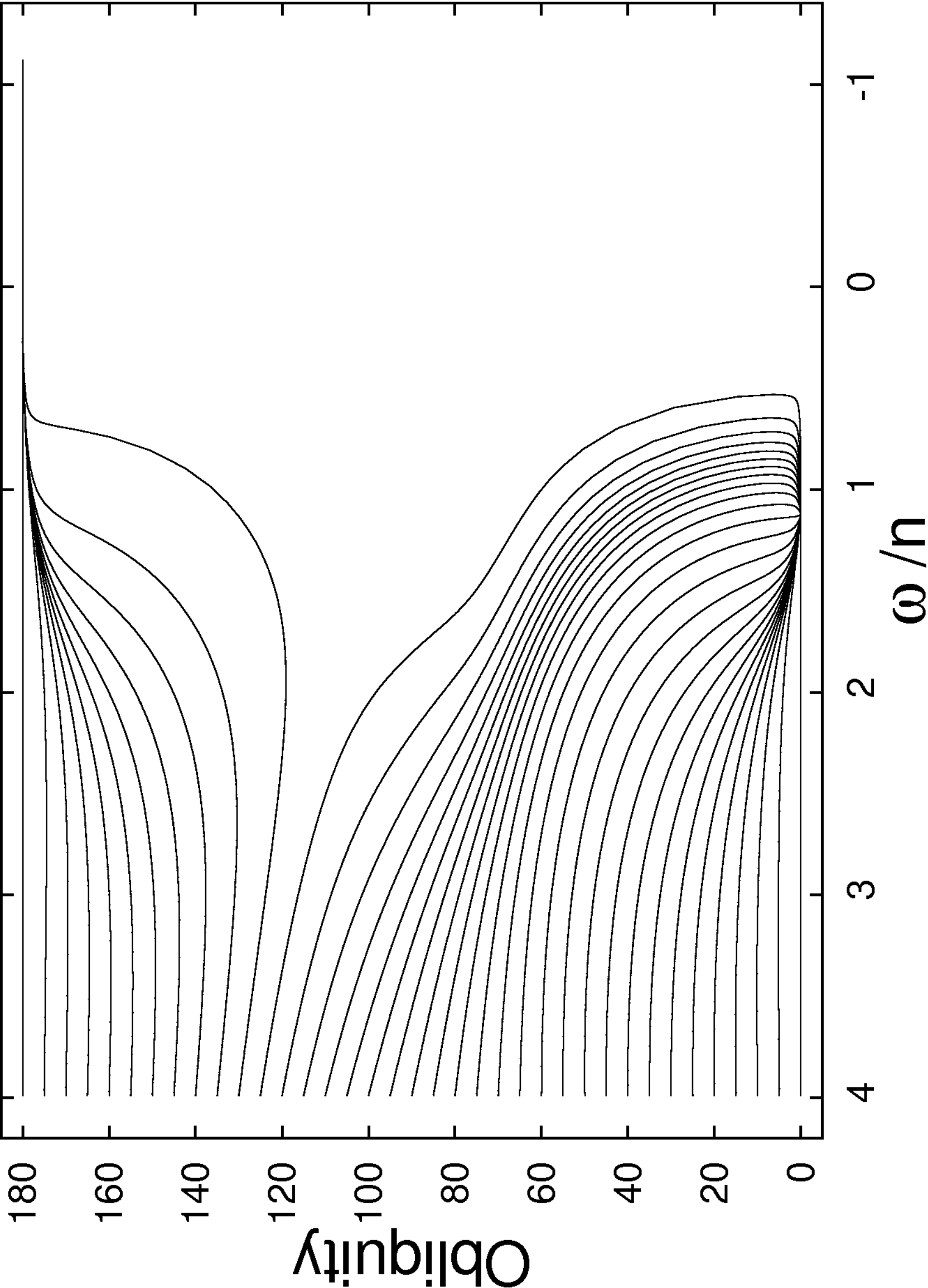}\put(23,100){\bf \large (i)}\end{overpic} \\
\begin{overpic}[angle=-90, width=0.62\columnwidth]{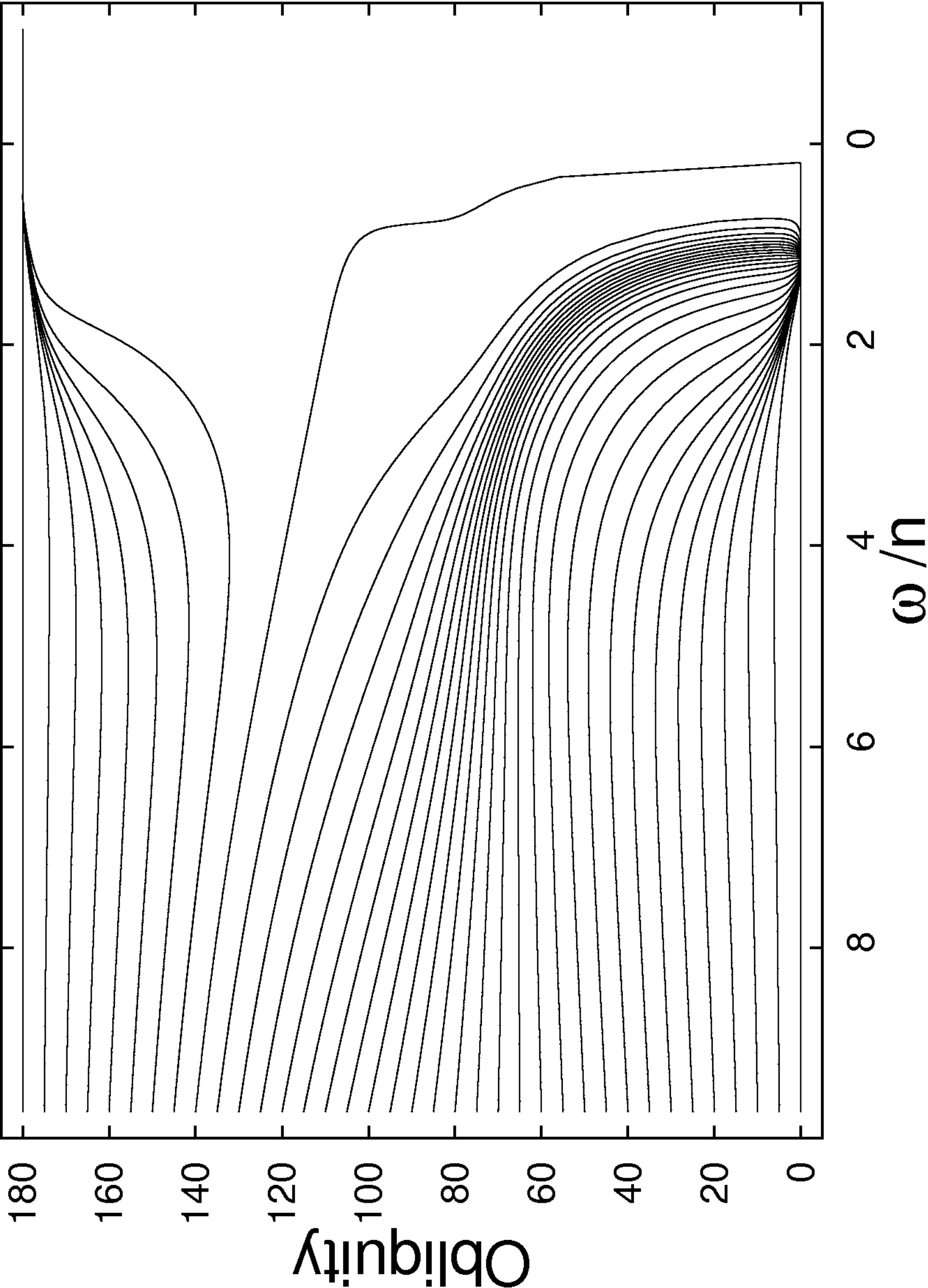}\put(23,100){\bf \large (j)}\end{overpic} &
\begin{overpic}[angle=-90, width=0.62\columnwidth]{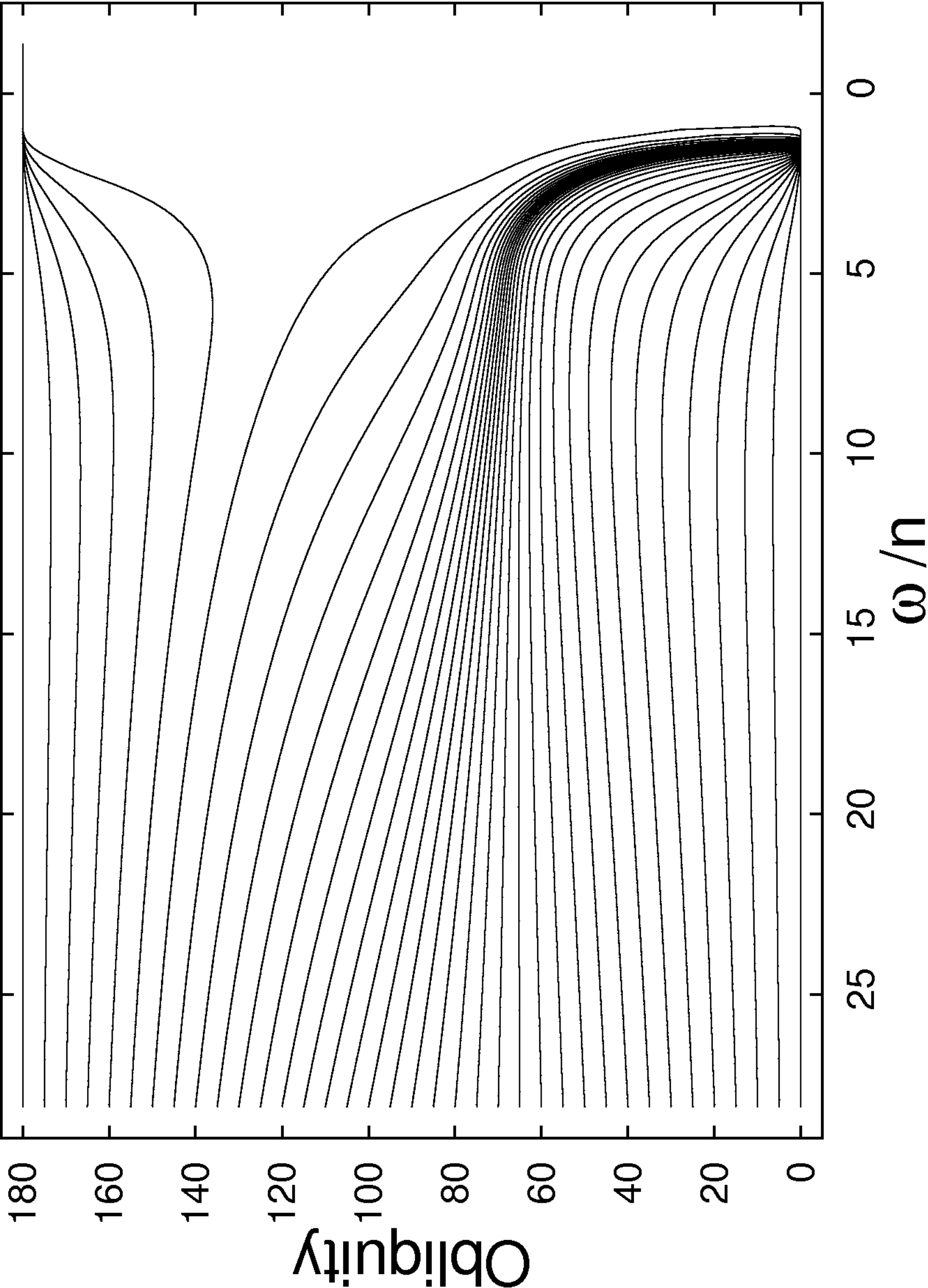}\put(23,100){\bf \large (k)}\end{overpic} &
\begin{overpic}[angle=-90, width=0.62\columnwidth]{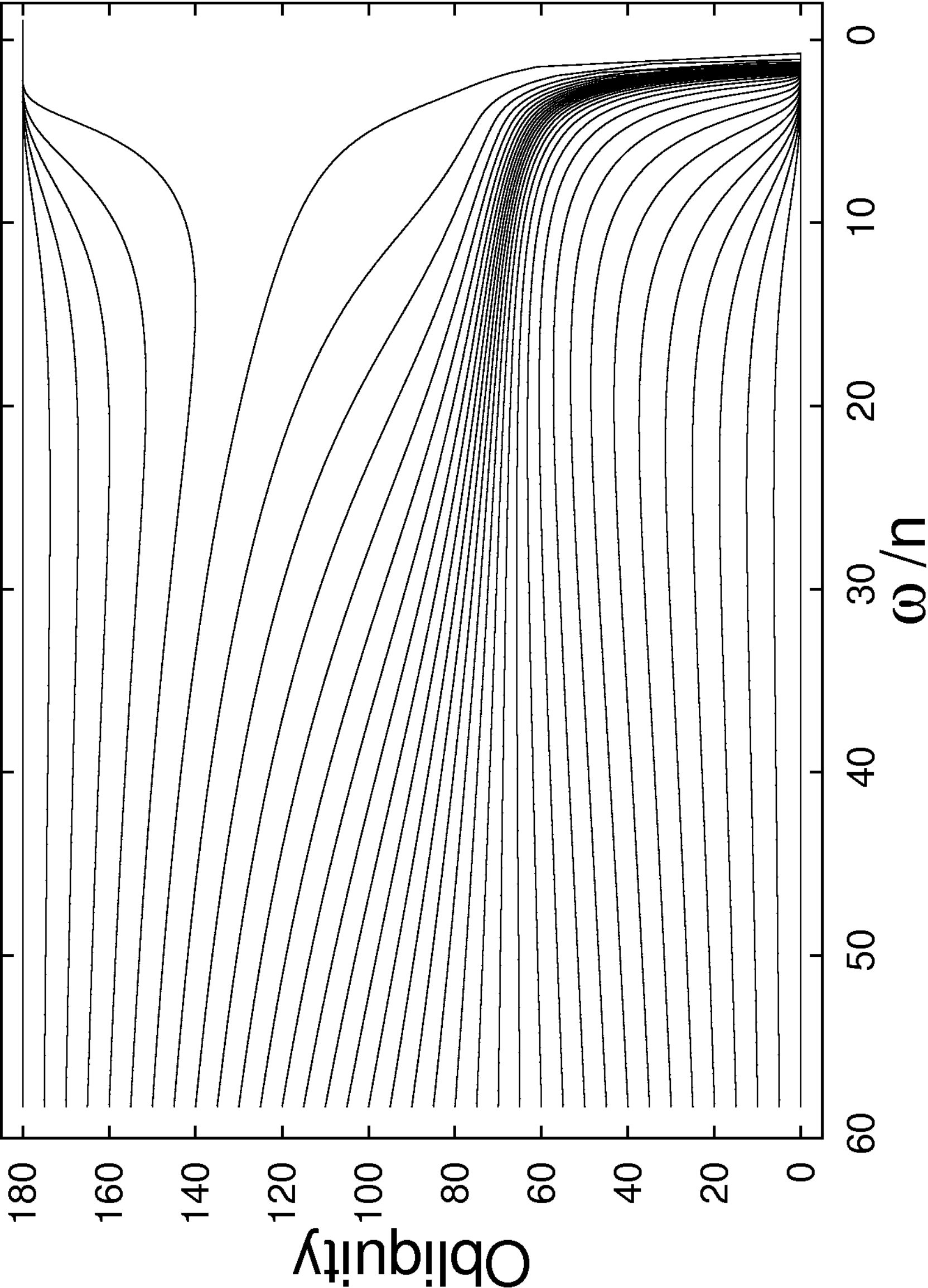}\put(23,100){\bf \large (l)}\end{overpic} 
\end{tabular}
\caption{\label{EvolucaoExoplanets1-1}Obliquity evolution with the rotation rate for several Earth-sized planets taken from Table~\ref{tbl:1} with an initial rotation period of $P_{in}=1$~day:
\textbf{(a)} to \textbf{(f)} HD\,40307\,$b$ to $g$;   
\textbf{(g)}  55\,Cnc\,$e$; \textbf{(h)} GJ\,1214\,$b$; \textbf{(i)} HD\,215497\,$b$;    \textbf{(j)} $\mu$\,Arae\,$c$;   \textbf{(k)} GJ\,667C\,$c$;  \textbf{(l)} HD\,85512\,$b$.}
\end{center}
\end{figure*}

\begin{figure*}
\begin{center}
\begin{tabular}{c c c}
\begin{overpic}[angle=-90, width=0.62\columnwidth]{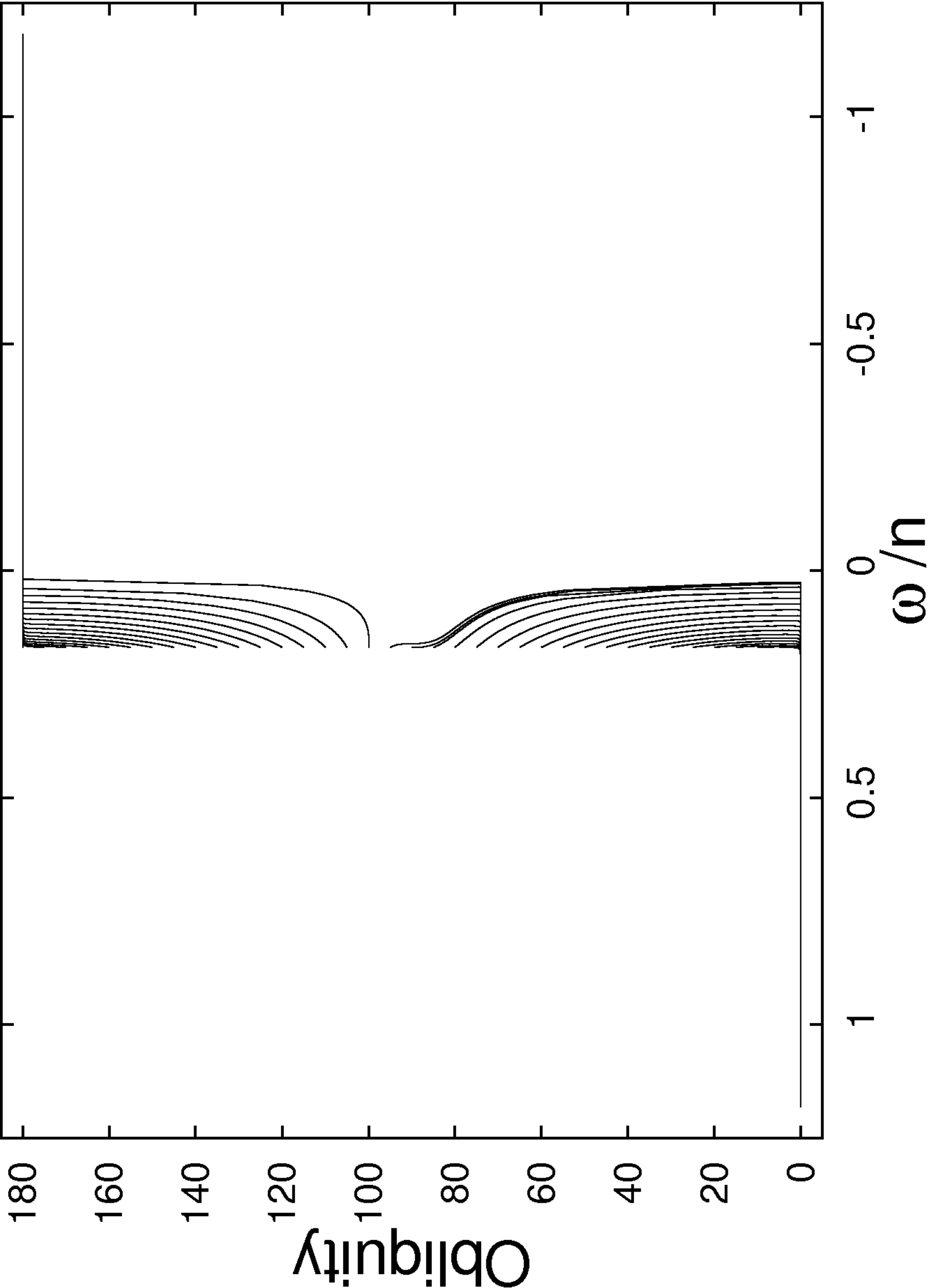}\put(23,100){\bf \large (a)}\end{overpic} &
\begin{overpic}[angle=-90, width=0.62\columnwidth]{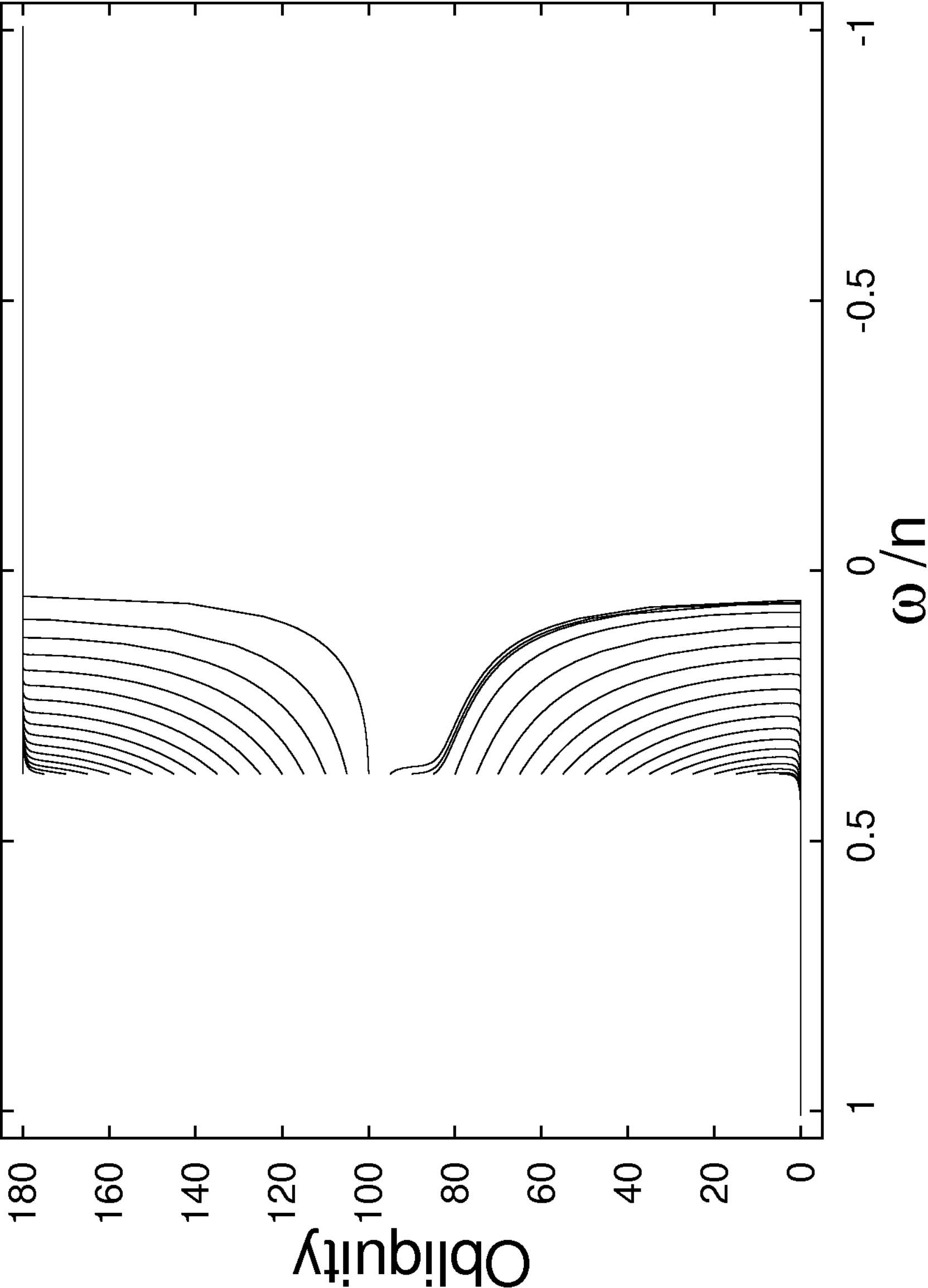}\put(23,100){\bf \large (b)}\end{overpic} &
\begin{overpic}[angle=-90, width=0.62\columnwidth]{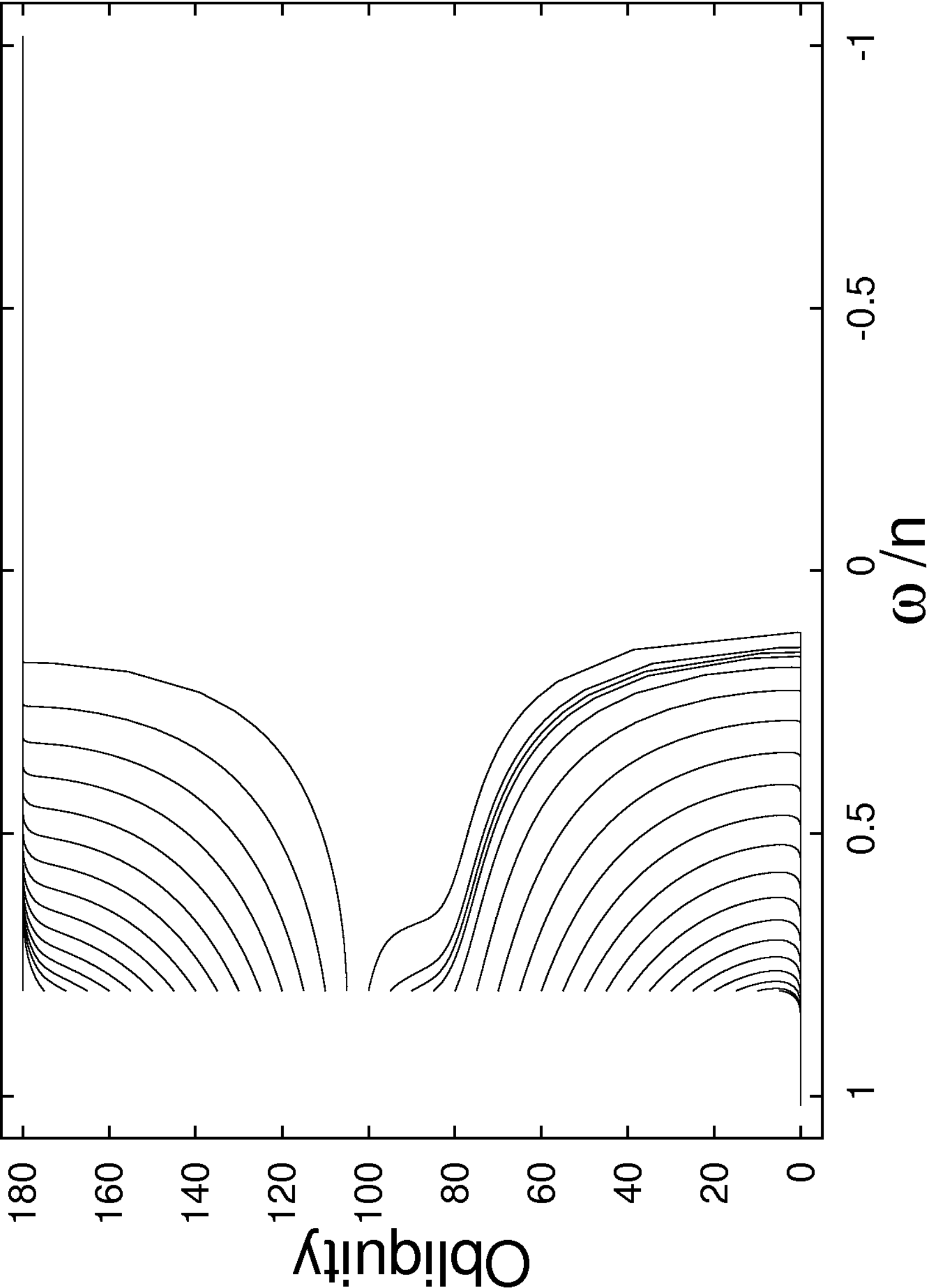}\put(23,100){\bf \large (c)}\end{overpic} \\
\begin{overpic}[angle=-90, width=0.62\columnwidth]{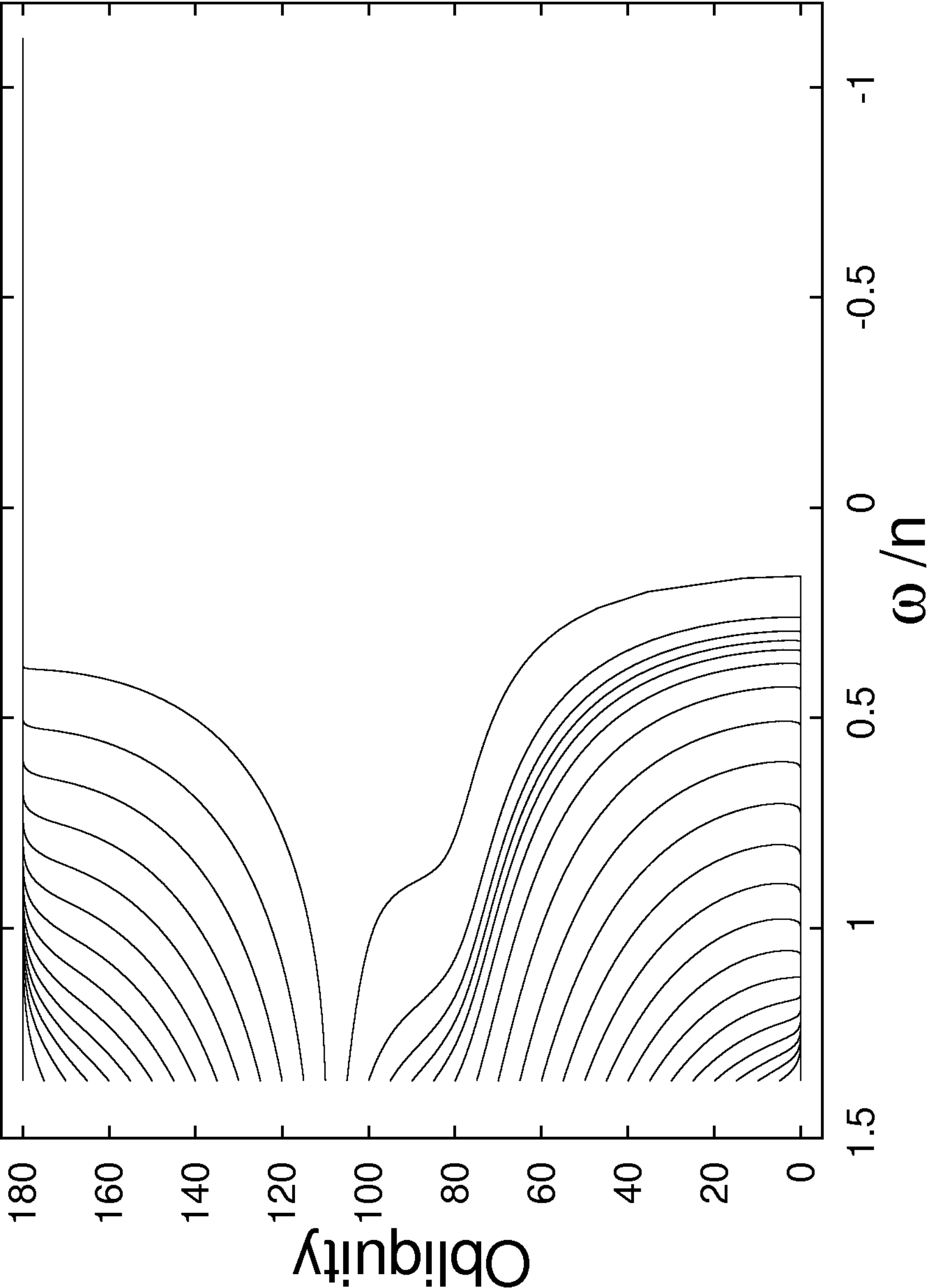}\put(23,100){\bf \large (d)}\end{overpic} &
\begin{overpic}[angle=-90, width=0.62\columnwidth]{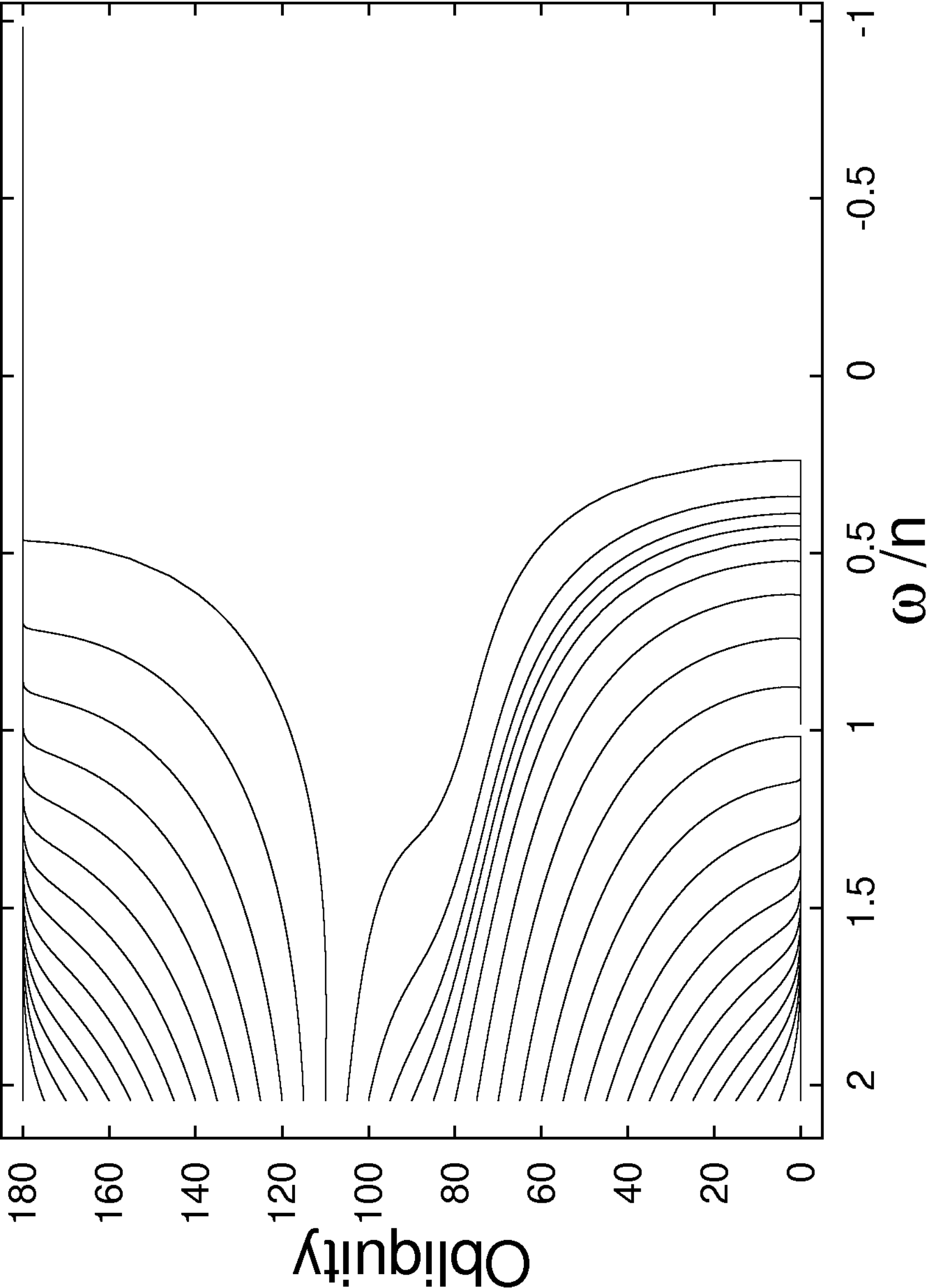}\put(23,100){\bf \large (e)}\end{overpic} &
\begin{overpic}[angle=-90, width=0.62\columnwidth]{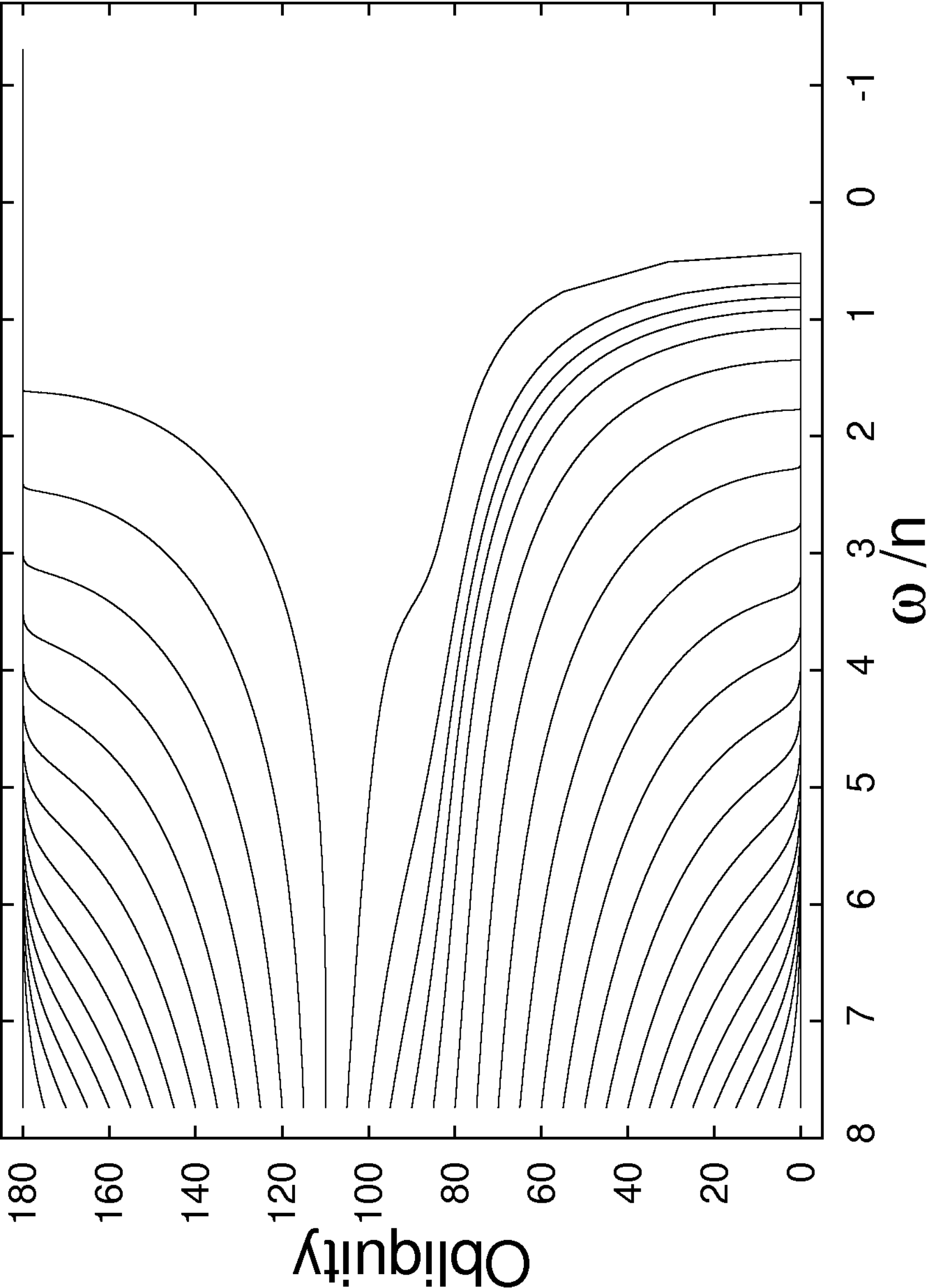}\put(23,100){\bf \large (f)}\end{overpic} \\
\begin{overpic}[angle=-90, width=0.62\columnwidth]{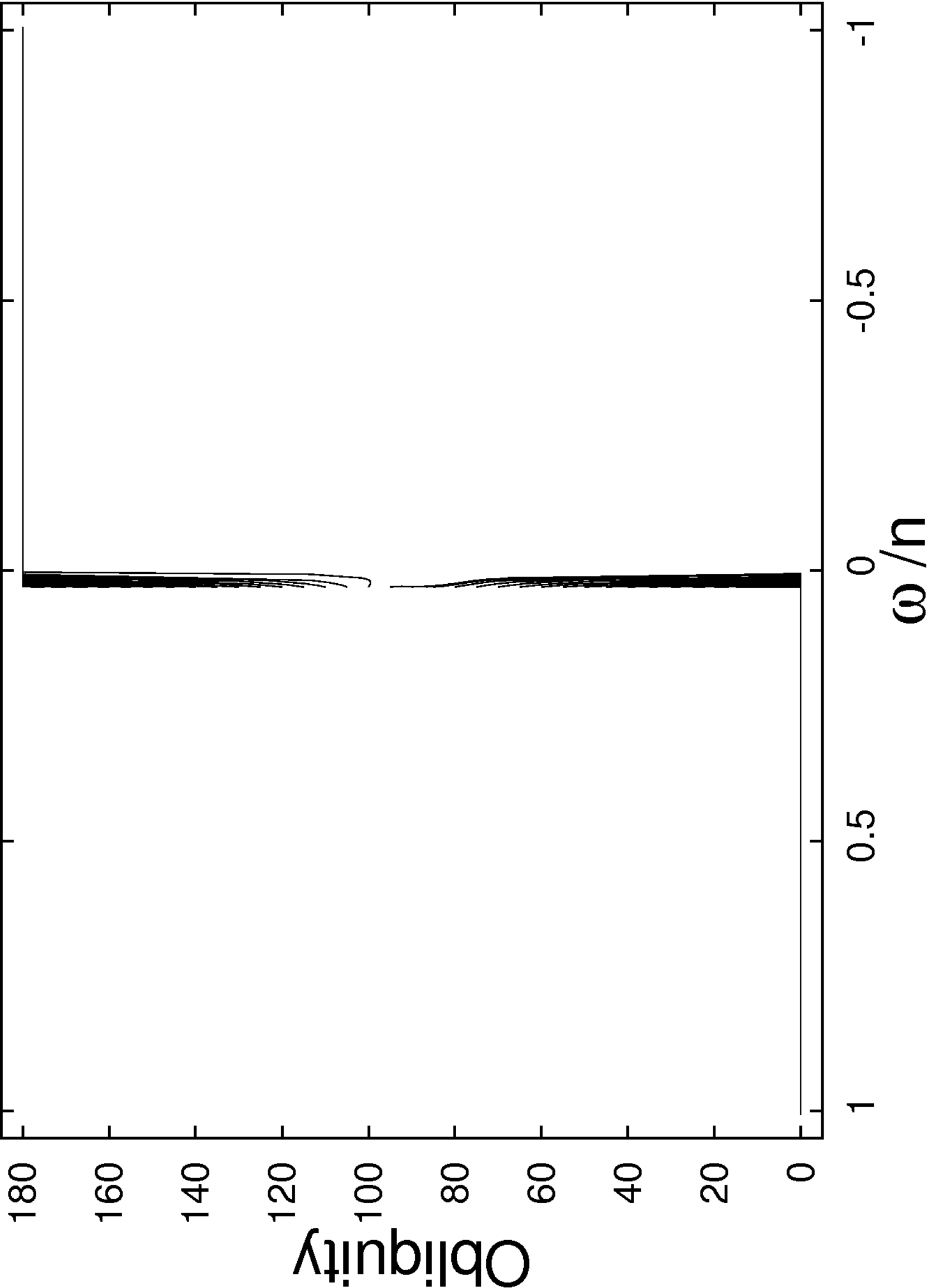}\put(23,100){\bf \large (g)}\end{overpic} &
\begin{overpic}[angle=-90, width=0.62\columnwidth]{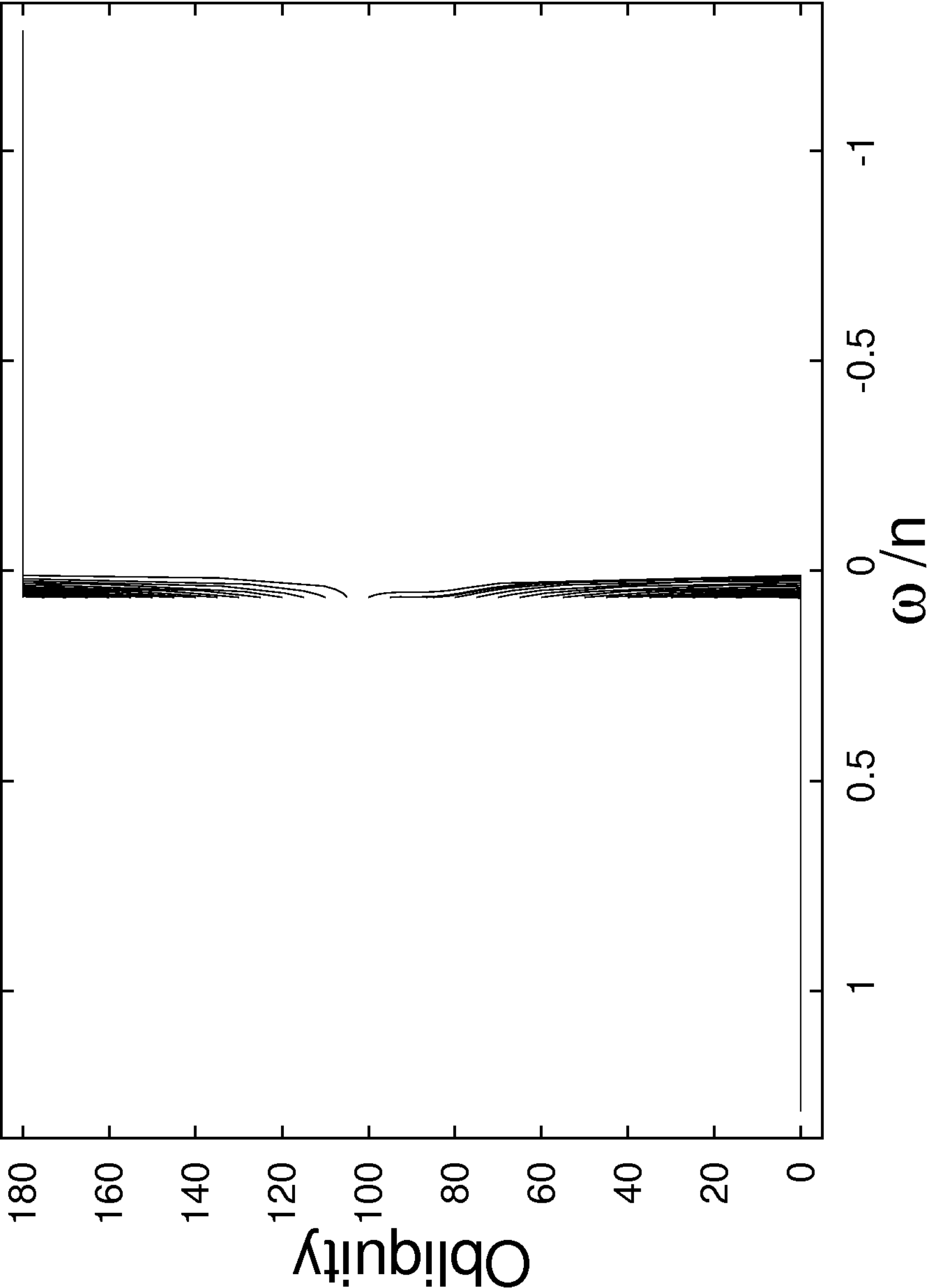}\put(23,100){\bf \large (h)}\end{overpic} &
\begin{overpic}[angle=-90, width=0.62\columnwidth]{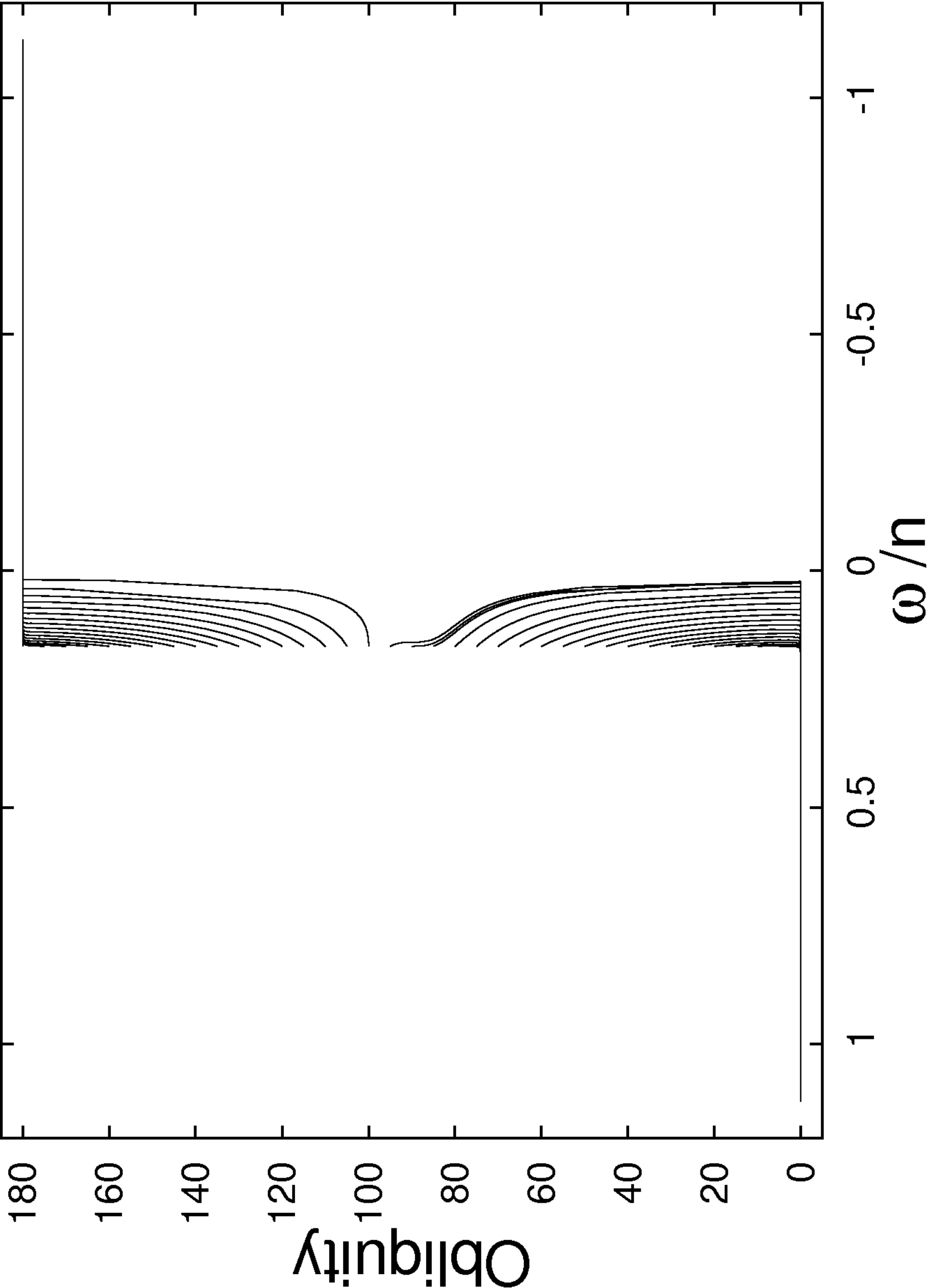}\put(23,100){\bf \large (i)}\end{overpic} \\
\begin{overpic}[angle=-90, width=0.62\columnwidth]{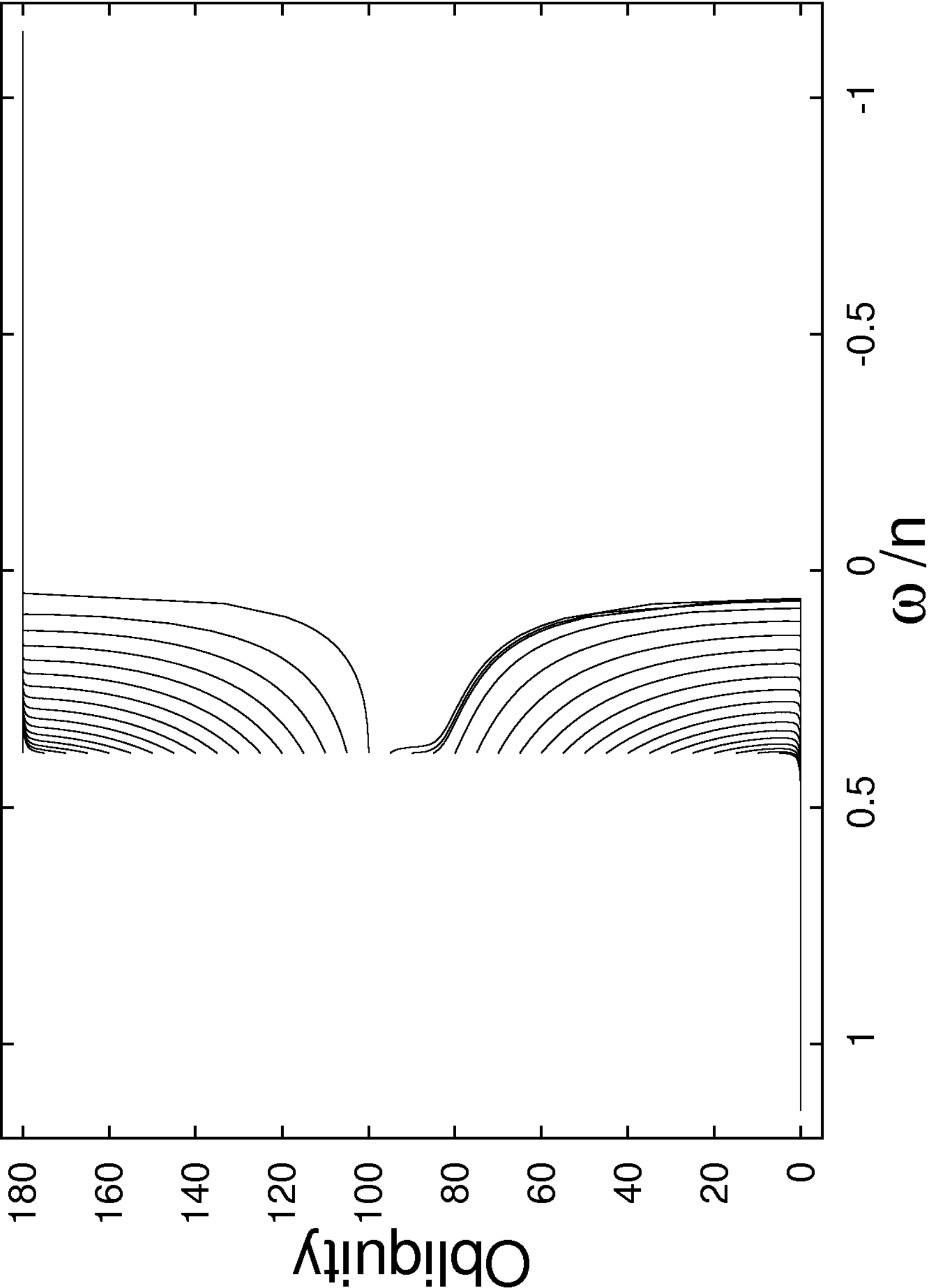}\put(23,100){\bf \large (j)}\end{overpic} &
\begin{overpic}[angle=-90, width=0.62\columnwidth]{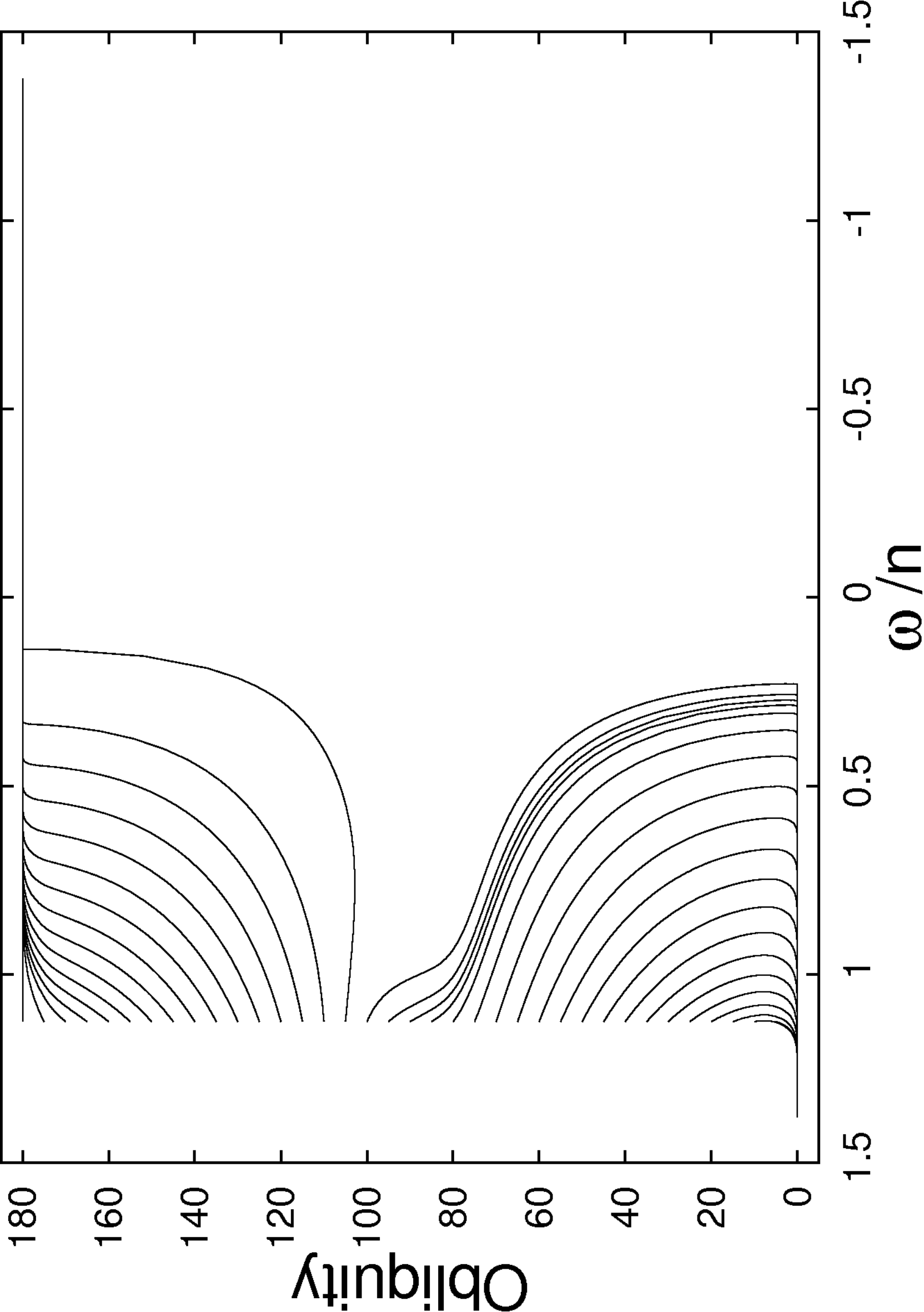}\put(23,100){\bf \large (k)}\end{overpic} &
\begin{overpic}[angle=-90, width=0.62\columnwidth]{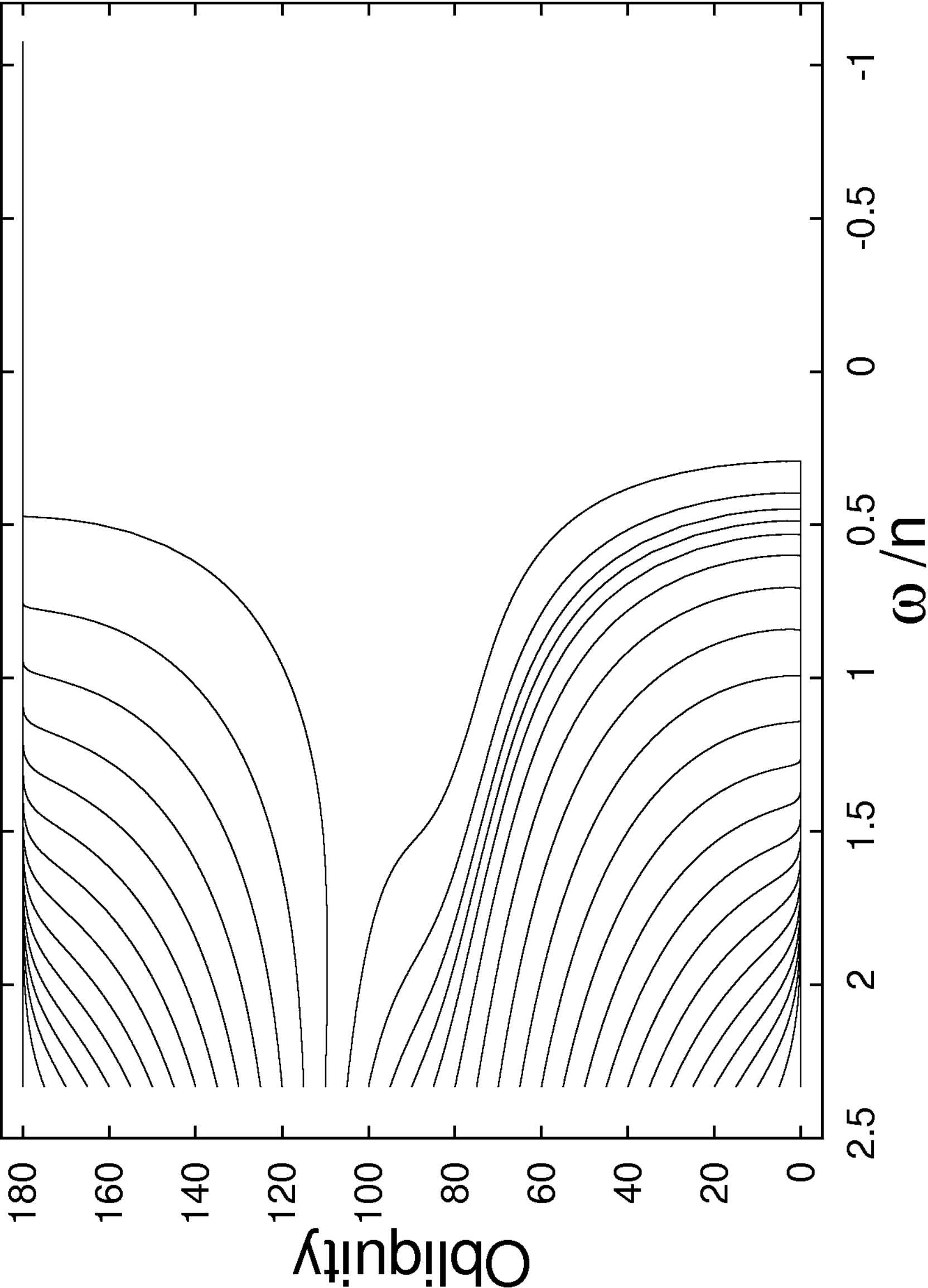}\put(23,100){\bf \large (l)}\end{overpic} 
\end{tabular}
\caption{\label{EvolucaoExoplanets25-1}Obliquity evolution with the rotation rate for several Earth-sized planets taken from Table~\ref{tbl:1} with an initial rotation period of $P_{in}=25$~day:
\textbf{(a)} to \textbf{(f)} HD\,40307\,$b$ to $g$;   
\textbf{(g)}  55\,Cnc\,$e$; \textbf{(h)} GJ\,1214\,$b$; \textbf{(i)} HD\,215497\,$b$;    \textbf{(j)} $\mu\,$Arae\,$c$;   \textbf{(k)} GJ\,667C\,$c$;  \textbf{(l)} HD\,85512\,$b$.}
\end{center}
\end{figure*}

\begin{figure*}
\begin{center}
\begin{tabular}{c c c}
$ e = 0.0 $ & $ e = 0.1 $ & $ e = 0.2 $ \\
\begin{overpic}[angle=-90, width=0.62\columnwidth]{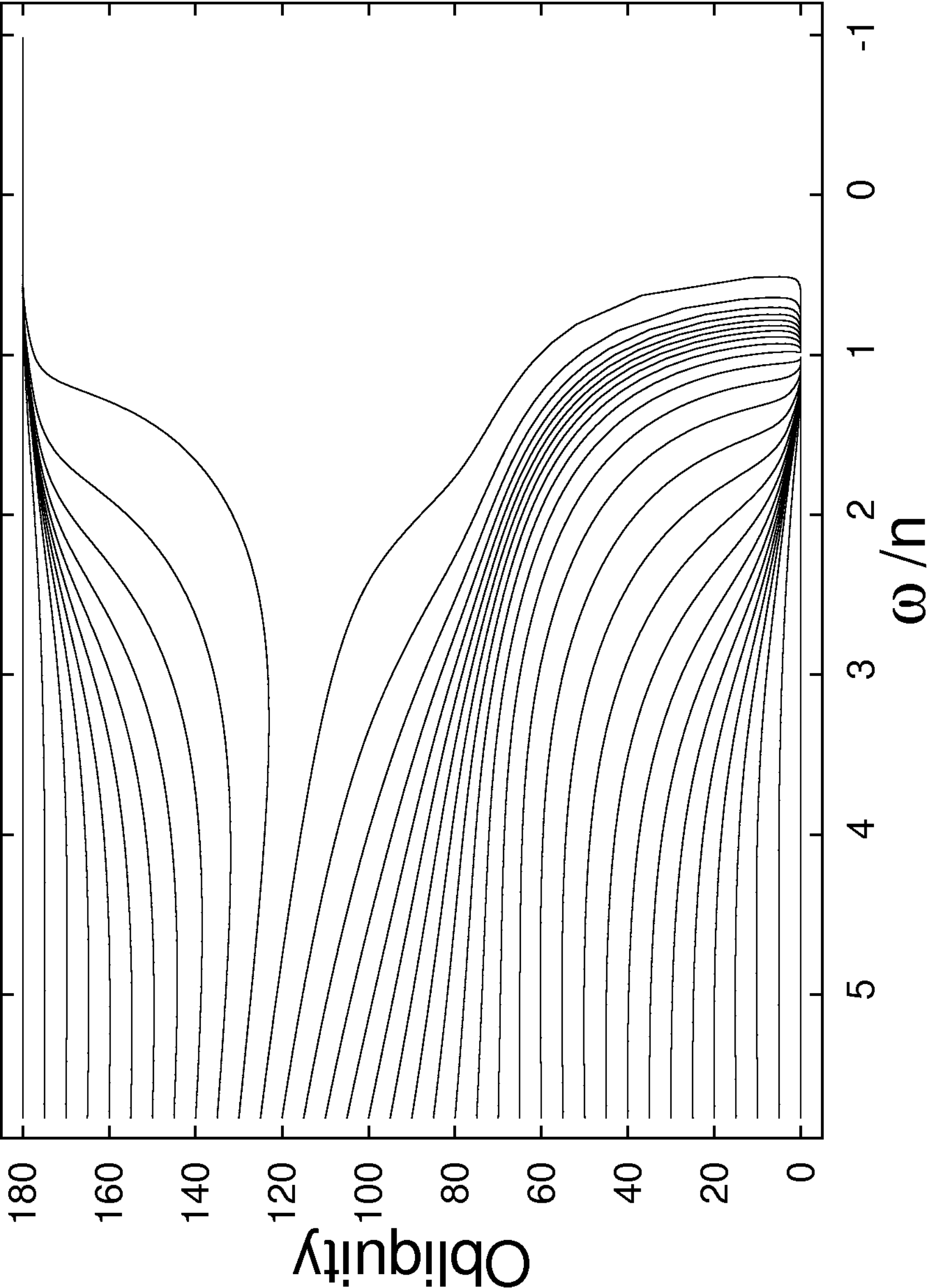}\put(23,100){\bf \large (a)}\end{overpic} &
\begin{overpic}[angle=-90, width=0.62\columnwidth]{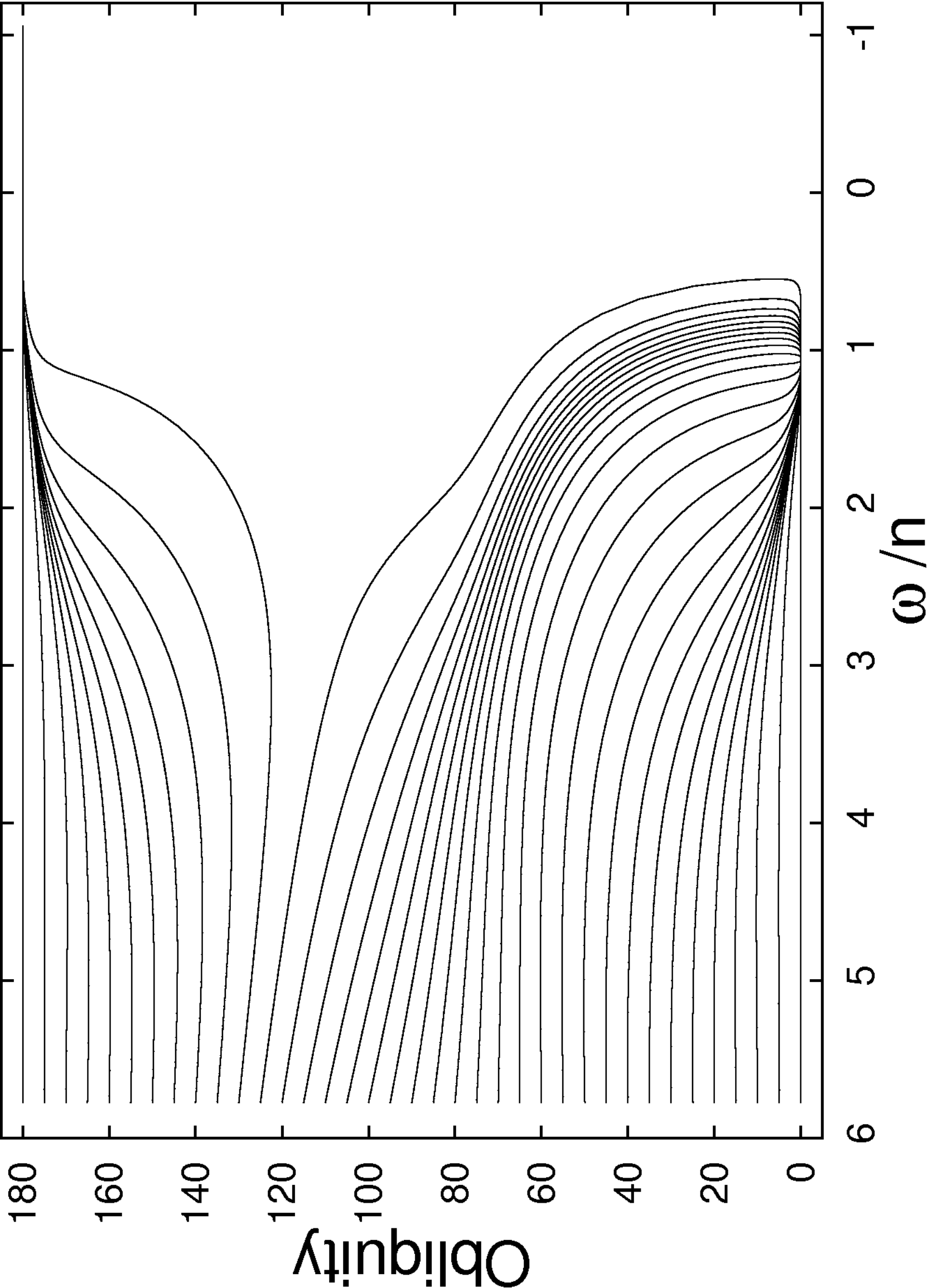}\put(23,100){\bf \large (b)}\end{overpic} &
\begin{overpic}[angle=-90, width=0.62\columnwidth]{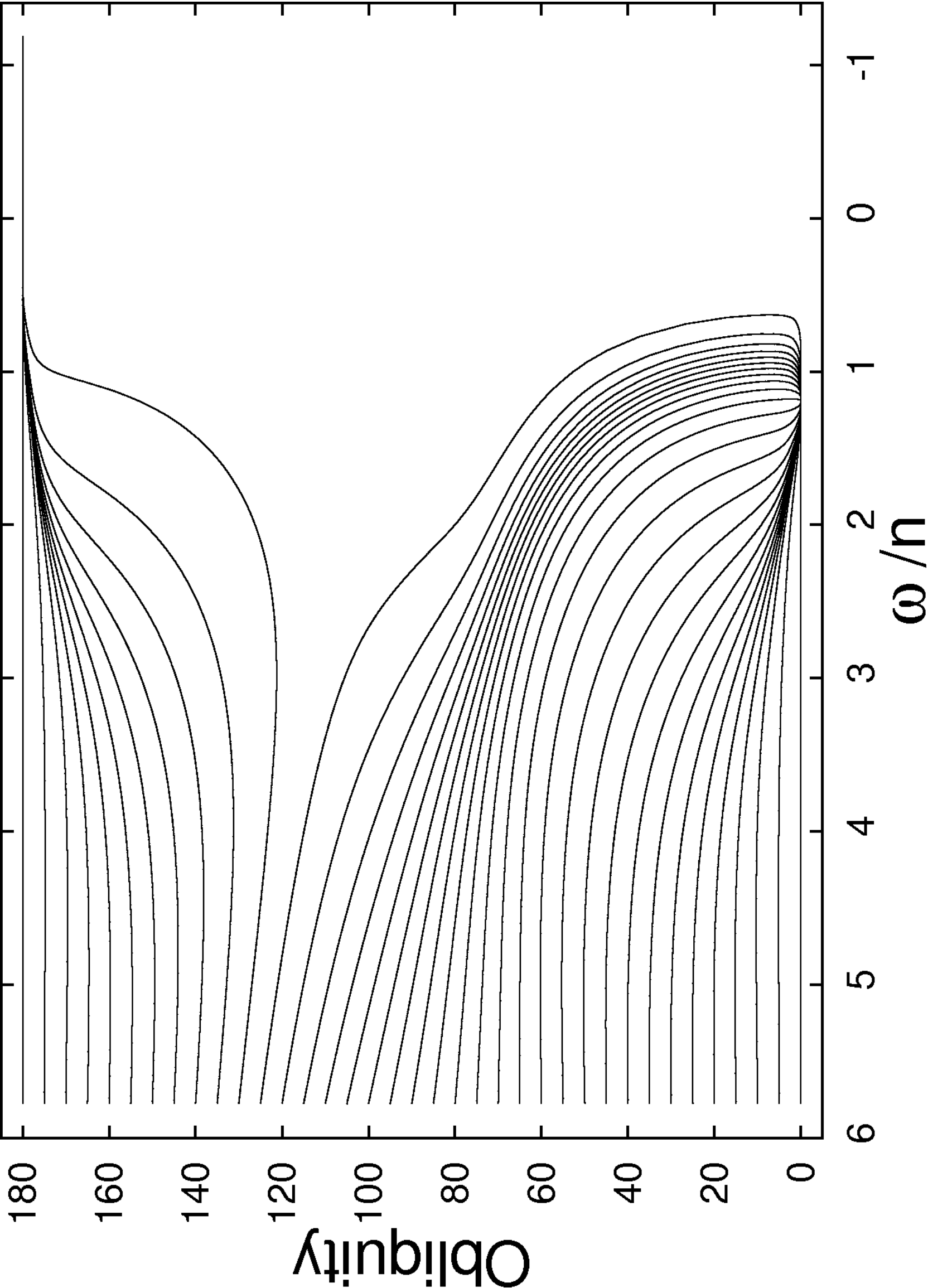}\put(23,100){\bf \large (c)}\end{overpic} \\
\begin{overpic}[angle=-90, width=0.62\columnwidth]{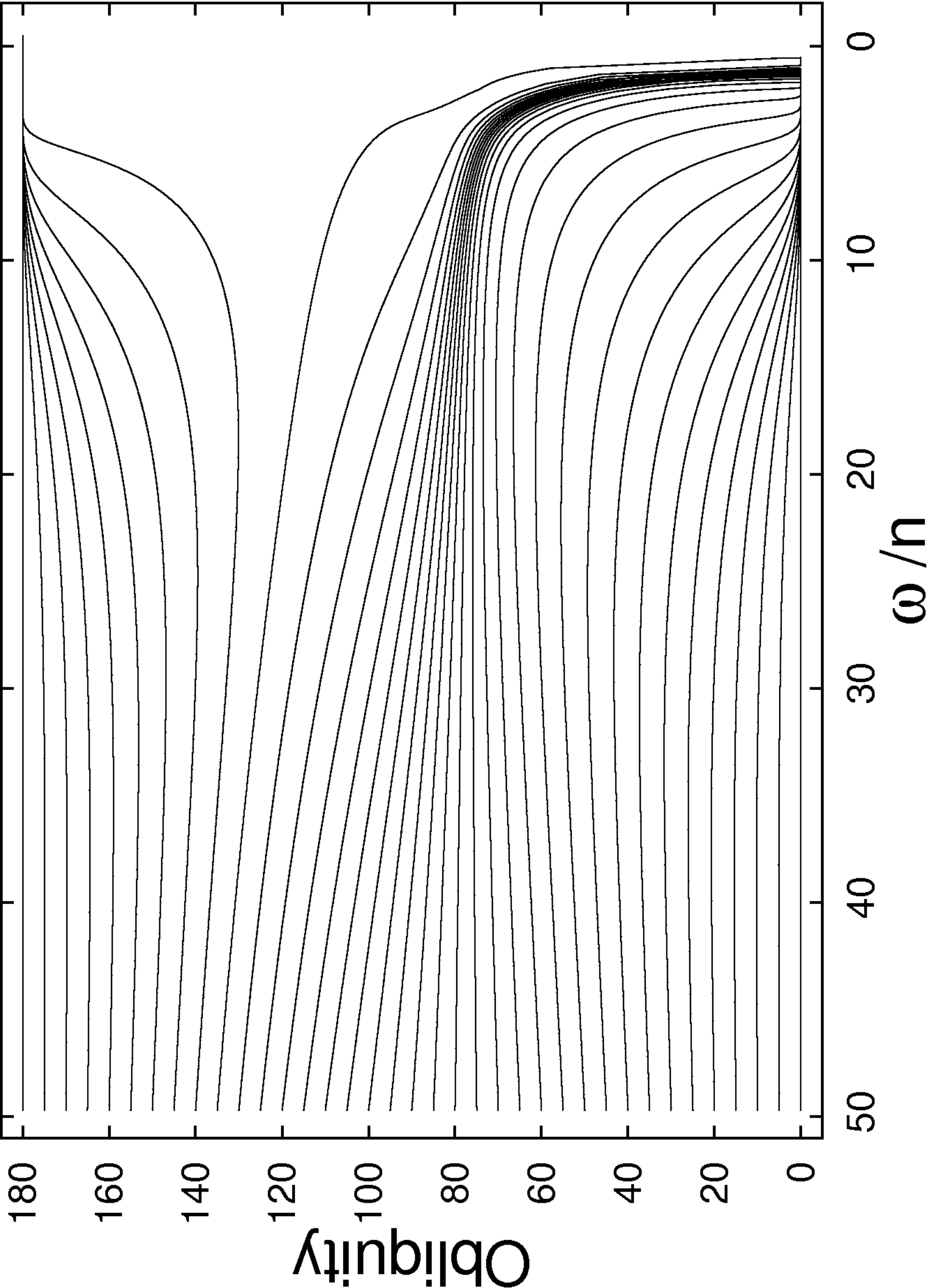}\put(23,100){\bf \large (d)}\end{overpic} &
\begin{overpic}[angle=-90, width=0.62\columnwidth]{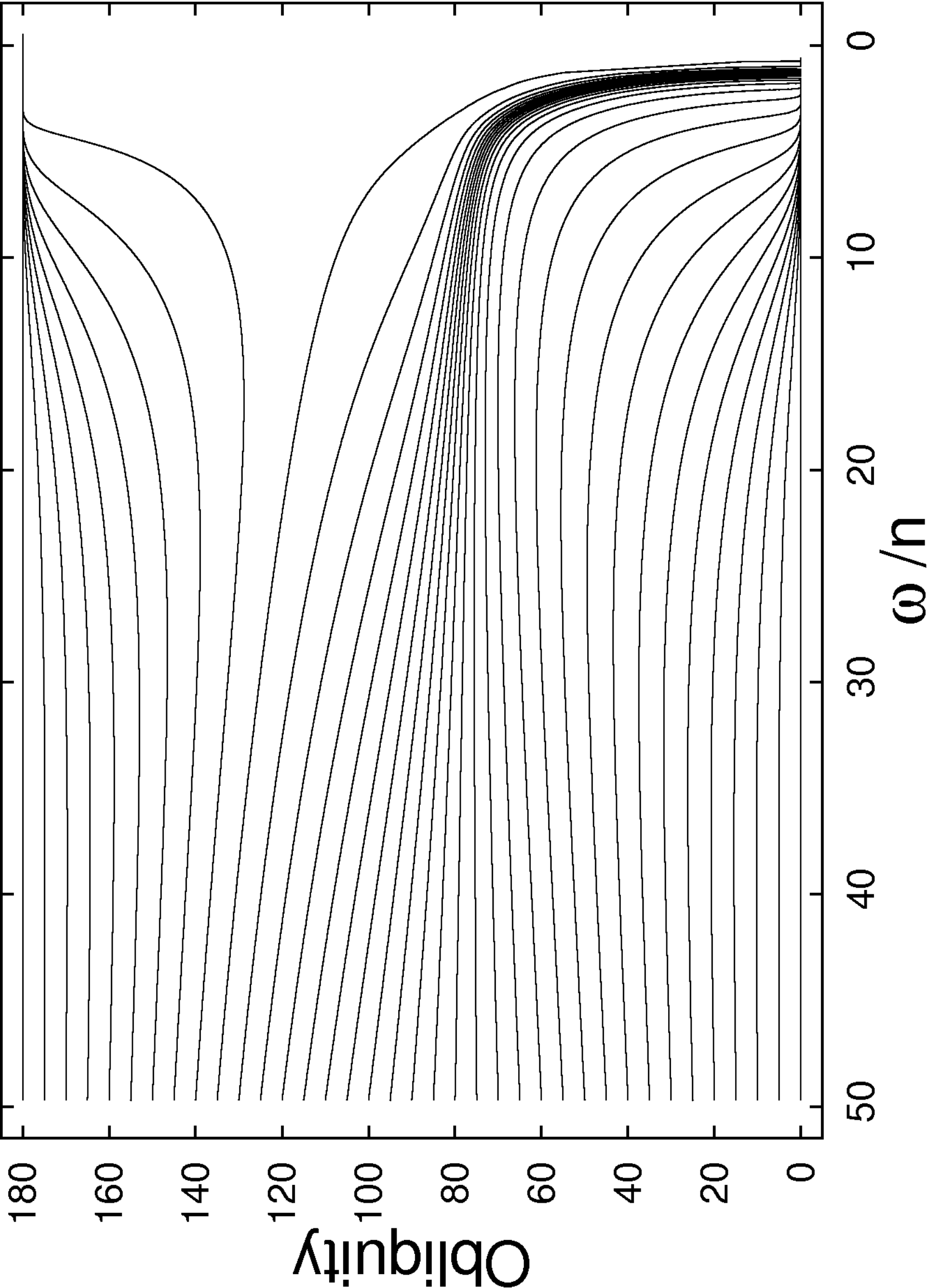}\put(23,100){\bf \large (e)}\end{overpic} &
\begin{overpic}[angle=-90, width=0.62\columnwidth]{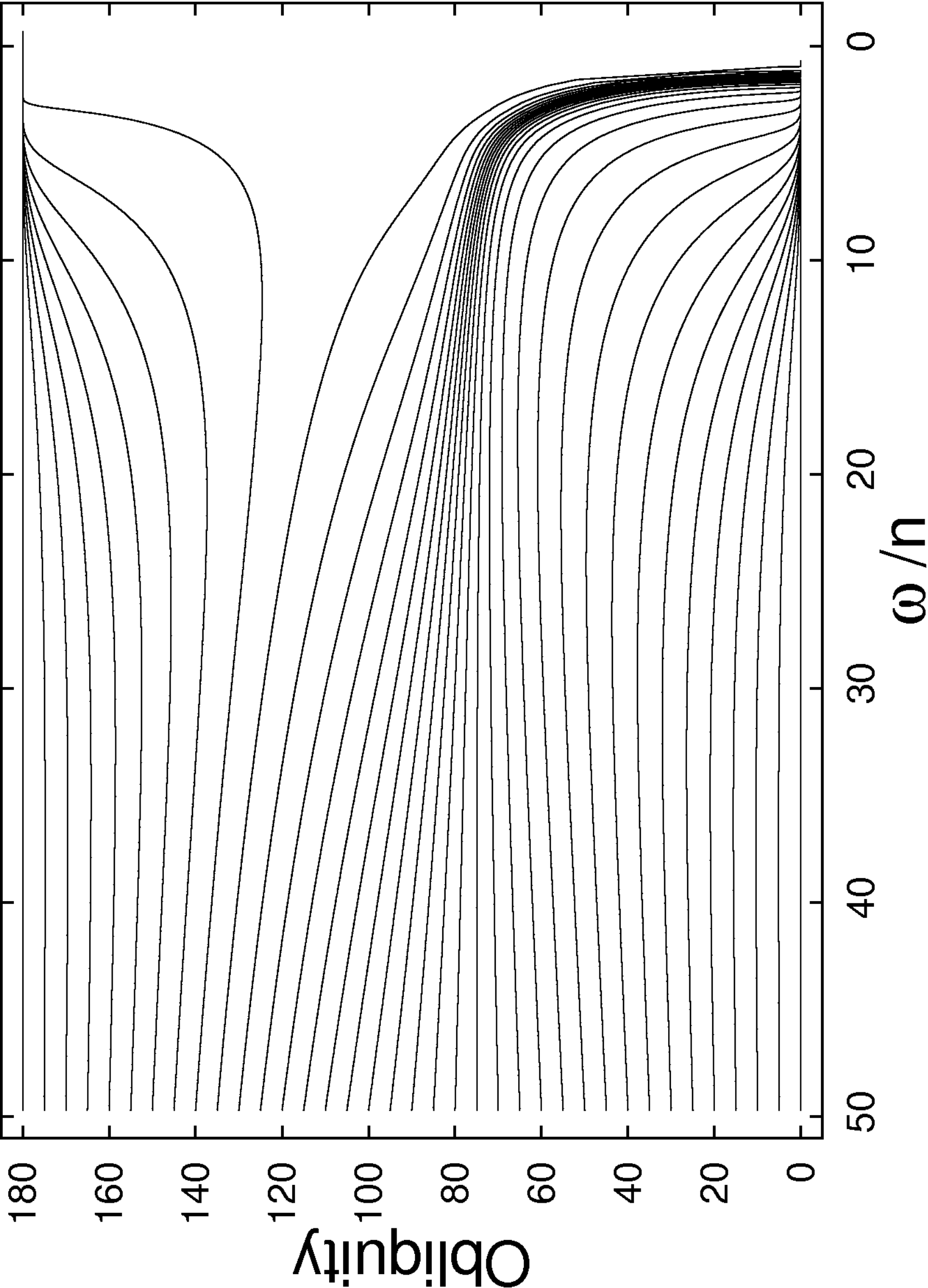}\put(23,100){\bf \large (f)}\end{overpic} \\
\begin{overpic}[angle=-90, width=0.62\columnwidth]{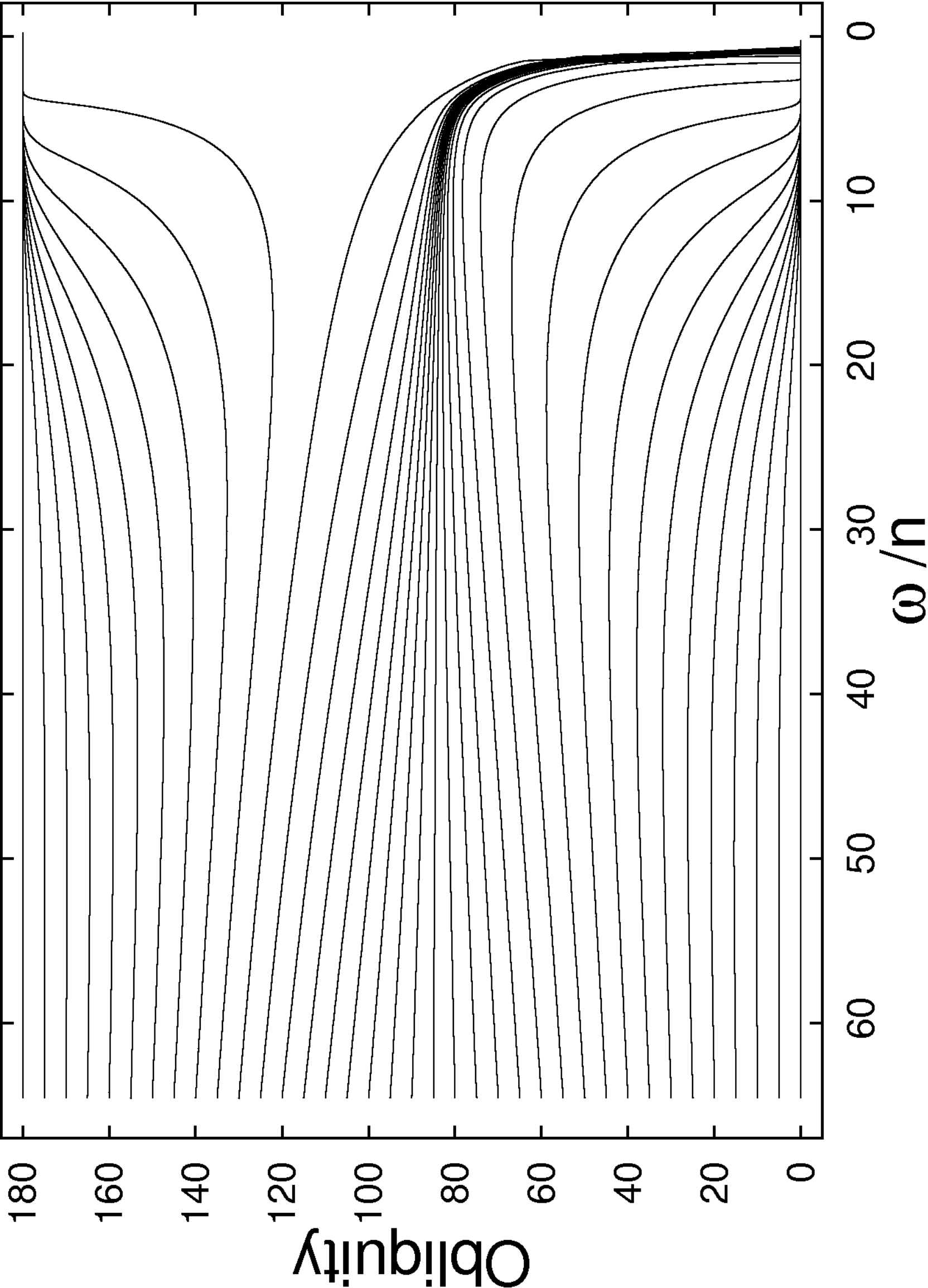}\put(23,100){\bf \large (g)}\end{overpic} &
\begin{overpic}[angle=-90, width=0.62\columnwidth]{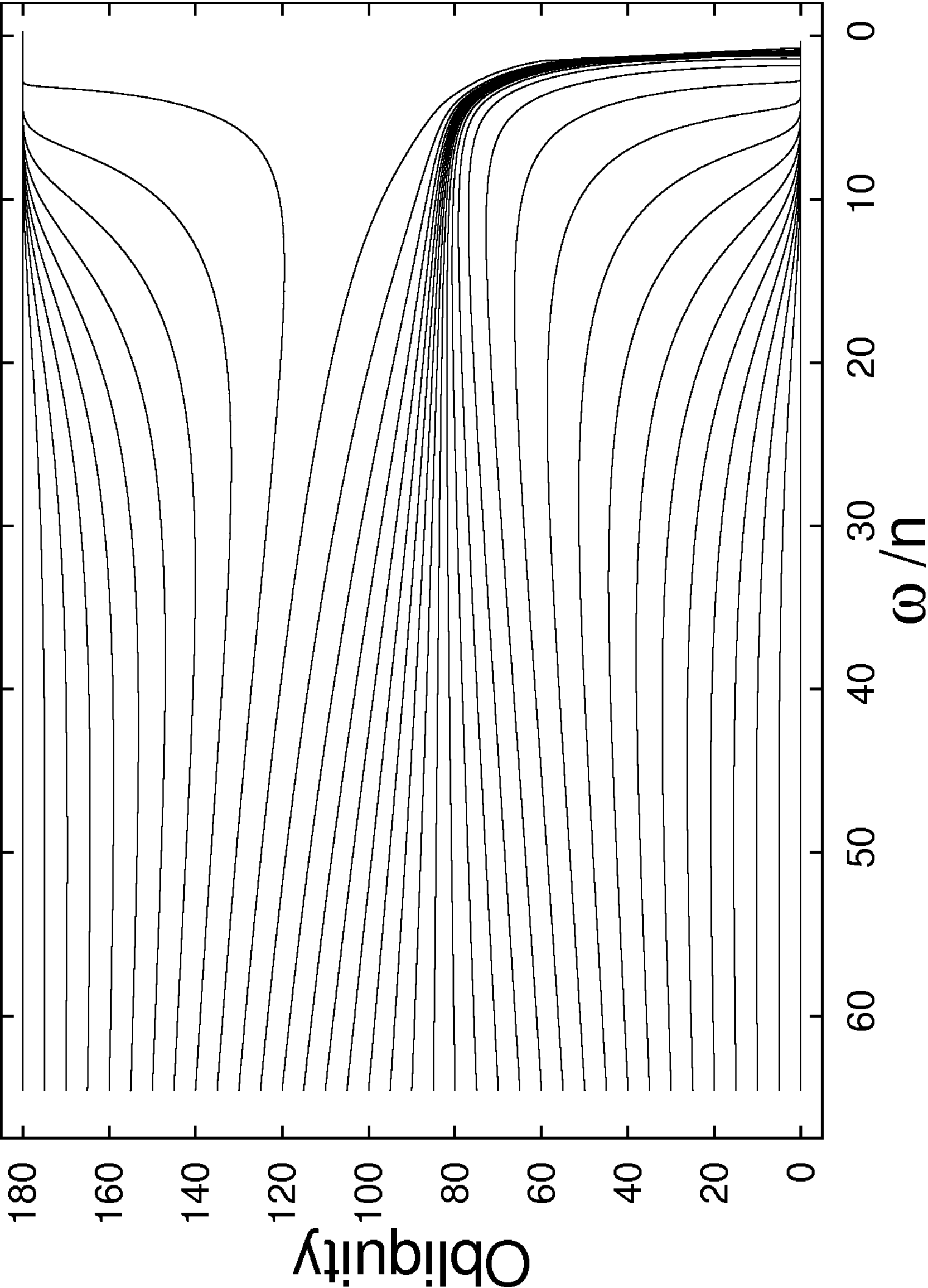}\put(23,100){\bf \large (h)}\end{overpic} &
\begin{overpic}[angle=-90, width=0.62\columnwidth]{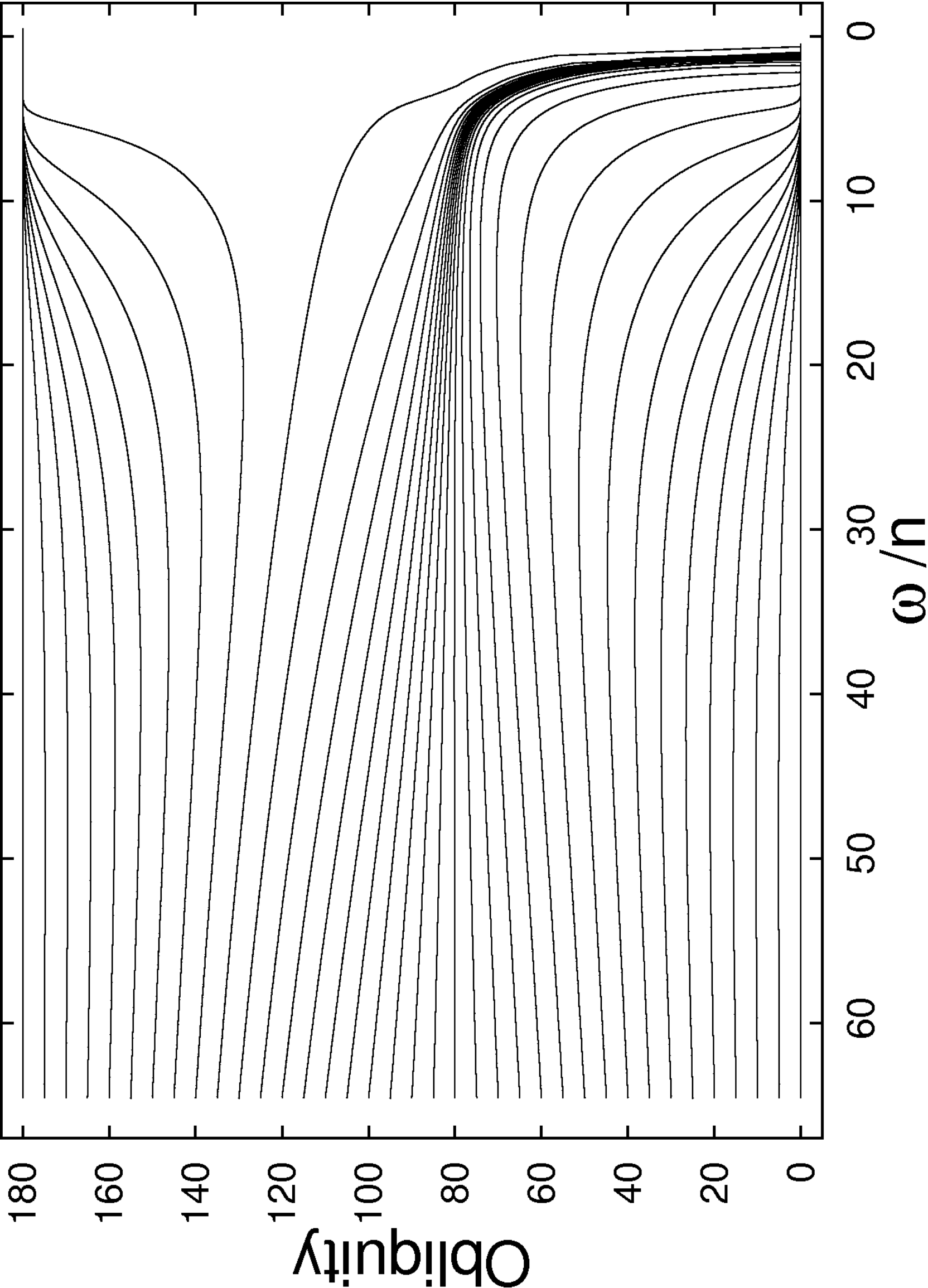}\put(23,100){\bf \large (i)}\end{overpic} \\
\begin{overpic}[angle=-90, width=0.62\columnwidth]{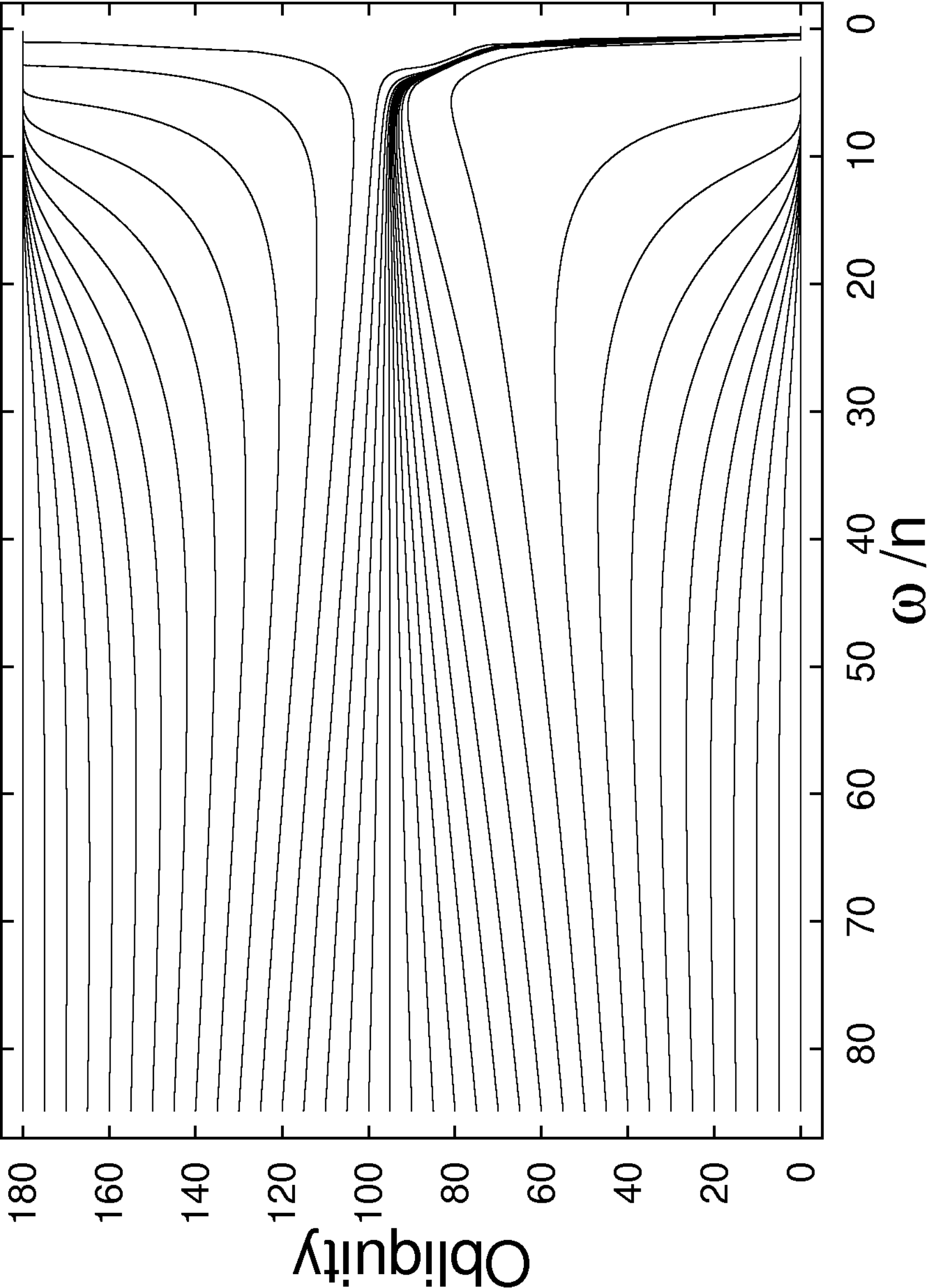}\put(23,100){\bf \large (j)}\end{overpic} &
\begin{overpic}[angle=-90, width=0.62\columnwidth]{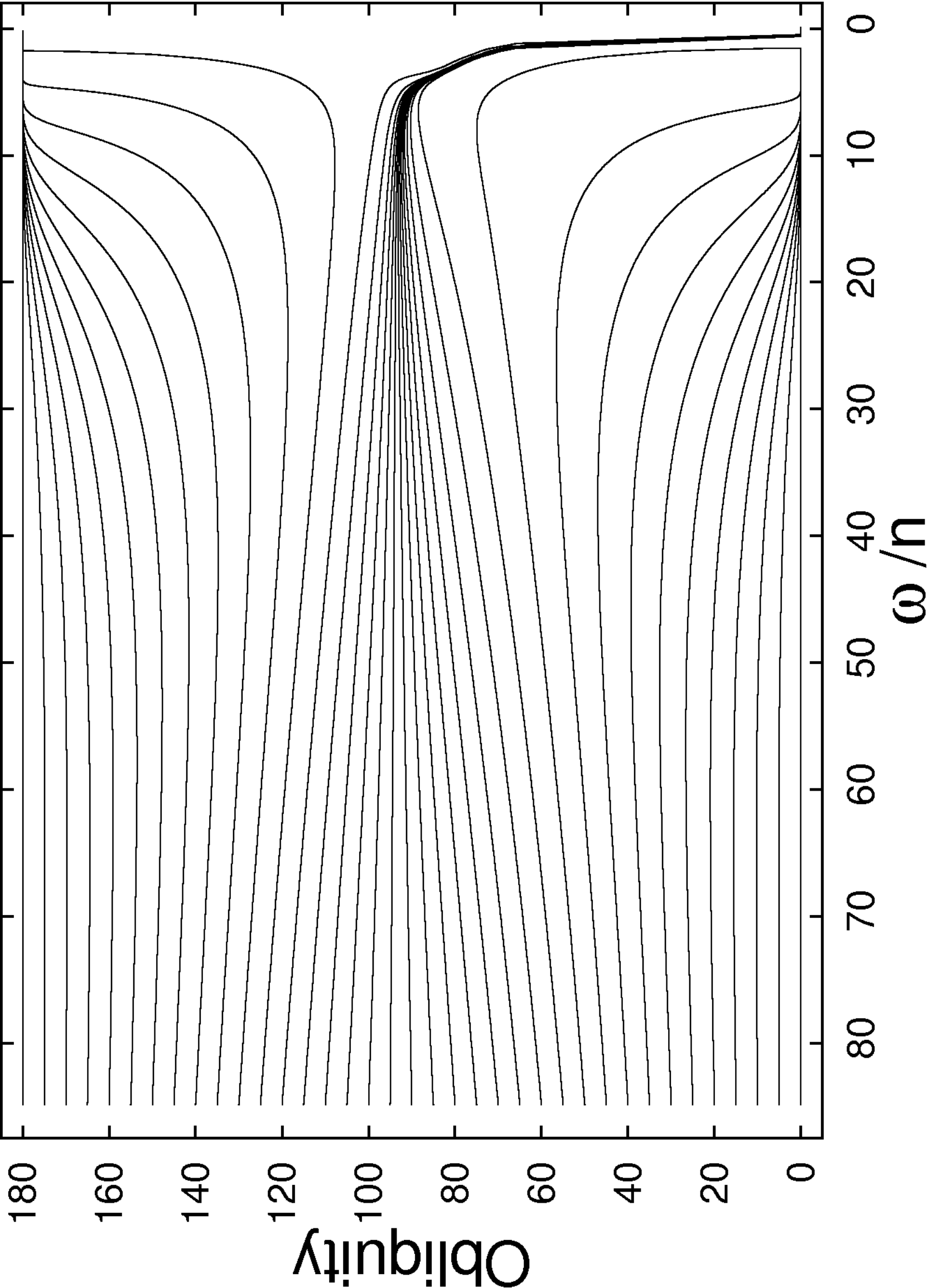}\put(23,100){\bf \large (k)}\end{overpic} &
\begin{overpic}[angle=-90, width=0.62\columnwidth]{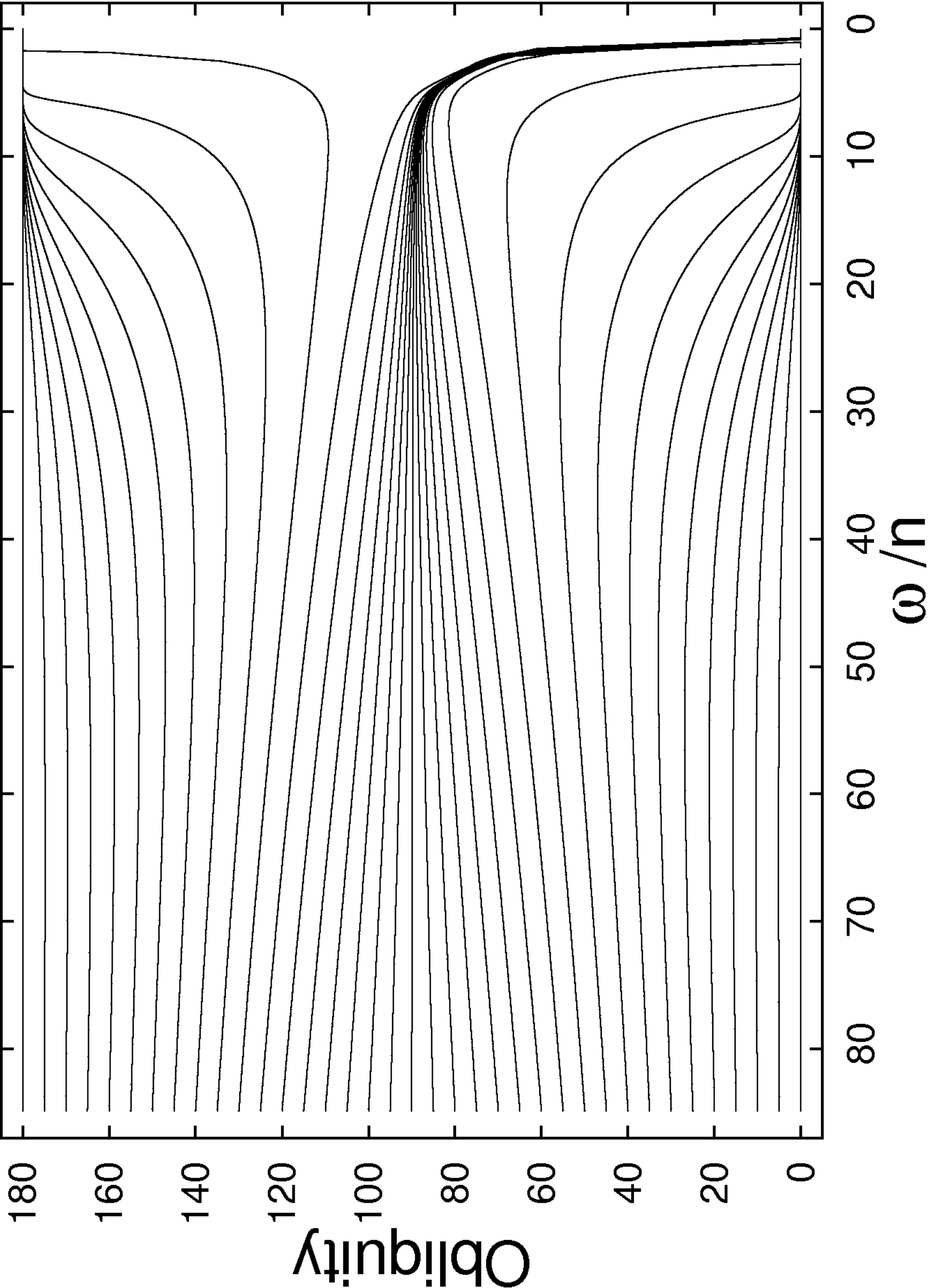}\put(23,100){\bf \large (l)}\end{overpic} 
\end{tabular}
\caption{\label{EvolucaoVenus0.0-2}Obliquity evolution with the rotation rate for several hypothetical Earth-sized planets taken from Table~\ref{tbl:2} with an initial rotation period of $P_{in}=2$~day.
The eccentricities are $e = 0.0$ {\bf (left)}, $e=0.1$ {\bf (middle)}, and $e=0.2$ {\bf (right)}.
The products semi-amjor axis times mass of the parent star are {\bf (from top to bottom)} $a \M =0.10$, $a \M =0.42$, $a \M =0.50$, and $a \M =0.60$ (units in  [AU M$_\odot$]).}
\end{center}
\end{figure*}

In Figure~\ref{EvolucaoExoplanets25-1} we show again the obliquity evolution with the rotation rate $\omega/n$ for the same systems in Figure~\ref{EvolucaoExoplanets1-1}, but now starting with a slower rotation period $P_{in}=25\,\mathrm{day}$.
Since most planets discovered so far are close-in planets with orbital periods smaller than 25\,day, we now have an initial rotation rates $ \omega < n $ for these planets.
As a consequence, we observe an identical behavior as the one described for 55\,Cnc\,$e$ with $P_{in}=1\,\mathrm{day}$ (Fig.~\ref{EvolucaoExoplanets1-1}g).

Although the spin evolution with $\omega < n$ (Fig.~\ref{EvolucaoExoplanets25-1}) is completely different from the situation with $\omega > n$ (Fig.~\ref{EvolucaoExoplanets1-1}), the final rotation states are exactly the same.
Different initial conditions lead to alternative evolutionary paths, but the final picture is the same. 
However, when $\omega < n$ we observe that there are much more trajectories increasing its obliquity to 180$^\circ$.
This behavior is related to the onset of core-mantle friction.
From equation (\ref{V32b}) we have that 1) this effect is stronger for slow rotation rates; 2) when this effect becomes dominating, the obliquity evolves into 180$^\circ$ if $\varepsilon > 90^\circ$ ($x_f = x_0 \omega_0 / \omega_f$). 
Indeed, in Figure~\ref{EvolucaoExoplanets25-1} we clearly observe that when $\omega/n \sim 0$ there is a sudden inversion of the obliquity trend, whose final evolution is dictated by the $90^\circ$ threshold. 

We conclude that gravitational tides control the initial stages of the evolution, and that thermal atmospheric tides are only important in the definition of the final equilibrium rotation.
Core-mantle friction only becomes important in the slow rotation regime ($\omega \sim n$), but for very slow rotations it rules over both tidal effects.

\subsection{Earth-sized planets in the Habitable Zone}

Since most existing terrestrial planets are very close to their parent stars and/or orbit low-mass stars, usually there is only one final equilibrium rotation rate, given by state $\omega_1^+$ (Fig.\,\ref{ecc}, Table\,\ref{tbl:1}).
In the examples from Figures~\ref{EvolucaoExoplanets1-1} and \ref{EvolucaoExoplanets25-1}, only for HD\,40703\,$f$ two final states are possible, but yet they are very close to each other and hardly distinguishable.
As a consequence, in the simulations from previous section, we were unable to observe evolutionary paths leading to very different configurations, such as retrograde rotation.

In section~\ref{App.Venus} we investigated the most favorable orbital parameters that lead to multi-final states configurations.
We saw that this correspond to planets not very close to the star (Eq.\,\ref{130422a}), which may coincide with the HZ of their systems.
Therefore, in order to observe more interesting evolutionary behaviors,
we have also performed simulations for hypothetical Earth-sized planets around a Sun-like star, starting with an orbital period of 2~day.
The final results from the simulations are in agreement with the equilibria states listed in Table\,\ref{tbl:2}.

In Figure~\ref{EvolucaoVenus0.0-2}  we plotted some representative examples for different eccentricity and semi-major axis values. 
Although the initial rotation period is the same in all simulations, the initial ratio $\omega/n$ increases with the semi-major axis (since the mean motion decreases as the planet is moved away from the star).
We still observe that the only possible final obliquities are $\varepsilon =0^\circ$ and $180^\circ$.

For different semi-major axis the main feature is the number of trajectories with increasing or decreasing obliquity. 
For close-in planets ($a < 0.1$~AU) the picture is very similar to the real Earth-sized planets studied in section~\ref{realplanets}, but as the planet moves away from the star (or the stellar mass increase), more obliquities evolve into $180^\circ$.
This is a direct consequence of thermal atmospheric tides, that push the obliquity toward $180^\circ$, and whose relative strength with respect to the gravitational tides increases for distant planets around Sun-like stars.

For a specific semi-major axis value, the global picture does not vary much for different eccentricities. However, the final evolution can be quite different, since the number of final states and their specific values can change completely (Table\,\ref{tbl:2}).
For planets in circular orbits ($e=0.0$) only two final states are possible, symmetrical around synchronous rotation ($\omega = n \pm \omega_s$) (Figs.~\ref{wsn} and \ref{ecc}). 
However, the rotation periods only become substantially different from the synchronous rotation when planets are sufficiently distant from the star, since the effect from thermal atmospheric tides is more significant there.
In particular, one of these states becomes retrograde for $ a \M > 0.59 \, \mathrm{AU}\,M_\odot$ (Eq.\,\ref{130422a}).
Indeed, when $a \M = 0.6 \, \mathrm{AU}\,M_\odot$ (Fig.\ref{EvolucaoVenus0.0-2}j), we can already observe a retrograde rotation state, $2 \pi / \omega_2^- = - 821 $~day, obtained whenever the initial obliquity is higher than $60^\circ$.
The detailed final evolution near $\varepsilon \simeq 0^\circ$ is shown in Figure~\ref{Deatail00}a.
Note, however,
that while for initial $60^\circ \le \varepsilon < 120^\circ $ the obliquity is reduced to zero and the final state is reached through negative rotation, for $\varepsilon \ge 120^\circ$ the obliquity evolves into $180^\circ$ and the rotation rate is stabilized at positive rotation of 821~day.
These final evolution scenarios are very similar to the case of Venus  ($a=0.72$~AU), for which the presently observed rotation is also retrograde \citep{Correia_Laskar_2001,Correia_Laskar_2003I}.

\begin{figure}[t!]
\begin{center}
\begin{tabular}{c}
\begin{overpic}[angle=-90, width=0.95\columnwidth]{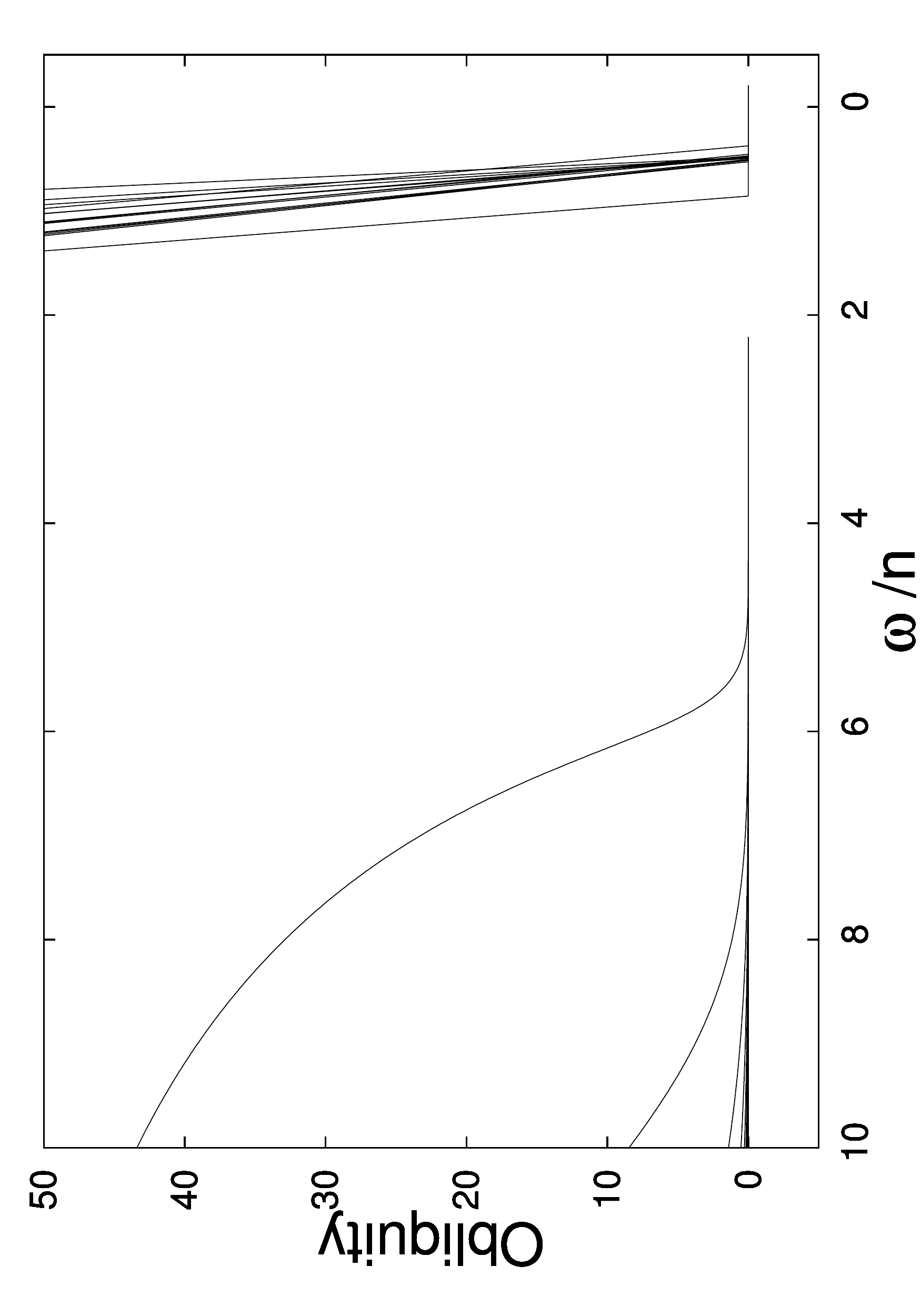}\put(37,145){\bf \Large (a)}\end{overpic} \\
\begin{overpic}[angle=-90, width=0.95\columnwidth]{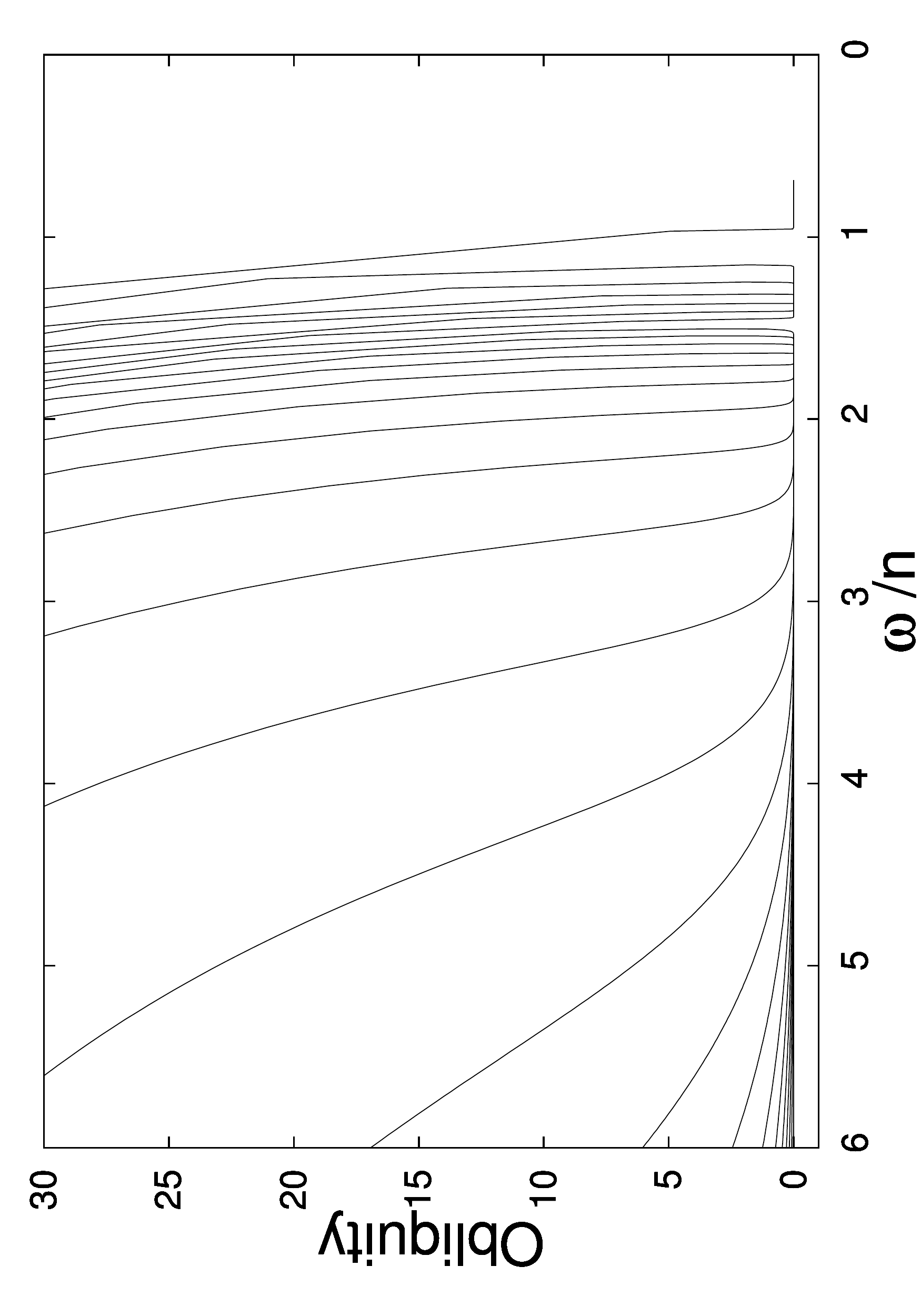}\put(37,145){\bf \Large (b)}\end{overpic} 
\end{tabular}
\caption{\label{Deatail00}Detail of the final obliquity evolution with the rotation rate for  \textbf{(a)} $a=0.60$~AU, $e= 0.0$, and \textbf{(b)} $a=0.42$~AU, $e=0.2$. 
Two and three final equilibrium rotation states can be distinguish, respectively.}
\end{center}
\end{figure}

For non-circular orbits, the final evolution of the planets become more interesting.
For instance, when $e=0.1 $, the simulation with $a = 0.42$~AU (Fig.~\ref{EvolucaoVenus0.0-2}e) shows three final possibilities for the spin (all prograde):
for initial obliquities $\varepsilon \le 40^\circ$ trajectories go to the final state $2 \pi / \omega_2^+ = 64 $~day; for $40^\circ < \varepsilon < 100^\circ$ they go to the final state $2 \pi / \omega_1^+ = 68 $~day, and for $\varepsilon \ge 100^\circ$ they go to the final state $2 \pi / \omega_1^- = 176 $~day.

Similarly, for $e=0.2$ and the same semi-major axis $a = 0.42$~AU (Fig.~\ref{EvolucaoVenus0.0-2}f), we still have the same three final states, but with slightly different periods. 
In Figure~\ref{Deatail00}b we show the detailed final evolution near $\varepsilon = 0^\circ$ in this case, where the three distinct equilibria can be  observed.
For $e=0.2$ with slightly higher semi-major axis or stellar mass ($a \M = 0.5$~AU~M$_\odot$) (Fig.~\ref{EvolucaoVenus0.0-2}i), we still observe three final states, but the final state $\omega_1^-$ is replaced by $\omega_2^-$, while for $a \M = 0.6$~AU~M$_\odot$ (Fig.~\ref{EvolucaoVenus0.0-2}l) we are back to two final states (the $\omega_1^+$ final state disappears). 

When we look at hypothetical Earth-sized planets with three possible final states,
we  observe that their sequence is always the same: lower values of initial obliquities correspond
 to the $\omega_2^+$ state,  intermediate initial values correspond to the $\omega_1^+$ state, and higher initial obliquities 
to the $\omega_1^-$ or $\omega_2^-$ final states. 
The final equilibrium of the planet depends on the value of the rotation rate  $\omega/n$ when the obliquity reaches 0$^\circ$. 
Indeed, in the final evolution zooms shown in Figure~\ref{Deatail00} we observe that
the rotation of the planet  evolves into to the equilibrium state that is closer to the rotation rate when it reaches $\varepsilon=0^\circ$.


\section{Discussion and Conclusion}
  
In this paper we have studied the long-term spin evolution of Earth-sized exoplanets.
Our study includes not only the commonly-used gravitational tides, but also thermal atmospheric tides and core-mantle friction.
In addition to previous works, we also considered the effect of the obliquity, and the effect of non-circular orbits with moderate eccentricity ($e < 0.3$).

Our model contains some uncertain parameters related to the dissipation within
the planets, but we can gather all this information in a single
parameter, $\omega_s$, which is a measurement of the relative strength between gravitational and thermal tides.
Therefore, by varying this parameter, we can cover all possibilities for the rotation of Earth-sized planets. 
For a planet with moderate eccentricity and low obliquity, at most four final
equilibrium positions are possible. 
For eccentricities higher than $ e \sim 0.5 $,
terms of higher degree in $e$ should be considered that may generate additional equilibrium positions, not included in the present study.

We have shown that gravitational tides control much of the evolution, in particular for initially fast rotating rates ($\omega \gg n $).
However, when the planet enters in a slow rotating regime ($\omega \sim n$), core-mantle friction drives the final evolution of the obliquity, that can only be stabilized at $0^\circ$ or $180^\circ$.
Thermal tides can then play an important role in determining the final equilibrium rotation rate, depending on $\omega_s / n$.
This ratio increases rapidly with the semi-major axis and mass of the star because $ \omega_s /n \propto (a \M)^{2.5} $. 
Thus, the effect of the atmosphere on the spin is more
pronounced for planets that orbit Sun-like stars at not very close distances.

The already discovered Earth-sized planets (Table\,\ref{tbl:1}) are mainly close-in planets around low-mass stars, since the radial velocity technique and transits are more sensitive to
the detection of short-period planets.
As a consequence, for these planets, the effect of atmospheric tides is extremely small with respect to the gravitational tides ($ \omega_s / n \sim 0 $). Indeed, the amplitude of the thermal tides 
varies as $(1/a^5)$ (Eqs.~\ref{eq:variacaospin1}$-$\ref{eq:p2}, \ref{eq:psigma}), while the gravitational tides amplitude varies as $(1/a^6)$ (Eqs.\,\ref{eq:variacaospinL}$-$\ref{130613a}). For close-in planets, the equilibrium rotation is thus essentially driven by the
 gravitational tidal torque.
Only few of these planets can be stabilized with a rotation rate $\omega < n $, because for $ e > 0.1 $ these final states only exist if $ a \M > 0.2 $\,AU\,$M_\odot$ (Fig.\,\ref{ecc}).

As the detection techniques improve, we expect that in the near future many Earth-sized planets will be found further away from the central star.
These planets are more interesting for habitability studies, since the surface temperature may sustain liquid water as on Earth.
However, thermal atmospheric tides also become more important for this special category of planets, so they cannot be neglected when we inspect their habitability.
They prevent the planet to evolve into synchronous rotation, which can help to redistribute the stellar flux over the surface, but on the other hand they can also develop life-unfriendly environments such as the retrograde rotation observed for Venus.

In future works, more complete tidal models should be tested \citep[e.g.][]{Efroimsky_Williams_2009, Remus_etal_2012b, Ferraz-Mello_2013}, as well as the effect from collisions \citep[e.g.][]{Correia_Laskar_2012} and spin-orbit resonances  \citep[e.g.][]{Correia_Laskar_2010}.

\bigskip
\textbf{ Acknowledgments.} The authors wish to thank Anthony Dobrovolskis and the referees Michael Efroimsky and Nader Haghighipour, who made very valuable suggestions.
We acknowledge support by PNP-CNRS, CS of Paris Observatory,
PICS05998 France-Portugal program, and Funda\c{c}\~ao
para a Ci\^encia e a Tecnologia, Portugal (PEst-C/CTM/LA0025/2011).

\bigskip

\parskip=0pt

\bibliographystyle{aa}
\bibliography{diana_spin}


\onecolumn
\begin{small}
\begin{center}
\begin{longtable}{lcccccc|cccccc}
\caption{Characteristics and equilibrium rotation rates of Earth-sized planets with masses lower than $10 \, M_{\oplus}$ (see text for notations). \label{tbl:1}} \\
\hline
Name & $\M$ & Age & $^* \tau_{eq}$ & $ m \sin i $ & $a$ & $ e $ &
 $ \omega_s / n $ & $ 2 \pi /n $ &$ 2 \pi / \omega^-_2 $ & $ 2 \pi / \omega^-_1 $ &
  $ 2 \pi / \omega^+_1 $ & $ 2 \pi / \omega^+_2 $ \\
 & [$M_\odot$] & [Gyr] & [Gyr] & $[m_\oplus]$ & [AU]  & & &
 [day] & [day]  & [day]  & [day] & [day] \\
\hline
\endfirsthead
\multicolumn{13}{c}%
{\tablename\ \thetable\ -- \textit{Continued from previous page}} \\
\hline
Name & $\M$ & Age & $^* \tau_{eq}$ & $ m \sin i $ & $a$ & $ e $ &
 $ \omega_s / n $ & $ 2 \pi /n $ &$ 2 \pi / \omega^-_2 $ & $ 2 \pi / \omega^-_1 $ &
  $ 2 \pi / \omega^+_1 $ & $ 2 \pi / \omega^+_2 $ \\
 & [$M_\odot$] & [Gyr] & [Gyr] & $[m_\oplus]$ & [AU]  & & &
 [day] & [day]  & [day]  & [day] & [day] \\
\hline
\endhead
\hline \multicolumn{13}{r}{\textit{Continued on next page}} \\
\endfoot
\hline
\endlastfoot

\noalign{\smallskip}
Venus & 1.00 & 4.5 & 2.3 & 0.82 & 0.723 & 0.007 &
1.9255 & 224.7 &$-$242.9   &  &   & 76.8 \\


Koi-1843\,b $^{1)}$ & 0.46 & --- & $10^{-11}$ & 0.44 & 0.0048& 0.0 &
$10^{-6}$ & 0.179 &   &  0.179 &  0.179  &  \\

Koi-55\,b $^{2)}$ & 0.5 & 0.02 & $10^{-12}$ & 0.44 & 0.0060 & 0.0 &
$10^{-6}$ & 0.241 &   &  0.241 &  0.241  &  \\

Koi-55\,c $^{2)}$& 0.5 & 0.02 & $10^{-11}$ & 0.66 & 0.0076 & 0.0 &
$10^{-6}$ & 0.345 &   &  0.345 &  0.245  &  \\

Koi-2700\,b $^{3)}$ & 0.632 & --- & $10^{-9}$ & 0.86 & 0.0158 & 0.0 &
$10^{-5}$ & 0.912 &   &  0.913 &  0912  &  \\

Kepler-42\,d $^{4,5)}$ & 0.158 & --- & $10^{-7}$ & 0.95 & 0.0154 & 0.0 &
$10^{-6}$ & 1.756 &   &  1.756 &  1.756  &  \\

$\alpha$ Cen B b $^{6)}$& 0.93 &--- & $10^{-7}$ & 1.1 & 0.04 & 0.0 &
$0.0009$ & 3.024 &   &  3.026 &  3.021  &  \\

Kepler-307\,c $^{7)}$ & 0.98 & --- & $10^{-6}$ & 1.5   & 0.108 & 0.0 &
$0.0101$ & 13.095 &   &  13.229 &  12.964  &  \\

Kepler-65\,d $^{8,9)}$& 1.25 & 2.9 & $10^{-6}$ & 1.7 & 0.084 & 0.0 &
0.0091 & 7.954 &  & 8.026 & 7.882 &  \\

Kepler-78\,b $^{10,11)}$ & 0.81 & --- & $10^{-11}$ & 1.9   & 0.0089 & 0.0 &
$10^{-5}$ & 0.341 &   &  0.341&  0.341  &  \\

Kepler-11 b $^{5,12,13)}$& 0.95 & $6-10$ & $10^{-6}$ & 1.9 & 0.091 & 0.0 &
0.0051 & 10.287 &   &  10.340 & 10.235  &  \\

Kepler-42\,c $^{4,5)}$& 0.13 &--- & $10^{-10}$ & 1.9& 0.0060 & 0.0 &
$10^{-7}$ & 0.427 &   &  0.427 &  0.427  &  \\
 
GJ\,581\,e $^{14)}$& 0.31 &$7-9$ & $10^{-7}$ & 1.9
& 0.028 & 0.32 &
$10^{-5}$ & 3.074 &   &           & 2.281 &   \\

Kepler-177\,b $^{7)}$ & 1.07 & --- & $10^{-4}$ & 2.0 & 0.2217 & 0.0 &
$0.0618$ & 36.86 &   &  39.288 &  34.714  &  \\

Kepler-11\,f $^{5,12,13)}$& 0.95 & $6-10$ & $10^{-3}$ & 2.0 & 0.2495 & 0.0 &
0.0617 & 46.703 &  & 49.773 & 43.989  &  \\

 HD20794\,c $^{15)}$& 0.70 & 5.8 & $10^{-3}$ & 2.4 & 0.2036 & 0.0 &
0.0152& 40.106 &  & 40.723 & 39.508  &  \\

HD20794\,b $^{15)}$& 0.70 & 5.8 & $10^{-4}$ & 2.7 & 0.1207 & 0.0 &
0.0038& 18.307 &  & 18.376 & 18.238  &  \\

GJ\,667C\,e $^{16)}$  & 0.33 &  $>2$ & $10^{-2}$ & 2.7 & 0.213 & 0.02 &
$0.0024$ & 62.5043 &   &         & 62.208 &   \\

GJ\,667C\,f $^{16)}$  & 0.33 &  $>2$ & $10^{-3}$ & 2.7 & 0.156 & 0.03 &
$0.0011$ & 39.176 &   &         & 38.926 &   \\

Koi-111\,c $^{8)}$& 0.796 & --- & $10^{-4}$ & 2.7 & 0.15 & 0.0 &
$0.0090$ & 23.739 &  & 23.955 & 23.527  &  \\

Koi-117\,b $^{8)}$& 1.142 & --- & $10^{-7}$ & 2.8 & 0.1044 & 0.0 &
$0.001$ & 3.155 &  & 3.158 & 3.151  &  \\

Kepler-42\,b $^{4,5)}$& 0.13 &--- & $10^{-8}$ & 2.9 & 0.0116 & 0.0 &
$10^{-7}$ & 1.148 &   &  1.148 &  1.148  &  \\

Koi-82\,c $^{8)}$& 0.799 & --- & $10^{-5}$ & 2.9 & 0.086 & 0.0 &
$0.0021$ & 10.306 &  & 10.328 & 10.284  &  \\

Kepler-11\,c $^{5,12,13)}$& 0.95 & $6-10$ & $10^{-6}$ & 2.9 & 0.106 & 0.0 &
$0.0055$ & 12.933 &  & 13.005 & 12.862  &  \\

Koi-115\,d $^{8)}$ & 1.105 & --- & $10^{-6}$ & 3.0 & 0.075 & 0.0 &
$0.0033$ & 7.137 &   &  7.161 &  7.113  &  \\

Kepler-307\,b $^{7)}$& 0.98 & --- & $10^{-6}$ & 3.1 & 0.0927 & 0.0 &
$0.0004$ & 10.414 &   &  10.456 &  10.371  &  \\

Kepler-20\,e $^{17)}$& 0.912 &8.8 & $10^{-6}$ & 3.1 & 0.0507 & 0.0 &
$10^{-3}$ & 4.366 &   &  4.37 &  4.363  &  \\

HD\,40307\,e $^{18)}$  & 0.77 & 4.5  &$10^{-3}$ & 3.5
& 0.1886& 0.15 &
0.0121            & 34.093&   &           & 30.364&   \\

HD\,85512\,b $^{15)}$  & 0.69 & 5.6  & $10^{-2}$ & 3.5
& 0.26 & 0.11 &
2.048             & 58.295&    &          & 53.858 &   \\

Koi-82\,d $^{8)}$& 0.799 & --- & $10^{-5}$ & 3.8 & 0.067 & 0.0 &
$0.0009$ & 7.087 &  & 7.093 & 7.08  &  \\

Kepler-114\,d $^{7)}$& 0.56 & --- & $10^{-6}$ & 3.9 & 0.0835 & 0.0 &
0.0007 & 11.777 &   &  11.785 & 11.769  &  \\

HD\,40307\,b $^{18)}$   & 0.77  & 4.5  & $10^{-7}$ & 4.0
& 0.047 & 0.2 &
0.0003            & 4.214 &    &          & 3.557 &   \\

Kepler-62\,c $^{19)}$& 0.69 & $3-11$ & $10^{-3}$ & 4.0 & 0.0929 & 0.0 &
$0.0014$ & 12.451 &   &  12.468 &  12.433  &  \\

Kepler-79\,e $^{20)}$& 1.165 & 3.4 & $10^{-3}$ & 4.1 & 0.386 & 0.012 &
$0.1816$ & 81.155 &  & 99.056 & 68.663  &  \\

HD156668 b $^{21)}$& 0.772 & $4-13$ & $10^{-6}$ & 4.2 & 0.0500 & 0.0 &
$0.0004$ & 4.648 &   &  4.65 &  4.46  &  \\

Kepler-36\,b $^{22)}$& 1.071 & $6-8$ & $10^{-4}$ & 4.5 & 0.1153 & 0.0 &
0.0068 & 13.818 &   &  13.912 & 13.725  &  \\

GJ\,676A\,d $^{23)}$  & 0.71 & --- & $10^{-7}$ & 4.4 & 0.0413 & 0.15 &
$0.0002$ & 3.638 &   &         & 3.2612 &   \\

GJ\,667C\,c $^{16)}$  & 0.33 &  $>2$ & $10^{-3}$ & 4.5 & 0.123 & 0.27 &
$0.0004$ & 27.428 &   &         & 21.381 &   \\


Kepler-10 b$^{24)}$ & 0.895 & $7-16$ & $10^{-9}$ & 4.6 & 0.0168 & 0.0 &
$10^{-5}$ & 0.863 &   &  0.863 & 0.863  &  \\

GJ\,667C\,g $^{16)}$  & 0.33 &  $>2$ & $6.39$ & 4.6 & 0.549 & 0.08 &
$0.0172$ & 258.64 &   &         & 245.92 &   \\

 HD20794 d $^{15)}$& 0.70 & 5.8 & $10^{-1}$ & 4.8 & 0.3499 & 0.0 &
0.0355& 90.357 &  & 93.679 & 87.262  &  \\

 Kepler-68 c $^{25)}$& 1.079 & $5-8$ & $10^{-4}$ & 4.8 & 0.0906 & 0.0 &
 0.0036 &9.587 &   &  9.622 &9.553  &  \\
 
Koi-115\,c $^{8)}$& 1.105 & --- & $10^{-7}$ & 5.1 & 0.062 & 0.0 &
0.0014 & 5.364 &   &  5.372 & 5.357  &  \\

61\,Vir\,b $^{26)}$  & 0.95 &  $6-12$ & $10^{-7}$ & 5.1
& 0.050 & 0.12 &
0.0006            & 4.215 &    &          & 3.909 &   \\

GJ\,667C\,d $^{16)}$  & 0.33 &  $>2$ & $10^{-1}$ & 5.1 & 0.276 & 0.03 &
$0.0029$ & 92.192 &   &         & 91.445 &   \\

HD\,40307\,f $^{18)}$   & 0.77  & 4.5  & $10^{-2}$ & 5.2
& 0.247 & 0.02 &
0.0178            &51.097 &    & 51.894  & 50.095 &   \\

GJ\,581\,c $^{14)}$  & 0.31 & $7-9$ & $10^{-5}$ &
5.4 & 0.073 & 0.07 &
$10^{-4}$ & 12.9349&   &   & 12.581 &   \\

Kepler-57\,c $^{27)}$& 0.83 &---& $10^{-5}$ & 5.4 & 0.1 & 0.0 &
0.0022 &12.678 &   &  12.706 &12.651  &  \\

GJ\,667C\,b $^{16)}$  & 0.33 &  $>2$ & $10^{-6}$ & 5.6 & 0.0505 & 0.13 &
$10^{-5}$ & 7.216 &   &         & 6.62 &   \\

GJ\,433\,b $^{28)}$  & 0.48 &  --- & $10^{-6}$ & 5.8
& 0.058 & 0.08 &
0.0001            & 7.364&    &           & 7.103 &   \\

 HD1461 c $^{29)}$ & 1.08 &--- & $10^{-5}$ & 5.9 & 0.1117 & 0.0 &
 0.0052 &13.121 &   &  13.189 &13.053  &  \\

Kepler-305\,c $^{7)}$& 0.76 &---& $10^{-6}$ & 6.0 & 0.0732 & 0.0 &
0.0007 & 8.298 &   &  8.304 & 8.292  &  \\

GJ\,581\,d $^{14)}$  & 0.31 & $7-9$ & $10^{-2}$& 6.0 
& 0.22 & 0.25 &
0.0012           & 67.692&   &             & 53.918 &   \\

Kepler-350\,c $^{7)}$& 1.00 &---& $10^{-5}$ & 6.1 & 0.1337 & 0.0 &
0.0066 &17.856&   &  17.974 &17.74  &  \\

Kepler-92\,c $^{7)}$& 1.21 &---& $10^{-4}$ & 6.1 & 0.1864 & 0.0 &
0.0242 &26.722&   &  27.386 & 26.09 &  \\

GJ\,1214\,b $^{30)}$  & 0.15 & $3-9$ &10$^{-9}$ & 6.3 & 0.0141 & 0.27 &
$10^{-7} $ & 1.581 &  &   & 1.232 &   \\

Kepler-87\,c $^{31)}$& 1.05 &---& $10^{-2}$ & 6.4 & 0.664 & 0.039 &
0.3934 &192.864&   &  313.124 & 138.723 &  \\

HD\,215497\,b $^{32)}$  & 0.87 & $<7$  & $10^{-7}$ & 6.4 & 0.047 & 0.16 &
0.0003  & 3.990 &  &      & 3.534 &   \\

Kepler-114\,b $^{7)}$& 0.56 &---& $10^{-6}$ & 6.5 & 0.048 & 0.0 &
0.0001 & 5.133&   &  5.133 & 5.132 &  \\

Koi-82\,b $^{8)}$& 0.799 &---& $10^{-5}$ & 6.6 & 0.116 & 0.0 &
0.0025 &16.144&   &  16.184 & 16.104 &  \\

HD\,40307\,c $^{18)}$ & 0.77 & 4.5  & $10^{-5}$ &6.6 & 0.0799 & 0.06 &
0.0009& 9.401 &  &       & 9.2 &   \\

Gl\,876\,d $^{33)}$  & 0.334 & $0.1-5$ &10$^{-8}$ & 6.7 & 0.0208 & 0.207 &
$10^{-6} $ & 1.897 &  &   & 1.588 &   \\

 Kepler-18 b $^{34)}$& 0.972 & $8-12$ & $10^{-7}$ & 6.9 & 0.0447 & 0.0 &
0.0004 & 3.501 &   &  3.503 & 3.5  &  \\

GJ\,3634\,b $^{35)}$  & 0.45 & --- &10$^{-7}$ & 7.0 & 0.0287 & 0.08 &
$1.723 $ & 2.647 &  &   & 2.554 &   \\

 Kepler-9 d $^{36,37)}$ & 1.0 &--- & $10^{-8}$ & 7.0 & 0.0273 & 0.0 &
0.0001 & 1.648 &   &  1.648 & 1.647  &  \\

HD\,40307\,g $^{18)}$  & 0.77 & 4.5  & 2.16 & 7.1 & 0.6 & 0.29 &
0.131& 193.452 &  &      & 148.585 &   \\

 Kepler-50 c $^{27)}$& 1.23 &--- & $10^{-6}$ & 7.1 & 0.0932 
 & 0.0 &
0.004 & 9.371 &   &  9.408 & 9.333  &  \\

GJ\,163\,c $^{35)}$  & 0.4 & $1-10$ &10$^{-3}$ & 7.3 & 0.1254 & 0.0094 &
$0.0005 $ & 25.645 &  &   & 25.619 &   \\

 Kepler-11\,d $^{5,12,13),*}$& 0.95 & $6-10$ & $10^{-4}$ & 7.3 & 0.159 & 0.0 &
0.0078 & 23.759 &   &  23.946 & 23.575  &  \\

CoRoT-7 b $^{5,12,38}$& 0.93 & $1-2$ & $10^{-9}$ & 7.4 & 0.0172 & 0.0 &
$10^{-5}$ &0.854 &   &  0.854 &0.854  &  \\

Kepler-177\,c $^{7)}$& 1.07 &---& $10^{-4}$ & 7.5 & 0.2695 & 0.0 &
0.0386 &249.402&   &  51.383 & 47.567 &  \\

HD\,181433\,b $^{39)}$  & 0.78 &--- & $10^{-5}$ & 7.6 & 0.08 & 0.396 &
0.0008 & 9.358 &   &   & 6.535 &   \\

HD\,1461\,b $^{29)}$ & 1.08 & 6.3 & $10^{-6}$ & 7.6 & 0.0634 & 0.14 &
0.0011 & 5.616 &   &   & 5.090 &   \\

Kepler-50\,b $^{27)}$ & 1.23 &--- & $10^{-6}$ & 7.6 & 0.826 
 & 0.0 &
0.0028 & 7.818 &   &  7.840 & 7.796  &  \\

HD\,97658\,b $^{5,12,40),*}$ & 0.85 & $3-11$  & $10^{-6}$ & 7.9 & 0.0797 & 0.13 &
0.0012  & 8.914 &  &      & 8.172 &   \\

 Kepler-11 e $^{5,12,13),*}$& 0.95 & $6-10$ & $10^{-4}$ & 8.0 & 0.194 & 0.0 &
0.01120 & 32.021 &   &  32.410 & 31.641  &  \\

 Kepler-36 c $^{22)}$& 1.071 & $6-8$ & $10^{-5}$ & 8.1 & 0.1283 & 0.0 &
0.0057 & 16.22 &   &  16.313 & 16.127  &  \\

 Kepler-68 b $^{25)}$& 1.079 & $5-8$ & $10^{-6}$ & 8.3 & 0.0617& 0.0 &
0.0009 & 5.389 &   &  5.394 & 5.384  &  \\

55\,Cnc\,e $^{41)}$  & 0.905 & $8-13$ & $10^{-10}$ &8.4 & 0.0156 & $<0.06$ &
$10^{-5}$ & 0.758 &   &   & 0.733&   \\

 CoRoT-7 c $^{42)}$& 0.93 & $1-2$ & $10^{-7}$ & 8.4 & 0.046 & 0.0 &
0.0003 & 3.737 &   &  3.738 & 3.736  &  \\

 GJ176 b $^{43)}$& 0.5 &--- & $10^{-5}$ & 8.4 & 0.066 & 0.0 &
0.0002 & 8.758 &   &  8.76 & 8.757  &  \\

 BD-061339 b $^{44)}$& 0.7 & $0.4-8$ & $10^{-7}$ & 8.5 & 0.0428 & 0.0 &
0.0001 & 3.866 &   &  3.866 & 3.865  &  \\

Kepler-20\,b $^{45)}$  & 0.912 & $7-12$ & $10^{-7}$ &8.6 & 0.0454 & $<0.32$ &
0.0003 & 3.696 &   &   & 2.743&   \\

HD\,41248\,c $^{46)}$& 0.92 &$<5$ & $10^{-3}$ & 8.6 & 0.172 & 0.0 &
0.0078 & 27.164 &   &  27.377 & 26.954  &  \\

Kepler-62 b $^{19)}$& 0.69 & $3-11$ & $10^{-6}$ &9.0 & 0.0553 & 0.0 &
0.0002 & 5.718 &   &  5.719 & 5.717  &  \\

HD\,7924\,b $^{47)}$  & 0.83 & --- & $10^{-6}$ &9.3 & 0.057 & 0.17 &
0.0004& 5.449 &  &   & 4.769 &   \\

HD\,40307\,d $^{18)}$ & 0.77 & 4.5  &  $10^{-4}$  &9.5 & 0.1321 & 0.07 &
0.0024& 19.985 &  &      & 19.393 &   \\

HD\,69830\,b $^{48)}$  & 0.86 & $4-10$ & $10^{-5}$ &10.2& 0.079 & 0.1 &
0.0008 & 8.6625 &   &   & 8.1995 &   \\

Kepler-21\,b $^{49)}$& 1.340 & $2-3$ & $10^{-7}$ & 10.5 & 0.0425 & 0.0 &
0.0005 & 2.765 &   &  2.767 & 2.764  &  \\

Kepler-89\,b $^{50)}$& 1.277 & 3.2 & $10^{-7}$ & 10.5 & 0.05 & 0.25 &
0.0007 & 3.614 &   &        & 2.879  &  \\

$\mu$\,Ara\,c $^{51)}$  & 1.08 & 6.4 & $10^{-5}$ &10.5 & 0.0909 & 0.172 &
0.0020 & 9.639 &   &   & 8.408 &   \\

%

%
%
%
%

 \noalign{\smallskip}
 \hline
 \noalign{\smallskip}

\multicolumn{13}{l}{$^*$ Using Eq.(\ref{eq:tau_eq}) with $ k_2 = 1/3 $ and $ \Delta t_g = 640 $\,s (Earth's values). } \\ 
\multicolumn{13}{l}{References:
\textbf{1)} \citet{Rappaport_2013a}; \textbf{2)}\citet{Charpinet_2011};\textbf {3)} \citet{Rappaport_2013b}; 
\textbf{4)} \citet{Muirhead_2012};} \\
\multicolumn{13}{l}{\textbf{5)} \citet{Callegari_2013}; \textbf {6)} \citet{Dumusque_2012};
\textbf{7)} \citet{Xie_2013}; \textbf{8)}\citet{Hadden_2013}; \textbf {9)} \citet{Chaplin_2013};}\\ 
\multicolumn{13}{l}{\textbf{10)} \citet{Sanchis_2013}; \textbf{11)} \citet{Pepe_etal_2013};\textbf {12)} \citet{Lopez_2013};
\textbf{13)} \citet{Lissauer_2011}; \textbf{14)}\citet{Forveille_etal_2011};}\\
\multicolumn{13}{l}{\textbf {15)} \citet{Pepe_etal_2011}; \textbf{16)} \citet{Anglada_2013}; \textbf{17)}\citet{Fressin_2012};
 \textbf {18)} \citet{Tuomi_etal_2013};\textbf{19)} \citet{Borucki_2013};}\\ 
\multicolumn{13}{l}{\textbf{20)} \citet{Jontof_2013}; \textbf {21)} \citet{Howard_2011};
\textbf{22)} \citet{Carter_2012}; \textbf{23)} \citet{Anglada_2012}; 
\textbf {24)} \citet{Batalha_2011};}\\
\multicolumn{13}{l}{\textbf{25)} \citet{Gilliland_2013}; \textbf{26)} \citet{Vogt_etal_2010}
\textbf{27)} \citet{Steffen_2013}; \textbf {28)} \citet{Delfosse_etal_2012};
\textbf{29)} \citet{Rivera_etal_2010b};}\\
\multicolumn{13}{l}{\textbf{30)} \citet{Harpsoe_etal_2013}; \textbf {31)} \citet{Ofir_2013};
\textbf{32)} \citet{LoCurto_etal_2010}; \textbf{33)} \citet{Rivera_etal_2010}; \textbf {34)} \citet{Cochran_2011};}\\
\multicolumn{13}{l}{\textbf{35)} \citet{Bonfils_etal_2011};  \textbf{36)} \citet{Holman_2010}; 
\textbf {37)} \citet{Torres_2011};\textbf {38)} \citet{Leger_2009};
\textbf{39)} \citet{Bouchy_etal_2009a};}\\
\multicolumn{13}{l}{\textbf {40)} \citet{Henry_etal_2011};  \textbf{41)} \citet{Endl_etal_2012}; 
\textbf {42)} \citet{Queloz_2009}; \textbf {43)} \citet{Forveille_2009};
\textbf{44)} \citet{LoCurto_2013};}\\
\multicolumn{13}{l}{\textbf {45)} \citet{Gautier_etal_2012}; \textbf{46)} \citet{Jenkins_2013}; 
\textbf {47)} \citet{Howard_etal_2009b};\textbf {48)} \citet{Lovis_etal_2006};
\textbf{49)} \citet{Howell_2012};}\\
\multicolumn{13}{l}{\textbf {50)} \citet{Weiss_2013}; \textbf {51)} \citet{Pepe_etal_2007}.}
\end{longtable}
\end{center}
\end{small}


\begin{table*}
\centering
\caption[]{Characteristics and equilibrium rotation rates of Venus-like planets
orbiting a Sun-like star ($M_* = M_\odot$).}
\label{tbl:2}
\begin{tabular}{cc|cccccc}
\hline
\noalign{\smallskip}
 $a \M$ & $ e $ &
 $ \omega_s / n $ & $ 2 \pi /n $ & $ 2 \pi / \omega^-_1 $ &
 $ 2 \pi / \omega^-_2 $ & $ 2 \pi / \omega^+_1 $ & $ 2 \pi / \omega^+_2 $ \\
 
 [AU $M_\odot$]  & & &
 [day] & [day]  & [day]  & [day] & [day] \\
 
\hline
\noalign{\smallskip}
0.1 & 0 &
0.014 & 11.550 & 11.711  &  &  11.394 &  \\
0.2 & 0 &
0.077 & 32.67 & 35.411  &  &  30.322 &  \\
0.3 & 0 &
0.213 & 60.018 & 76.291  &  &  49.466 &  \\
0.42 & 0 &
0.495 & 99.419 & 196.749  &  &  66.515 &  \\
0.5 & 0 &
0.765 & 129.138 &   & 549.406 &   & 73.168 \\
0.6 & 0 &
1.207 & 169.756 &   & $-$821.407 &   & 76.929 \\
0.72 & 0 &
1.903 & 223.15 &   & $-$247 &   & 76.857 \\
0.1 & 0.1 &
0.014 & 11.550 &   &  &  10.826 &  \\
0.2 & 0.1 &
0.077 & 32.67 & 33.378  &  &  29.209 &  \\
0.3 & 0.1 &
0.213 & 60.018 & 71.154 &  &  48.857 &  \\
0.42 & 0.1 &
0.495 & 99.419 & 176.455  &  &  68.275 & 64.600  \\
0.5 & 0.1 &
0.765 & 129.138 &   & 418.391 &   & 71.625 \\
0.6 & 0.1 &
1.207 & 169.756 &   & $-$1382.504 &   & 75.973 \\
0.72 & 0.1 &
1.903 & 223.15 &   & $-$277.77 &   & 76.552 \\
0.1 & 0.2 &
0.014 & 11.550 &  &  &  9.709 &  \\
0.2 & 0.2 &
0.077 & 32.67 &  &  &  26.928 &  \\
0.3 & 0.2 &
0.213 & 60.018 & 61.583 &  &  47.506 &  \\
0.42 & 0.2 &
0.495 & 99.419 & 142.529  &  &  72.714 & 60.575  \\
0.5 & 0.2 &
0.765 & 129.138 &   & 269.876 & 88.025 & 68.301 \\
0.6 & 0.2 &
1.207 & 169.756 &   & 2398.654 &   & 73.856 \\
0.72 & 0.2 &
1.903 & 223.15 &   &$-$389.844 &   & 75.859 \\

 \noalign{\smallskip}
 \hline
 \noalign{\smallskip}
 \end{tabular}


\end{table*}

\end{document}